\documentclass[12pt,a4paper]{article}
\usepackage{amssymb}

\usepackage{graphicx}
\usepackage{amsmath}
\usepackage{flafter}

\input{tcilatex}

\begin{document}

\title{Review of analytical treatments of barrier-type problems in plasma theory}
\author{F. Spineanu$^{1,2}$, M. Vlad$^{1,2}$, K. Itoh$^{1}$, S.-I. Itoh$^{3}$ \\
$^{1}$ National Institute for Fusion Science\\
322-6 Oroshi-cho, Toki-shi, Gifu-ken 509-5292, Japan\\
$^{2}$ Association EURATOM-MECT Romania\\
NILPRP, P.O.Box MG-36 Magurele, Bucharest, Romania\\
$^{3}$ Research Institute for Applied Mechanics, Kyushu University, \\
Kasuga 816-8580, Japan}
\maketitle

\begin{abstract}
We review the analytical methods of solving the stochastic equations for
barrier-type dynamical behavior in plasma systems. The path-integral
approach is examined as a particularly efficient method of determination of the
statistical properties.
\end{abstract}

\tableofcontents

\section{Introduction}

A large number of physical systems, in particular in plasma physics, are
described in terms of a variable evolving in a deterministic velocity field
under the effect of a random perturbation. This is described by a stochastic
differential equation of the form 
\begin{equation}
\overset{\cdot }{x}=V\left( x\right) +\sqrt{2D}\xi \left( t\right) 
\label{eqcanoi}
\end{equation}
where $V\left( x\right) $ is the velocity field and the perturbation is the
white noise 
\begin{equation}
\left\langle \xi \left( t\right) \xi \left( t^{\prime }\right) \right\rangle
=\delta \left( t-t^{\prime }\right)   \label{csicor}
\end{equation}

This general form has been invoked in several applications. More recently,
in a series of works devoted to the explanation of the intermittent behavior
of the statistical characteristics of the turbulence in magnetically
confined plasma it has been developed a formalism based on barrier crossing.
Previous works that have discussed subcritical excitation of plasma
instabilities are Refs. \cite{Itoh1}, \cite{Itoh2}, \cite{Itoh21}, \cite{Itoh3}, \cite{Itoh4}, \cite{Itoh5}, \cite{Itoh6}, \cite{Itoh7}, \cite{ItohCuNoi}.

We offer a comparative presentation of two functional integral approaches to the
determination of the statistical properties of the system's variable $%
x\left( t\right) $ for the case where the space dependence of $V\left(
x\right) $ is characterized by the presence of three equilibrium points, $%
V\left( x\right) =0$. We will take 
\begin{equation}
V\left( x\right) =ax-bx^{3}\;,\text{with}\;a,b>0  \label{barrier}
\end{equation}

\section{The functional approach}

\subsection{Overview of the functional methods}

In studying the stochastic processes the functional methods can be very
useful and obtain systematic results otherwise less accessible to
alternative methods. The method has been developed initially in quantum
theory and it is now a basic instrument in condensed matter, field theory,
statistical physics, etc. In general it is based on the formulation of the
problem in terms of an action functional. There are two distinct advantages
from this formulation: (1) the system's behavior appears to be determined by all
classes of trajectories that extremize the action and their contributions
are summed after appropriate weights are applied; (2) the method naturally
includes the contributions from states close to the extrema, so that
fluctuations can be accounted for.

There are technical limitations to the applicability of this method. In the
statistical problems (including barrier-type problems) it is simpler to
treat cases with white noise, while colored noise can be treated
perturbatively. In the latter case, the procedure is however useful since
the diagrammatic series can be formulated systematically.

The colored noise can be treated by extending the space of variables: the
stochastic variable with finite correlation is generated by integration of a
new, white noise variable.

The one dimensional version can be developed up to final explicit result. Since
however the barrier type problem is frequently formulated in two-dimensions,
one has to look for extrema of the action and ennumerate all possible
trajectories. It is however known that, in these cases, the behavior of the
system is dominated by a particular path, ``the optimum escape path'', and a
reasonnable approximation is to reduce the problem to a one-dimensional one
along this system's trajectory.

\subsection{The path integral with the MSR action}

We will briefly mention the steps of constructiong the MSR action
functional, in the Jensen path integral reformulation. We begin by choosing
a particular realisation of the noise $\xi \left( t\right) $. All the
functions and derivatives can be discretised on a lattice of points in the
time interval $\left[ -T,T\right] $ (actually one can take the limits to be $%
\pm \infty $). The solution of the equation (\ref{eqcanoi}) is a
``configuration'' of the field $x\left( t\right) $ which can be seen as a
point in a space of functions. We extend the space of configurations $%
x\left( t\right) $ to this space of functions, including all possible forms
of $x\left( t\right) $, not necessarly solutions. In this space the solution
itself will be individualised by a functional Dirac $\delta $ function. 
\begin{equation*}
\delta \left( \overset{\cdot }{x}-V\left( x\right) -\sqrt{2D}\xi \left(
t\right) \right)
\end{equation*}
Any functional of the system's real configuration (\emph{i.e.} solutions of
the equations) can be formally expressed by taking as argument an arbitrary
functional variable, multiplying by this $\delta $ functional and
integrating over the space of all functions.

We will skip the discretization and the Fourier representation of the $%
\delta $ functions, followed by reverting to the continuous functions. The
result is the following functional 
\begin{equation*}
\emph{Z}_{\xi }=\int \emph{D}\left[ x\left( t\right) \right] \emph{D}\left[
k\left( t\right) \right] \exp \left[ i\int_{-T}^{T}dt\left( -k\overset{\cdot 
}{x}+kV\left( x\right) +\sqrt{2D}k\xi \right) \right]
\end{equation*}
The label $\xi $ means that the functional is still defined by a choice of a
particular realization of the noise. The generating functional is obtained
by averaging over $\xi $. 
\begin{equation}
\emph{Z}=\left\langle \emph{Z}_{\xi }\right\rangle =\int \emph{D}\left[
x\left( t\right) \right] \emph{D}\left[ k\left( t\right) \right] \exp \left[
i\int dt\left( -k\overset{\cdot }{x}+kV\left( x\right) +iDk^{2}\right) %
\right]  \label{zfaraj}
\end{equation}
We add a formal interaction with two currents 
\begin{equation}
\emph{Z}_{J}=\int \emph{D}\left[ x\left( t\right) \right] \emph{D}\left[
k\left( t\right) \right] \exp \left[ i\int dt\left( -k\overset{\cdot }{x}%
+kV\left( x\right) +iDk^{2}+J_{1}x+J_{2}k\right) \right]  \label{zj}
\end{equation}
in view of future use to the determination of correlations. This functional
integral must be determined explicitely. The standard way to proceed to the
calculation of $\emph{Z}_{J}$ is to find the saddle point in the function
space and then expand the action around this point to include the
fluctuating trajectories. This requires first to solve the Euler-Lagrange
equations 
\begin{equation}
\left\{ 
\begin{array}{l}
\overset{\cdot }{k}=-ak+3bkx^{2}-J_{1} \\ 
\overset{\cdot }{x}=ax-bx^{3}+2iDk+J_{2}
\end{array}
\right.  \label{eqEL}
\end{equation}
The simplest case should be examined first. We assume there is no
deterministic velocity ($a=b=0$) in order to see how the purely diffusive
behavior is obtained in this framework 
\begin{eqnarray*}
\overset{\cdot }{x}-2iDk &=&J_{2} \\
\overset{\cdot }{k} &=&-J_{1}
\end{eqnarray*}
The equations can be trivially integrated \cite{Flmadi1}, \cite{Flmadi2}
\begin{eqnarray*}
x^{\left( 0\right) }\left( t\right) &=&\int_{-T}^{T}dt\Delta _{11}\left(
t,t^{\prime }\right) J_{1}\left( t^{\prime }\right) +\int_{-T}^{T}dt\Delta
_{12}\left( t,t^{\prime }\right) J_{2}\left( t^{\prime }\right) \\
k^{\left( 0\right) }\left( t\right) &=&\int_{-T}^{T}dt\Delta _{21}\left(
t,t^{\prime }\right) J_{1}\left( t^{\prime }\right) +\int_{-T}^{T}dt\Delta
_{22}\left( t,t^{\prime }\right) J_{2}\left( t^{\prime }\right)
\end{eqnarray*}
where 
\begin{eqnarray*}
\Delta _{11}\left( t,t^{\prime }\right) &=&2iD\left[ t\Theta \left(
t^{\prime }-t\right) +t^{\prime }\Theta \left( t-t^{\prime }\right) \right]
\\
\Delta _{12}\left( t,t^{\prime }\right) &=&\Theta \left( t-t^{\prime }\right)
\\
\Delta _{21}\left( t,t^{\prime }\right) &=&\Theta \left( t^{\prime }-t\right)
\\
\Delta _{22}\left( t,t^{\prime }\right) &=&0
\end{eqnarray*}
with the symmetry 
\begin{equation*}
\Delta _{ij}\left( t,t^{\prime }\right) =\Delta _{ji}\left( t^{\prime
},t\right)
\end{equation*}

The lowest approximation to the functional integral $\emph{Z}_{J}$ is
obtained form this saddle point solution, by calculating the action along
this system's trajectory. We insert this solutions in the expression of the
generating functional, for $V\left( x\right) \equiv 0$ 
\begin{eqnarray*}
\emph{Z}_{J}^{\left( 0\right) } &=&\left. \int \emph{D}\left[ x\left(
t\right) \right] \emph{D}\left[ k\left( t\right) \right] \exp \left[
i\int_{-T}^{T}dt\left( -k\overset{\cdot }{x}+iDk^{2}+J_{1}x+J_{2}k\right) %
\right] \right| _{x^{\left( 0\right) },k^{\left( 0\right) }} \\
&=&\exp \left[ \frac{1}{2}i\int_{-T}^{T}dt\int_{-T}^{T}dt^{\prime
}J_{i}\left( t\right) \Delta _{ij}\left( t,t^{\prime }\right) J_{j}\left(
t\right) \right]
\end{eqnarray*}
The dispersion of the stochastic variable $x\left( t\right) $ can be
obtained by a double functional derivative followed by taking $J_{i}\equiv 0$%
. We obtain 
\begin{eqnarray*}
\left\langle x\left( t\right) x\left( t^{\prime }\right) \right\rangle
&=&\left. \frac{1}{\emph{Z}_{J}^{\left( 0\right) }}\frac{\delta ^{2}}{\delta
J_{1}\left( t\right) \delta J_{1}\left( t^{\prime }\right) }\exp \left[ 
\frac{1}{2}i\int_{-T}^{T}dt\int_{-T}^{T}dt^{\prime }J_{i}\left( t\right)
\Delta _{ij}\left( t,t^{\prime }\right) J_{j}\left( t\right) \right] \right|
_{J_{1,2}=0} \\
&=&D\min \left( t,t^{\prime }\right)
\end{eqnarray*}
which is the diffusion. The same mechanism will be used in the following,
with the difference that the equations cannot be solved in explicit form due
to the nonlinearity.

In general the nonlinearity can be treated by perturbation expansion, if the
amplitude can be considered small. This is an analoguous procedure as that
used in the field theory and leads to a series of terms represented by
Feynman diagrams. We can separate in the Lagrangian the part that can be
explicitely integrated and make a perturbative treatment for the
non-quadratic term. This is possible when we assume a particular
(polynomial) form of the deterministic velocity, $V\left( x\right) $.
Obviously, this term is $k\left( t\right) x\left( t\right) ^{3}$ in Eq.(\ref
{zj}). The functional integral can be written, taking account of this
separation 
\begin{equation}
\emph{Z}_{J}=\exp \left[ i\int_{0}^{T}dt\left( -b\right) \frac{\delta }{%
i\delta J_{2}\left( t\right) }\frac{\delta }{i\delta J_{1}\left( t\right) }%
\frac{\delta }{i\delta J_{1}\left( t\right) }\frac{\delta }{i\delta
J_{1}\left( t\right) }\right] \emph{Z}_{J}^{\left( q\right) }  \label{zjnl}
\end{equation}
where the remaining part in the Lagrangian is \emph{quadratic} 
\begin{equation}
\emph{Z}_{J}^{\left( q\right) }=\int \emph{D}\left[ x\left( t\right) \right] 
\emph{D}\left[ k\left( t\right) \right] \exp \left[ i\int dt\left( -k%
\overset{\cdot }{x}+akx+iDk^{2}+J_{1}x+J_{2}k\right) \right]  \label{Zq}
\end{equation}
The Euler-Lagrange equations are 
\begin{eqnarray*}
\overset{\cdot }{k} &=&-ak-J_{1} \\
\overset{\cdot }{x} &=&ax+2iDk+J_{2}
\end{eqnarray*}
The solutions can be expressed as follows 
\begin{eqnarray}
x\left( t\right) &=&x_{0}\exp \left( at\right) +\int_{0}^{T}dt^{\prime
}\Delta _{1j}\left( t,t^{\prime }\right) J_{j}\left( t^{\prime }\right)
\label{solxk} \\
k\left( t\right) &=&\int_{0}^{T}dt^{\prime }\Delta _{2j}\left( t,t^{\prime
}\right) J_{j}\left( t^{\prime }\right)  \notag
\end{eqnarray}
with 
\begin{eqnarray}
\Delta _{11}\left( t,t^{\prime }\right) &=&\frac{iD}{\left( -a\right) }\exp
\left( at\right)  \label{soldeltas} \\
&&\times \left\{ \left[ \exp \left( -2at\right) -1\right] \Theta \left(
t^{\prime }-t\right) +\left[ \exp \left( -2at^{\prime }\right) -1\right]
\Theta \left( t-t^{\prime }\right) \right\}  \notag \\
&&\times \exp \left( at^{\prime }\right)  \notag \\
\Delta _{21}\left( t,t^{\prime }\right) &=&\exp \left( -at\right) \Theta
\left( t^{\prime }-t\right) \exp \left( at^{\prime }\right)  \notag \\
\Delta _{12}\left( t,t^{\prime }\right) &=&\exp \left( at\right) \Theta
\left( t-t^{\prime }\right) \exp \left( -at^{\prime }\right)  \notag \\
\Delta _{22}\left( t,t^{\prime }\right) &=&0  \notag
\end{eqnarray}
The form of the generating functional derived from the quadratic part is 
\begin{equation}
\emph{Z}_{J}^{\left( q\right) }=\exp \left\{ i\int_{0}^{T}dtx_{0}\exp \left(
at\right) J_{1}\left( t\right) +\frac{i}{2}\int_{0}^{T}dt\int_{0}^{T}dt^{%
\prime }J_{i}\left( t\right) \Delta _{ij}\left( t,t^{\prime }\right)
J_{j}\left( t^{\prime }\right) \right\}  \label{zjq}
\end{equation}
The occurence of the first term in the exponent is the price to pay for not
making the expansion around $x_{0}$. However, such an expansion would have
produced two non-quadratic terms in the Lagrangian density: $\left(
-3bx_{0}\right) \varepsilon ^{2}k$ and $\left( -b\right) \varepsilon ^{3}k$.
This would render the perturbative expansion extremly complicated since we
would have to introduce two vertices : one, of order four, is that shown in
Eq.(\ref{zjnl}) and another, of order three, related to the first of the
nonlinearities mentioned above.

Even in the present case, the calculation appears very tedious. We have to
expand the vertex part of $\emph{Z}_{J}$ as an exponential, in series of
powers of the vertex operator. In the same time we have to expand the
exponential in Eq.(\ref{zjq}) as a formal series. Then we have to apply term
by term the first series on the second series. The individual terms can be
represented by diagrams. In this particular case we have a finite
contribution even at the zero-loop order (the ``tree'' graph). It is however
much more difficult to extract the statistics since we will need at least
the diagrams leaving two free ends with currents $J_{1}$.

In the case we examine here, the perturbative treatment is not particularly
useful since the form of the potential (from which the velocity field is
obtained) supports topologically distinct classes of saddle point solutions
and this cannot be represented by a series expansion.

\bigskip

\subsection{The Onsager-Machlup functional}

To make comparison with other approaches, we take $J_{1,2}=0$ and integrate
over the functional variable $k$. 
\begin{eqnarray*}
\emph{Z} &=&\int \emph{D}\left[ x\left( t\right) \right] \emph{D}\left[
k\left( t\right) \right] \exp \left\{ \int_{0}^{T}dt\left[ -Dk^{2}+i\left( -k%
\overset{\cdot }{x}+kV\right) \right] \right\} \\
&=&\int \emph{D}\left[ x\left( t\right) \right] \emph{D}\left[ k\left(
t\right) \right] \exp \left\{ -\int_{0}^{T}dt\left[ \sqrt{D}k+\frac{i%
\overset{\cdot }{x}-iV}{2\sqrt{D}}\right] ^{2}+\int_{0}^{T}dt\frac{\left( i%
\overset{\cdot }{x}-iV\right) ^{2}}{4D}\right\}
\end{eqnarray*}
\begin{equation*}
\emph{Z}=N_{1}\int \emph{D}\left[ x\left( t\right) \right] \exp \left\{ -%
\frac{1}{4D}\int_{0}^{T}dt\left( \overset{\cdot }{x}-V\right) ^{2}\right\}
\end{equation*}
In other notations 
\begin{equation*}
\emph{Z}=N_{1}\int \emph{D}\left[ x\left( t\right) \right] \exp \left[ -%
\frac{S}{D}\right]
\end{equation*}
where 
\begin{equation*}
S=\int_{0}^{T}dt\frac{1}{4}\left( \overset{\cdot }{x}-V\right) ^{2}
\end{equation*}
This is Eq.(25) of the reference Lehmann, Riemann and H\"{a}nggi,
PRE62(2000)6282. In this reference it is called the Onsager-Machlup action
functional and the analysis is based on this formula.

However, we can go further and we will find inconsistencies. We now take
account of the fact that the velocity is derived form a potential 
\begin{eqnarray*}
V\left[ x\left( t\right) \right] &=&-\frac{dU\left[ x\left( t\right) \right] 
}{dx} \\
&\equiv &-U^{\prime }\left[ x\left( t\right) \right]
\end{eqnarray*}
\begin{eqnarray*}
S &=&\int_{0}^{T}dt\frac{1}{4}\left( \overset{\cdot }{x}^{2}+2\overset{\cdot 
}{x}U^{\prime }+U^{\prime 2}\right) \\
&=&\int_{0}^{T}dt\left[ \frac{1}{4}\left( \overset{\cdot }{x}^{2}+U^{\prime
2}\right) +\frac{1}{2}\frac{dU}{dt}\right] \\
&=&\frac{1}{2}\left[ U\left( T\right) -U\left( 0\right) \right]
+\int_{0}^{T}dt\left[ \frac{1}{4}\left( \overset{\cdot }{x}^{2}+U^{\prime
2}\right) \right]
\end{eqnarray*}
This leads to the form of the generating functional 
\begin{equation*}
\emph{Z}=\frac{\exp \left[ -\frac{U\left( T\right) }{2D}\right] }{\exp \left[
-\frac{U\left( 0\right) }{2D}\right] }K\left( x,t;x_{i},t_{i}\right)
\end{equation*}
with 
\begin{equation*}
K\left( x,t;x_{i},t_{i}\right) =\int \emph{D}\left[ x\left( t\right) \right]
\exp \left( -\frac{S}{D}\right)
\end{equation*}
\begin{equation}
S=\int_{0}^{T}dt\left[ \frac{1}{4}\left( \overset{\cdot }{x}^{2}+U^{\prime
2}\right) \right]  \label{sfara}
\end{equation}
These are almost identical to the formulas (2-5) of the reference \cite
{Weiss1} (except that $\varepsilon \rightarrow 2D$). Also, it is quite close
of the Eqs.(7a-7c) of the ref.\cite{Caroli}.

\textbf{However there is an important difference}.

There is a term missing in Eq.(\ref{sfara}) which however is present in the
two above references. The full form of the action $S$, instead of Eq.(\ref
{sfara}) is 
\begin{equation*}
S=\int_{0}^{T}dt\left( \frac{1}{4}\overset{\cdot }{x}^{2}+W\right)
\end{equation*}
\begin{equation*}
W=\frac{U^{\prime 2}}{4}-\frac{D}{2}U^{\prime \prime }
\end{equation*}

This term comes from the \emph{Jacobian} that is hidden in the functional $%
\delta $ integration.

\subsection{Connection between the MSR formalism and Onsager-Machlup}

In our approach the most natural way of proceeding with a stochastic
differential equation is to use the MSR type reasonning in the Jensen
reformulation. The equation is discretized in space and time and selectd
with $\delta $ functions in an ensemble of functions (actually in sets of
arbitrary numbers at every point of discretization). The result is a
functional integral. There is however a particular aspect that needs careful
analysis, as mentioned in the previous Subsection. It is the problem of the Jacobian associated with the $\delta $
functions. This problem is discussed in Ref.\cite{DeDoPel}.

The equation they analyse is presented in most general form as 
\begin{equation*}
\frac{\partial \phi _{j}\left( t\right) }{\partial t}=-\left( \Gamma
_{0}\right) _{jk}\frac{\delta H}{\delta \phi _{k}\left( t\right) }+V_{j}%
\left[ \phi \left( t\right) \right] +\theta _{j}
\end{equation*}
where the number of stochastic equations is $N$ , $H$ is functional of the
fields, $V_{j}$ is the streaming term which obeys a current-conserving type
relation 
\begin{equation*}
\frac{\delta }{\delta \phi _{j}}V_{j}\left[ \phi \right] \exp \left\{ -H%
\left[ \phi \right] \right\} =0
\end{equation*}
The noise is $\theta _{j}$.

The following generating functional can be written 
\begin{equation*}
Z_{\theta }=\int \emph{D}\left[ \phi _{j}\left( t\right) \right] \exp \int dt%
\left[ l_{j}\phi _{j}\left( t\right) \right] \prod_{j,t}\delta \left( \frac{%
\partial \phi _{j}\left( t\right) }{\partial t}+K_{j}\left[ \phi \left(
t\right) \right] -\theta _{j}\right) J\left[ \phi \right]
\end{equation*}
the functions $l_{j}\left( t\right) $ are currents, 
\begin{equation*}
K_{j}\left[ \phi \left( t\right) \right] \equiv -\left( \Gamma _{0}\right)
_{jk}\frac{\delta H}{\delta \phi _{k}\left( t\right) }+V_{j}\left[ \phi
\left( t\right) \right]
\end{equation*}
and $J\left[ \phi \right] $ is the Jacobian associated to the Dirac $\delta $
functions in each point of discretization.

The Jacobian can be written 
\begin{equation*}
J=\det \left[ \left( \delta _{jk}\frac{\partial }{\partial t}+\frac{\delta
K_{j}\left[ \phi \right] }{\delta \phi _{k}}\right) \delta \left(
t-t^{\prime }\right) \right]
\end{equation*}
Up to a multiplicative constant 
\begin{equation*}
J=\exp \left( Tr\ln \left[ \left( \frac{\partial }{\partial t}+\frac{\delta K%
}{\delta \phi }\right) \frac{\delta \left( t-t^{\prime }\right) }{\frac{%
\partial }{\partial t}\delta \left( t-t^{\prime }\right) }\right] \right)
\end{equation*}
or 
\begin{equation*}
J=\exp \left( Tr\ln \left[ 1+\left( \frac{\partial }{\partial t}\right) ^{-1}%
\frac{\delta K\left( t\right) }{\delta \phi \left( t^{\prime }\right) }%
\right] \right)
\end{equation*}
Since the operator $\left( \frac{\partial }{\partial t}\right) ^{-1}$ is
retarded, only the lowest order term survives after taking the trace 
\begin{equation*}
J=\exp \left[ -\frac{1}{2}\int dt\frac{\delta K_{j}\left[ \phi \left(
t\right) \right] }{\delta \phi _{j}\left( t\right) }\right]
\end{equation*}
The factor $1/2$ comes from value of the $\Theta $ function at zero.

In the treatment which preserves the dual function $\widehat{\phi }$
associated to $\phi $ in the functional, there is a part of the action 
\begin{equation*}
\widehat{\phi }K\left[ \phi \right]
\end{equation*}
Then a $\widehat{\phi }$ and a $\phi $ of the same coupling term from $%
\widehat{\phi }K\left[ \phi \right] $ close onto a loop.Since $G_{\widehat{%
\phi }\phi }$ is retarded, all these contributions vanish except the one
with a single propagator line. This cancels exactly, in all orders, the part
coming from the Jacobian.

Then it is used to ignore all such loops and together with the Jacobian.

\bigskip

We can now see that in our notation this is precisely the term needed in the
expression of the action. 
\begin{eqnarray*}
\phi \left( t\right) &\rightarrow &x\left( t\right) \\
K_{j}\left[ \phi \left( t\right) \right] &\rightarrow &U^{\prime }\left[
x\left( t\right) \right] \\
\frac{\delta K_{j}\left[ \phi \left( t\right) \right] }{\delta \phi
_{j}\left( t\right) } &\rightarrow &-U^{\prime \prime }\left[ x\left(
t\right) \right]
\end{eqnarray*}
and the action (\ref{sfara}) is completed with the new term 
\begin{equation*}
\int_{0}^{T}dt\left( -\frac{D}{2}U^{\prime \prime }\right)
\end{equation*}
Now the generating functional is 
\begin{equation*}
\emph{Z}=\frac{\exp \left[ -\frac{U\left( T\right) }{2D}\right] }{\exp \left[
-\frac{U\left( 0\right) }{2D}\right] }K\left( x,t;x_{i},t_{i}\right)
\end{equation*}
with 
\begin{equation*}
K\left( x,t;x_{i},t_{i}\right) =\int \emph{D}\left[ x\left( t\right) \right]
\exp \left( -\frac{S}{D}\right)
\end{equation*}
and 
\begin{equation}
S=\int_{0}^{T}dt\left( \frac{1}{4}\overset{\cdot }{x}^{2}+W\right)
\label{news}
\end{equation}
\begin{equation}
W=\frac{U^{\prime 2}}{4}-\frac{D}{2}U^{\prime \prime }  \label{w}
\end{equation}

Now the two expressions are identical with those in the references cited.
This will be the starting point of our analysis.

In conclusion we have compared the two starting points in a functional
approach: The one that uses \emph{dual functions} $x\left( t\right) $ and $%
k\left( t\right) $, closer in spirit to MSR; And the approach based on
Onsager-Machlup functional, traditionally employed for the determination of
the probabilities \cite{Caroli}, \cite{Weiss1}. Either we keep $k\left(
t\right) $ and ignore the Jacobian (the first approach) or integrate over $%
k\left( t\right) $ and include the Jacobian. The approaches are equivalent
and, as we will show below, lead to the same results.

A final observation concerning the choice of one or another method: in the
MSR method, the trajectories include the diffusion from the direct solution
of the Euler-Lagrange equations. In the Onsager-Machlup method the paths
extremizing the action are deterministic and the diffusion is introduced by
integrating on a neighborhood in the space of function, around the
deterministic motion.

\section{The transition solutions (instantons)}

\subsection{Numerical trajectories}

The equations for the saddle point trajectory are in complex so we extend
also the variable in complex space 
\begin{eqnarray*}
x &\rightarrow &\left( x_{R},x_{I}\right) \\
k &\rightarrow &\left( k_{R},k_{I}\right)
\end{eqnarray*}
The obtain a system of four nonlinear ordinary differential equations which
can be integrated numerically.

A typical form of the solution $x\left( t\right) $ is similar to the \emph{%
kink instanton} (\emph{i.e.} the $\tanh $ function). The function $x\left(
t\right) $ spends very much time in the region close to the equilibrium
point; then it performs a fast transition to the neighbour equilibrium
point, where it remains for the rest of the time interval. 
\begin{figure}[tbh]
\centering
\includegraphics[width=10cm]{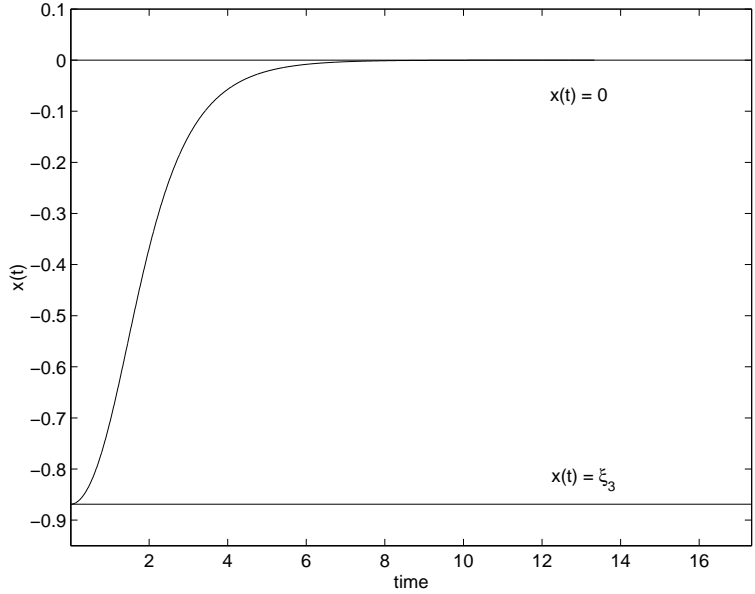}
\caption{The kink-like instanton connecting a position close to the left
(stable) equilibrium point to the middle $x=0$ (unstable) equilibrium point
in the effective potential $-W(x)$}
\label{FigKink1}
\end{figure}
\begin{figure}[tbh]
\centering
\includegraphics[width=10cm]{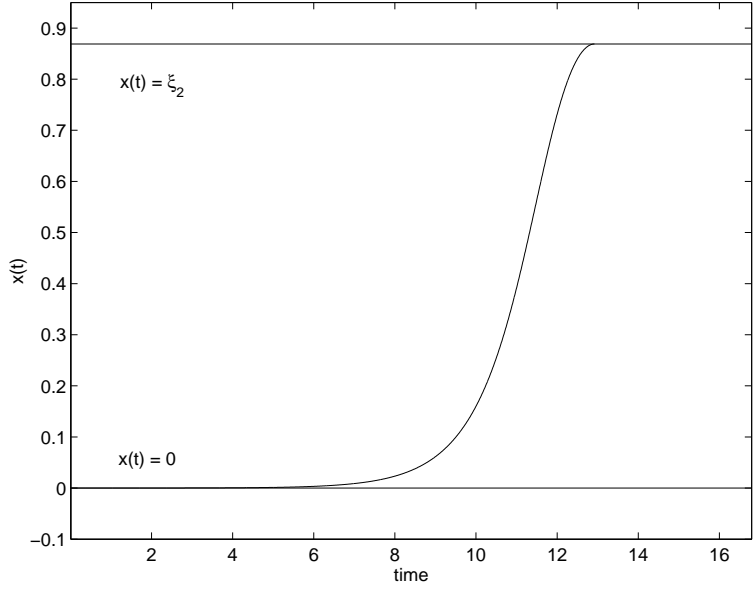}
\caption{The kink-like instanton connecting the middle $x=0$ (unstable)
equilibrium point with a position close to the right (stable) equilibrium
point to in the effective potential $-W(x)$}
\label{FigKink2}
\end{figure}

\subsection{Elliptic functions instantons (from Onsager-Machlup action)}

The action functional Eqs.(\ref{news}) and (\ref{w}) leads to the following
differential equation (which replaces Eqs.(\ref{eqEL})) 
\begin{equation*}
\frac{1}{2}\overset{\cdot \cdot }{x}=\left( \frac{3b^{2}}{2}\right)
x^{5}+\left( -2ab\right) x^{3}+\left( \frac{a^{2}}{2}-3Db\right) x
\end{equation*}
Multiplying by $\overset{\cdot }{x}$ and integrating we have 
\begin{equation}
\overset{\cdot }{x}=\pm \left[ b^{2}x^{6}+\left( -2ab\right) x^{4}+\left(
a^{2}-6Db\right) x^{2}+c_{1}\right] ^{1/2}  \label{eqxpoint}
\end{equation}
We are interested in the functions $x\left( t\right) $ that has the
following physical property: they stay for very long time stuck to the
equilibrium points and perform a fast jump between them at a certain moment
of time. Then we can take $c_{1}=0$. The solution can be obtained form the
integration 
\begin{equation*}
\int_{x}^{x_{q}}\frac{d\xi }{\xi \sqrt{\xi ^{4}+\left( -2a/b\right) \xi
^{2}+a^{2}/b^{2}-6D/b}}=\pm bt+c_{2}
\end{equation*}
The upper limit $x_{q}$ will be specified later. For the next calculation it
will be taken as the smallest of the roots of the polynomial under the
square root.

The details of the calculations in terms of elliptic functions can be found
in \emph{Byrd and Friedman} \cite{Byrd}.

The roots of the forth degree polynomial will be noted 
\begin{equation*}
\xi ^{4}+\left( -2a/b\right) \xi ^{2}+a^{2}/b^{2}-6D/b=\left( \xi _{1}-\xi
\right) \left( \xi _{2}-\xi \right) \left( \xi _{3}-\xi \right) \left( \xi
_{4}-\xi \right)
\end{equation*}
where 
\begin{equation*}
\begin{array}{cc}
\xi _{1}=\sqrt{\alpha _{1}+\alpha _{2}} & \xi _{3}=-\sqrt{\alpha _{1}-\alpha
_{2}} \\ 
\xi _{2}=\sqrt{\alpha _{1}-\alpha _{2}} & \xi _{4}=-\sqrt{\alpha _{1}+\alpha
_{2}}
\end{array}
\end{equation*}
such as to have $\xi _{1}>\xi _{2}>\xi _{3}>\xi _{4}$; then we will use $%
x_{q}\equiv \xi _{4}$. The notations are 
\begin{equation*}
\alpha _{1}\equiv a/b\;,\;\alpha _{2}\equiv \sqrt{6D/b}
\end{equation*}
The following substitutions are required 
\begin{eqnarray*}
sn^{2}u &=&\frac{\left( \xi _{1}-\xi _{3}\right) \left( \xi _{4}-\xi \right) 
}{\left( \xi _{1}-\xi _{4}\right) \left( \xi _{3}-\xi \right) } \\
k^{2} &=&\frac{\left( \xi _{2}-\xi _{3}\right) \left( \xi _{1}-\xi
_{4}\right) }{\left( \xi _{1}-\xi _{3}\right) \left( \xi _{2}-\xi
_{4}\right) } \\
g &=&\frac{2}{\sqrt{\left( \xi _{1}-\xi _{3}\right) \left( \xi _{2}-\xi
_{4}\right) }}
\end{eqnarray*}
a new variable is introduced identifying the lower limit of the integral, $%
x\rightarrow u_{1}$ 
\begin{eqnarray*}
\alpha ^{2} &=&\frac{\xi _{1}-\xi _{4}}{\xi _{1}-\xi _{3}}>1 \\
\varphi &=&am\,u_{1}=\arcsin \sqrt{\frac{\left( \xi _{1}-\xi _{3}\right)
\left( \xi _{4}-x\right) }{\left( \xi _{1}-\xi _{4}\right) \left( \xi
_{3}-x\right) }} \\
sn\,u_{1} &=&\sin \varphi
\end{eqnarray*}
The integral can be written 
\begin{eqnarray*}
&&\int_{x}^{\xi _{4}}\frac{d\xi }{\xi \sqrt{\left( \xi _{1}-\xi \right)
\left( \xi _{2}-\xi \right) \left( \xi _{3}-\xi \right) \left( \xi _{4}-\xi
\right) }} \\
&=&\frac{g}{\xi _{4}}\int_{0}^{u_{1}}\frac{1-\alpha ^{2}sn^{2}\left(
u\right) }{1-\frac{\xi _{3}\alpha ^{2}}{\xi _{4}}sn^{2}\left( u\right) }du
\end{eqnarray*}
This integral can be expressed in terms of \emph{elliptic} functions. We
take $p\equiv \frac{\xi _{3}\alpha ^{2}}{\xi _{4}}$%
\begin{eqnarray*}
&&\int_{0}^{u_{1}}\frac{1-\alpha ^{2}sn^{2}\left( u\right) }{%
1-p\,sn^{2}\left( u\right) }du \\
&=&\frac{1}{p^{6}}\left[ \alpha ^{6}u+3\alpha ^{4}\left( p^{2}-\alpha
^{2}\right) V_{1}+3\alpha ^{2}\left( p^{2}-\alpha ^{2}\right)
^{2}V_{2}+\left( p^{2}-\alpha ^{2}\right) ^{3}V_{3}\right]
\end{eqnarray*}
Here the notations are 
\begin{equation*}
V_{1}=\int \frac{du}{1-p^{2}sn^{2}\left( u\right) }=\Pi \left( \varphi
,p^{2},k\right)
\end{equation*}
\begin{eqnarray*}
V_{2} &=&\int \frac{du}{\left( 1-p^{2}sn^{2}\left( u\right) \right) ^{2}} \\
&=&\frac{1}{2\left( p^{2}-1\right) \left( k^{2}-p^{2}\right) }\left[
p^{2}E\left( u\right) +\left( k^{2}-p^{2}\right) u+\right. \\
&&\left. +\left( 2p^{2}k^{2}+2p^{2}-p^{4}-3k^{2}\right) \Pi \left( \varphi
,p^{2},k\right) -\frac{p^{4}sn\left( u\right) \,cn\left( u\right) \,dn\left(
u\right) }{1-p^{2}sn^{2}\left( u\right) }\right]
\end{eqnarray*}
\begin{eqnarray*}
V_{3} &=&\frac{1}{4\left( 1-p^{2}\right) \left( k^{2}-p^{2}\right) }\left[
k^{2}V_{0}+\right. \\
&&+2\left( p^{2}k^{2}+p^{2}-3k^{2}\right) V_{1}+ \\
&&+3\left( p^{4}-2p^{2}k^{2}-2p^{2}+3k^{2}\right) V_{2}+ \\
&&\left. -\frac{p^{4}sn\left( u\right) \,cn\left( u\right) \,dn\left(
u\right) }{\left( 1-p^{2}sn^{2}\left( u\right) \right) ^{2}}\right]
\end{eqnarray*}
and 
\begin{equation*}
V_{0}=\int du=u=F\left( \varphi ,k\right) =\int \frac{d\varphi }{\sqrt{%
1-k^{2}\sin ^{2}\varphi }}
\end{equation*}
The symbol $\Pi \left( \varphi ,p^{2},k\right) \equiv \Pi \left(
u,p^{2}\right) $ represents the Legendre 's incomplete elliptic integral of
the third kind and $\varphi =am\left( u\right) $ is the amplitude of $u$.
The symbols $sn$, $cn$, $dn$ represent the Jacobi elliptic functions.

\subsection{Typical instanton solutions}

There are several well known examples of instantons. They appear in physical
systems whose lowest energy state is degenerate and the minima of the action
functional (or the energy, for stationary solutions) are separated by energy
barriers. Instantons connect these minima by performing transitions which
are only possible in imaginary time (the theory is expressed in Euclidean
space, with uniform positive metric). It is only by including these
instantons that the action functional is correctly calculated and real
physical quantities can be determined.

From this calculation we can obtain the explicit trajectories that extremize
the action functional and in the same time reproduce the jump of the system
between the two distant equilibrium positions. These trajectories will be
necessary in the calculation of the functional integral. However, since we
have eliminated the external currents and integrated over the dual
functional variable $k\left( t\right) $ we cannot derive the statistical
properties of $x\left( t\right) $ from a generating functional.

\subsection{Approximations of the instanton form}

In the approach based on the Onsager-Machlup action the instanton is not
used in its explicit form (elliptic functions) in the calculation of the
action. The reason is that the result can be proved to depend essentially on
local properties of the potential $V\left( x\right) $. This will be shown
later.

In the approach with dual functions, one can reduce the instanton to its
simplest form, an instantaneous transition between two states, a jump
appearing at an arbitrary moment of time. Using this form as a first
approximation we will calculate the solutions of the Euler-Lagrange
equations and then the action.

\section{Fluctuations around the saddle point (instanton) solution}

\subsection{The expansion of the action functional around the extrema}

Using the Onsager-Machlup action we have 
\begin{equation*}
P\left( x,t;x_{i},t_{i}\right) =\exp \left[ -\frac{U\left( x\right) -U\left(
x_{i}\right) }{2D}\right] K\left( x,t;x_{i},t_{i}\right)
\end{equation*}
The new function $K$ has the expression 
\begin{equation}
K\left( x,t;x_{i},t_{i}\right) =\int_{x_{i}}^{x}\emph{D}\left[ x\left( \tau
\right) \right] \exp \left( -\frac{1}{D}\int_{t_{i}}^{t}d\tau \left[ \frac{%
\overset{\cdot }{x}^{2}}{4}+W\left( x\right) \right] \right)  \label{Kcar}
\end{equation}
The integrand at the exponent can be considered as the Lagrangean density
for a particle of mass $1/2$ moving in a potential given by 
\begin{equation*}
\text{potential\ =}\;-W\left( x\right)
\end{equation*}

In a semiclassical treatment (similar to the quantum problem, where $\hbar $
is the \emph{small} diffusion coefficient $D$ of the present problem), the
most important contribution comes from the neighborhood of the classical
trajectories, $x_{c}\left( \tau \right) $ that extremalizes the action $S$.

The ``classical'' equation of motion is 
\begin{eqnarray}
\frac{1}{2}\overset{\cdot \cdot }{x}_{c} &=&\left. \frac{dW\left( x\right) }{%
dx}\right| _{x=x_{c}}  \label{eqcar} \\
x_{c}\left( t_{i}\right) &=&x_{i}\;,\;x_{c}\left( t\right) =x  \notag
\end{eqnarray}
To take into account the trajectories in a functional neighborhood around $%
x_{c}\left( \tau \right) $ we expand the action to second order introducing
the new variables 
\begin{equation*}
y\left( \tau \right) =x\left( \tau \right) -x_{c}\left( \tau \right)
\end{equation*}
This gives 
\begin{eqnarray*}
K\left( x,t;x_{i},t_{i}\right) &=&\exp \left[ -\frac{1}{D}S_{c}\left(
x,t;x_{i},t_{i}\right) \right] \\
&&\times \int_{y\left( t_{i}\right) =0}^{y\left( t\right) =0}\emph{D}\left[
y\left( \tau \right) \right] \exp \left\{ -\frac{1}{D}\int_{t_{i}}^{t}d\tau %
\left[ \frac{1}{4}\overset{\cdot }{y}^{2}+\frac{1}{2}y^{2}W^{\prime \prime
}\left( x_{c}\left( \tau \right) \right) \right] \right\}
\end{eqnarray*}
The deviation of the action from that obtained at the extremum $x_{c}\left(
\tau \right) $, can be rewritten 
\begin{equation*}
\delta S=\frac{1}{2}\int_{t_{i}}^{t}d\tau y\left( \tau \right) \left[ -\frac{%
1}{2}\frac{d^{2}}{d\tau ^{2}}+W^{\prime \prime }\left( x_{c}\left( \tau
\right) \right) \right] y\left( \tau \right)
\end{equation*}
The functional integration can be done since it is Gaussian and the result
is 
\begin{equation*}
\sim \frac{1}{\left[ \det \left( -\frac{1}{2}\frac{d^{2}}{d\tau ^{2}}%
+W^{\prime \prime }\left( x_{c}\left( \tau \right) \right) \right) \right]
^{1/2}}
\end{equation*}
In order to calculate the determinant, one needs to solve the eigenvalue
problem for this operator 
\begin{equation*}
\left[ -\frac{1}{2}\frac{d^{2}}{d\tau ^{2}}+W^{\prime \prime }\left(
x_{c}\left( \tau \right) \right) \right] y_{n}\left( \tau \right) =\lambda
_{n}y_{n}\left( \tau \right)
\end{equation*}
withe the eigenfunctions verifying the conditions 
\begin{equation*}
y_{n}\left( t_{i}\right) =y_{n}\left( t\right) =0
\end{equation*}
\begin{equation*}
\int_{t_{i}}^{t}d\tau y_{n}\left( \tau \right) y_{m}\left( \tau \right)
=\delta _{nm}
\end{equation*}
The formal result for $K$ is (also Van Vleck) 
\begin{equation*}
K\left( x,t;x_{i},t_{i}\right) =N\frac{1}{\left( \prod_{n}\lambda
_{n}\right) ^{1/2}}\exp \left[ -\frac{S_{c}}{D}\right]
\end{equation*}
where $N$ is a constant that will be calculated by normalizing $P$. Another
way to calculate $N$ is to fit this result to the known \emph{harmonic
oscillator} problem.

It has been shown (Coleman) that the factor arising from the determinant can
be written in the form 
\begin{equation*}
N\frac{1}{\left( \prod_{n}\lambda _{n}\right) ^{1/2}}\simeq \frac{1}{\left[
4\pi D\psi \left( t\right) \right] ^{1/2}}
\end{equation*}
where the function $\psi $ is the solution of 
\begin{equation*}
-\frac{1}{2}\frac{d^{2}\psi }{d\tau ^{2}}+W^{\prime \prime }\left(
x_{c}\left( \tau \right) \right) \psi =0
\end{equation*}
with the boundary conditions 
\begin{eqnarray*}
\psi \left( t_{i}\right) &=&0 \\
\left. \frac{d\psi }{d\tau }\right| _{t=t_{i}} &=&1
\end{eqnarray*}

\bigskip

In the case where there are degenerate minima in $W\left( x\right) $ the
particle can travel from one minimum to another. These solutions are called
instantons. Consider for example the potential with two degenerate maxima of 
$-V\left( x\right) $ at $\pm a$ and with a minimum at $x=0$. We want to
calculate the probability $P\left( a,t/2;-a,-t/2\right) $.

The classical solution connecting the point $-a$ to the point $a$ is a \emph{%
kinklike} instanton. The energy of this solution is exponentially small 
\begin{eqnarray*}
E &=&\frac{1}{4}\overset{\cdot }{x}^{2}-W \\
&\simeq &2W_{a}^{\prime \prime }a^{2}\exp \left[ -t\left( 2W_{a}^{\prime
\prime }\right) ^{1/2}\right]
\end{eqnarray*}
This solution spends quasi-infinite time in both harmonic regions around $%
\pm a$ where it has very small velocity; and travels very fast, in a short
time $\Delta t$ between these points (this is the time-width of the
instanton).

\bigskip

The special effect of the translational symmetry in time is seen in the
presence of the parameter representing the center of the instanton. It can
be any moment of time between $-t/2$ and $t/2$. This case must be treated
separately and we note that this corresponds to the lowest eigenvalue in the
spectrum, since the range of variation of the coefficient in the expansion
of any solution in terms of eigenfunction is the inverse of the eigenvalue 
\begin{equation*}
\delta c_{n}=\left( \frac{D}{\lambda _{n}}\right) ^{1/2}
\end{equation*}
The widest interval, for the variation of the center of the instanton, must
be associated with the smallest eigenvalue and this and its eigenfunction
must be known explicitely. Instead of a precise knowledge of the lowest
eigenvalue and its corresponding eigenfunction we will use an approximation,
exploiting the fact the function $\overset{\cdot }{x}_{c}\left( t\right) $
is very close of what we need.

We start by noting that $\overset{\cdot }{x}_{c}\left( t\right) $ is a
solution of the eigenvalue problem for the operator of second order
functional expansion around the instanton. This eigenfunction corresponds to
the eigenvalue $0$%
\begin{equation*}
-\frac{1}{2}\frac{d^{2}}{d\tau ^{2}}\overset{\cdot }{x}_{c}\left( t\right)
+W^{\prime \prime }\left( x_{c}\left( \tau \right) \right) \overset{\cdot }{x%
}_{c}\left( t\right) =0
\end{equation*}
and has boundary conditions 
\begin{equation*}
\overset{\cdot }{x}_{c}\left( \pm \frac{t}{2}\right) \sim \text{exponential
small }
\end{equation*}
very close to $0$, which is be the exact boundary condition we require from
the eigenfunctions of the operator. So the difference between $\overset{%
\cdot }{x}_{c}\left( t\right) $ and the true eigenfunction are very small.
Since $\overset{\cdot }{x}_{c}\left( t\right) $ corresponds to eigenvalue $0$
we conclude that, by continuity, the true eigenfunction will have an
eigenvalue $\lambda _{0}\left( t\right) $ very small, exponentially small.
Then the range of important values of the coefficient $c_{0}$ is very large
and the Gaussian expansion is invalid since the departure of such a solution
from the classical one (the instanton) cannot be considered small.

The degeneracy in the moment of time where the center of the instanton is
placed (\emph{i.e.} the moment of transition) can be solved treating this
parameter as a colective coordinate. The result is 
\begin{eqnarray*}
K\left( a,t/2;-a,-t/2\right) &=&\left[ \frac{\lambda _{0}\left( t\right) }{%
4\pi D\psi \left( t/2\right) }\right] ^{1/2}\int_{-t/2}^{t/2}d\theta \left\{ 
\frac{S\left[ x_{I}\left( \tau -\theta \right) \right] }{4\pi D}\right\}
^{1/2} \\
&&\times \exp \left\{ -\frac{1}{DS\left[ x_{I}\left( \tau -\theta \right) %
\right] }\right\}
\end{eqnarray*}
the parameter $\tau $ in the expression of the instanton solution shows is
the current time variable along the solution that is used to calculate the
action. The integartion is performed on the intermediate transition moment $%
\theta $. We also note that 
\begin{equation*}
S\left[ x_{I}\left( \tau -\theta \right) \right] =\underset{t\rightarrow
\infty }{\lim }S\left[ x_{I}\left( \tau \right) \right] =S_{0}
\end{equation*}
since, except for the very small intervals (approx. the width of the
instanton) at the begining and the end of the interval $\left( -\frac{t}{2},%
\frac{t}{2}\right) $, the value of the action $S$ is not sensitive to the
position of the transition moment. 
\begin{equation*}
K\left( a,t/2;-a,-t/2\right) =\left[ \frac{\lambda _{0}\left( t\right) }{%
4\pi D\psi \left( t/2\right) }\right] ^{1/2}\left( \frac{S_{0}}{4\pi D}%
\right) ^{1/2}t\exp \left( -\frac{S_{0}}{D}\right)
\end{equation*}
The instanton degeneracy introduces a \emph{linear} time dependence of the
probability.

In general, for a function $U\left( x\right) $ that has two minima separated
by a barrier (a maximum) the potential $W\left( x\right) $ calculated form
the action will have three minima and these are not degenerate. The inverse
of this potential, $-W\left( x\right) $ , which is appears in the equation
of motion, will have three maxima in general nondegenerate and the
differences in the values of $-W\left( x\right) $ at these maxima is
connected with the presence of the term containing $D$. Since we assume that 
$D$ is small, the non-degeneracy is also small. The previous discussion in
which the notion of instanton was introduced and $K$ was calculated, take
into consideration the degenerate maxima and the instanton transition at
equal initial and final $W$.

Let us consider the general shape for $-W\left( x\right) $ with three
maxima, at $x=b$, $0$ and $a$. The heigths of these maximas are 
\begin{eqnarray}
\frac{DU_{\alpha }^{\prime \prime }}{2}\;,\;\alpha &=&b,0,a  \label{dupp} \\
U_{b}^{\prime \prime } &>&0,\;U_{a}^{\prime \prime }>0,\;U_{0}^{\prime
\prime }<0  \notag
\end{eqnarray}
This is because the extrema of $-W\left( x\right) $ 
\begin{equation*}
\frac{d\left[ -W\left( x\right) \right] }{dx}=0
\end{equation*}
coincides according to the equation of motion to the pointes where 
\begin{equation*}
\overset{\cdot \cdot }{x}=0\;,\text{or}\;\overset{\cdot }{x}\left( t\right) =%
\text{const}
\end{equation*}
and the constant cannot be taken other value but zero 
\begin{equation*}
\overset{\cdot }{x}\left( t\right) =\text{const}=0
\end{equation*}
Then, since we have approximately that $\overset{\cdot }{x}\approx V\left(
x\right) $ (for small $D$) then we have that at these extrema of $-W\left(
x\right) $ we have $V\left( x\right) =0$ and only the second term in the
expression of $W\left( x\right) $ remains. This justifies Eq.(\ref{dupp}).

It will also be assumed that $U_{b}^{\prime \prime }>U_{a}^{\prime \prime }$.

\subsection{The harmonic region (diffusion around the equilibrium points)}

\subsubsection{Contribution from the trivial fixed point solutions}

We want to calculate, on a Kramers time scale, $\tau _{K}\sim \exp \left(
\Delta U/D\right) $ the probability 
\begin{equation*}
P\left( b,t;b,0\right) =P\left( b,\frac{t}{2};b,-\frac{t}{2}\right) =K\left(
b,\frac{t}{2};b,-\frac{t}{2}\right)
\end{equation*}
We have to find the classical solution of the equation of motion connecting $%
\left( b,-\frac{t}{2}\right) $ with $\left( b,\frac{t}{2}\right) $. This is
the trivial solution, particle sitting at $b$%
\begin{equation*}
x_{c}\left( t\right) =b
\end{equation*}
We have to calculate explicitely the form of the propagator in this case 
\begin{equation*}
K\left( x,t;x_{i},t_{i}\right) =N\frac{1}{\left( \prod_{n}\lambda
_{n}\right) ^{1/2}}\exp \left[ -\frac{S_{c}}{D}\right]
\end{equation*}
We use the formulas given before 
\begin{equation}
N\frac{1}{\left( \prod_{n}\lambda _{n}\right) ^{1/2}}\simeq \frac{1}{\left[
4\pi D\psi \left( t\right) \right] ^{1/2}}  \label{Colem}
\end{equation}
where the function $\psi $ is the solution of 
\begin{equation}
-\frac{1}{2}\frac{d^{2}\psi }{d\tau ^{2}}+W^{\prime \prime }\left(
x_{c}\left( \tau \right) \right) \psi =0  \label{psieq}
\end{equation}
with the boundary conditions 
\begin{eqnarray}
\psi \left( t_{i}\right) &=&0  \label{psibo} \\
\left. \frac{d\psi }{d\tau }\right| _{t=t_{i}} &=&1  \notag
\end{eqnarray}
We use simply $U^{\prime }$ and $U^{\prime \prime }$ for the respective
functions calculated at the fixed point $b$.

We have 
\begin{eqnarray*}
W^{\prime \prime } &=&\frac{d^{2}}{dx^{2}}\left[ \frac{1}{4}\left( U^{\prime
}\right) ^{2}-\frac{D}{2}U^{\prime \prime }\right] \\
&\approx &\frac{1}{2}\left( U^{\prime \prime }\right) ^{2}\;+\;\text{terms
of order }D
\end{eqnarray*}
The equation for the eigenvalues becomes 
\begin{equation*}
\psi ^{\prime \prime }-\left( U^{\prime \prime }\right) ^{2}\psi =0
\end{equation*}
\begin{equation*}
\psi \left( t\right) =a_{1}\exp \left( \left| U^{\prime \prime }\right|
t\right) +a_{2}\exp \left( -\left| U^{\prime \prime }\right| t\right)
\end{equation*}
and it results form the boundary conditions 
\begin{eqnarray*}
a_{1} &=&\frac{1}{2\left| U^{\prime \prime }\right| }\exp \left[ -\left|
U^{\prime \prime }\right| t_{i}\right] \\
a_{2} &=&-\frac{1}{2\left| U^{\prime \prime }\right| }\exp \left[ \left|
U^{\prime \prime }\right| t_{i}\right]
\end{eqnarray*}
Then 
\begin{equation*}
\psi \left( t\right) =\frac{1}{2\left| U^{\prime \prime }\right| }\left\{
\exp \left[ \left| U^{\prime \prime }\right| \left( t-t_{i}\right) \right]
-\exp \left[ -\left| U^{\prime \prime }\right| \left( t-t_{i}\right) \right]
\right\}
\end{equation*}
\textbf{NOTE}. This is the $\sinh $ which is obtained in the calculation of
the ground level splitting by quantum tunneling for a particle in two-well
potential.

Now we can calculate 
\begin{equation*}
\frac{1}{\left[ 4\pi D\psi \left( t\right) \right] ^{1/2}}=\frac{1}{\left(
4\pi D\right) ^{1/2}}\left[ 2\left| U^{\prime \prime }\right| \right] ^{1/2}%
\frac{\exp \left[ -\frac{1}{2}\left| U^{\prime \prime }\right| \left(
t-t_{i}\right) \right] }{\left( 1-\exp \left[ -2\left| U^{\prime \prime
}\right| \left( t-t_{i}\right) \right] \right) ^{1/2}}
\end{equation*}
It remains to calculate the action for this trivial trajectory $x_{c}\left(
t\right) =b$%
\begin{eqnarray*}
S_{c} &=&\int_{t_{i}}^{t}dt\left[ \frac{1}{4}\overset{\cdot }{x}%
_{c}^{2}+W\left( x_{c}\right) \right] \\
&=&\int_{t_{i}}^{t}dtW\left( x_{c}\right) \\
&=&\int_{t_{i}}^{t}dt\left[ \frac{1}{4}U^{\prime 2}\left( x_{c}\right) -%
\frac{D}{2}U^{\prime \prime }\left( x_{c}\right) \right]
\end{eqnarray*}
Since the position of the extremum of $W$ is very close (to order $D$) of
the position where $V\left( x\right) $ is zero, and since $V=-U^{\prime }$,
we can take with good approximation the first term in the integrand zero.
Then 
\begin{equation*}
S_{c}=-\frac{D}{2}\int_{t_{i}}^{t}dtU^{\prime \prime }\left( x_{c}\right) =-%
\frac{D}{2}U^{\prime \prime }\left( t-t_{i}\right)
\end{equation*}
We have to put together the two factors of the propagator and take into
account that at $b$ we have $U^{\prime \prime }>0$%
\begin{eqnarray*}
K\left( x,t;x_{i},t_{i}\right) &=&\frac{1}{\left[ 4\pi D\psi \left( t\right) %
\right] ^{1/2}}\exp \left[ -\frac{S_{c}}{D}\right] \\
&=&\left( \frac{U^{\prime \prime }}{2\pi D}\right) ^{1/2}\frac{\exp \left[ -%
\frac{1}{2}U^{\prime \prime }\left( t-t_{i}\right) \right] }{\left( 1-\exp %
\left[ -2U^{\prime \prime }\left( t-t_{i}\right) \right] \right) ^{1/2}}\exp
\left( \frac{1}{2}U^{\prime \prime }\left( t-t_{i}\right) \right)
\end{eqnarray*}
\begin{equation*}
K\left( x,t;x_{i},t_{i}\right) =\left( \frac{U^{\prime \prime }}{2\pi D}%
\right) ^{1/2}\frac{1}{\left( 1-\exp \left[ -2U^{\prime \prime }\left(
t-t_{i}\right) \right] \right) ^{1/2}}
\end{equation*}

\bigskip

The contribution to the action is 
\begin{eqnarray*}
K^{0}\left( b,\frac{t}{2};b,-\frac{t}{2}\right) &=&\left( \frac{%
U_{b}^{\prime \prime }}{2\pi D}\right) ^{1/2}\frac{1}{\left[ 1-\exp \left(
-2U_{b}^{\prime \prime }t\right) \right] ^{1/2}} \\
&\simeq &\left( \frac{U_{b}^{\prime \prime }}{2\pi D}\right) ^{1/2}\;\;\text{%
(at large }t\text{)}
\end{eqnarray*}

It should be noticed that no other classical solution exists since there are
no turning points permitting the solution to come back to $b$.

\subsubsection{Contribution from the trivial fixed point solution in the
dual function approach}

We apply the procedure described for the purely diffusive case to the case $%
V\left( x\right) \neq 0$ and for this we need the solution of the
Euler-Lagrange equations. An approximation is possible if the diffusion
coefficient is small. In this case the diffusion will take place around the
equilibrium positions $-\sqrt{a/b}$ and $+\sqrt{a/b}$. Taking the
equilibrium $x\left( t\right) =x_{0}=\pm \sqrt{a/b}$ in the equation for $%
k\left( t\right) $ we have 
\begin{equation}
k\left( t\right) =\int_{0}^{T}dt^{\prime }\Delta _{21}\left( t,t^{\prime
}\right) J_{1}\left( t^{\prime }\right)  \label{k1}
\end{equation}
where 
\begin{equation}
\Delta _{21}\left( t,t^{\prime }\right) =\exp \left[ \left(
-a+3bx_{0}^{2}\right) t\right] \Theta \left( t^{\prime }-t\right) \exp \left[
-\left( -a+3bx_{0}^{2}\right) t^{\prime }\right]  \label{delta21}
\end{equation}
The symbol $\Theta $ stands for the Heaviside function. In the equation for $%
x\left( t\right) $ we expand around the equilibrium position 
\begin{equation}
x=x_{0}+\varepsilon \left( t\right)  \label{expx}
\end{equation}
and solve the equation 
\begin{equation}
\overset{\cdot }{\varepsilon }=\left( a-3bx_{0}^{2}\right) \varepsilon
+2iDk+J_{2}  \label{eqeps}
\end{equation}
taking account of Eq.(\ref{k1}) 
\begin{equation}
x\left( t\right) =x_{0}+\int_{0}^{T}dt^{\prime }\Delta _{11}\left(
t,t^{\prime }\right) J_{1}\left( t^{\prime }\right) +\int_{0}^{T}dt^{\prime
}\Delta _{12}\left( t,t^{\prime }\right) J_{2}\left( t^{\prime }\right)
\label{x1}
\end{equation}
where 
\begin{eqnarray}
\Delta _{11}\left( t,t^{\prime }\right) &=&\left( 2iD\right) \exp \left[
\left( a-3bx_{0}^{2}\right) t\right]  \label{delta11} \\
&&\hspace*{-1cm}\times \left\{ \frac{1}{2\left( -a+3bx_{0}^{2}\right) }%
\left( \exp \left[ 2\left( -a+3bx_{0}^{2}\right) t\right] -1\right) \Theta
\left( t^{\prime }-t\right) \right.  \notag \\
&&\hspace*{-1cm}\left. +\frac{1}{2\left( -a+3bx_{0}^{2}\right) }\left( \exp %
\left[ 2\left( -a+3bx_{0}^{2}\right) t^{\prime }\right] -1\right) \Theta
\left( t-t^{\prime }\right) \right\}  \notag \\
&&\times \exp \left[ \left( a-3bx_{0}^{2}\right) t^{\prime }\right]  \notag
\end{eqnarray}
and 
\begin{equation}
\Delta _{12}\left( t,t^{\prime }\right) =\exp \left[ \left(
a-3bx_{0}^{2}\right) t\right] \Theta \left( t-t^{\prime }\right) \exp \left[
\left( -a+3bx_{0}^{2}\right) t^{\prime }\right]  \label{delta12}
\end{equation}

\bigskip

We note that at the limit where no potential would be present, $\left(
-a+3bx_{0}^{2}\right) \rightarrow 0$, the propagators $\Delta $ become 
\begin{equation*}
\Delta _{11}\left( t,t^{\prime }\right) \rightarrow 2iD\left[ t\Theta \left(
t^{\prime }-t\right) +t^{\prime }\Theta \left( t-t^{\prime }\right) \right]
\end{equation*}
\begin{equation*}
\Delta _{12}\left( t,t^{\prime }\right) \rightarrow \Theta \left(
t-t^{\prime }\right)
\end{equation*}
\emph{i.e.} the propagators of a purely diffusive process (see \cite{Flmadi1}%
).

Using the solutions Eqs.(\ref{k1}) and (\ref{x1}) we can calculate the
action along this path. 
\begin{equation*}
\emph{S}_{J}=\int_{0}^{T}dt\left[ -k\overset{\cdot }{x}+kV\left( x\right)
+iDk^{2}+J_{1}x+J_{2}k\right]
\end{equation*}
We will insert the expansion Eq.(\ref{expx}), perform an integration by
parts over the first term and take into account the equations, \emph{i.e.}
the first line of Eq.(\ref{eqEL}) and Eq.(\ref{eqeps}) 
\begin{eqnarray*}
\emph{S}_{J} &=&\int_{0}^{T}dt\left\{ -\frac{1}{2}k\overset{\cdot }{%
\varepsilon }+kV^{\prime }\left( x_{0}\right) \varepsilon
+iDk^{2}+J_{1}x_{0}+J_{1}\varepsilon \right. \\
&&\left. +\frac{1}{2}k\left[ -V^{\prime }\left( x_{0}\right) -J_{1}\right]
\varepsilon +J_{2}k\right\}
\end{eqnarray*}
\begin{equation*}
\emph{S}_{J}=x_{0}\int_{0}^{T}dtJ_{1}+\frac{1}{2}\int_{0}^{T}dt\left(
\varepsilon J_{1}+kJ_{2}\right)
\end{equation*}
Now we express the two solutions, for $k\left( t\right) $ and $\varepsilon
\left( t\right) $ in terms of the propagators $\Delta $%
\begin{equation}
\emph{S}_{J}=x_{0}\int_{0}^{T}dtJ_{1}\left( t\right) +\frac{1}{2}%
\int_{0}^{T}dt\int_{0}^{T}dt^{\prime }J_{i}\left( t\right) \Delta
_{ij}\left( t,t^{\prime }\right) J_{j}\left( t^{\prime }\right)
\label{salongx0}
\end{equation}
where summation over $i,j=1,2$ is assumed, and $\Delta _{22}\left(
t,t^{\prime }\right) \equiv 0$.

We now dispose of the generating functional of the system \emph{when this is
in a region around }$x_{0}=\pm \sqrt{a/b}$, the fixed equilibrium points. To
see what is the effect of the diffusion in this case we calculate for the
variable $x\left( t\right) $ the average and the dispersion. 
\begin{equation*}
\left\langle x\left( t\right) \right\rangle =\left. \frac{\delta }{i\delta
J_{1}\left( t\right) }\emph{Z}_{J}\right| _{J_{1,2}=0}=x_{0}
\end{equation*}
which was to be expected. And 
\begin{equation*}
\left\langle x\left( t\right) x\left( t^{\prime }\right) \right\rangle
=\left. \frac{\delta }{i\delta J_{1}\left( t\right) }\frac{\delta }{i\delta
J_{1}\left( t^{\prime }\right) }\emph{Z}_{J}\right| _{J_{1,2}=0}
\end{equation*}
\begin{eqnarray*}
&&\frac{\delta }{i\delta J_{1}\left( t\right) }\frac{\delta }{i\delta
J_{1}\left( t^{\prime }\right) }\emph{Z}_{J} \\
&=&\frac{1}{2}\frac{1}{i}\Delta _{11}\left( t,t^{\prime }\right) \exp \left(
i\emph{S}_{J}\right) \\
&&\hspace*{-1cm}+\left[ x_{0}+\frac{1}{2}\int_{0}^{T}dt^{\prime }\Delta
_{11}\left( t,t^{\prime }\right) J_{1}\left( t^{\prime }\right) +\frac{1}{2}%
\int_{0}^{T}dt^{\prime }\Delta _{12}\left( t,t^{\prime }\right) J_{2}\left(
t^{\prime }\right) +\frac{1}{2}\int_{0}^{T}dt^{\prime \prime }J_{2}\left(
t^{\prime \prime }\right) \Delta _{21}\left( t^{\prime \prime },t\right) %
\right] \\
&&\hspace*{-1cm}\times \left[ x_{0}+\frac{1}{2}\int_{0}^{T}dt^{\prime \prime
}\Delta _{11}\left( t^{\prime },t^{\prime \prime }\right) J_{1}\left(
t^{\prime \prime }\right) +\frac{1}{2}\int_{0}^{T}dt^{\prime \prime }\Delta
_{12}\left( t^{\prime },t^{\prime \prime }\right) J_{2}\left( t^{\prime
\prime }\right) +\frac{1}{2}\int_{0}^{T}dtJ_{2}\left( t\right) \Delta
_{21}\left( t,t^{\prime }\right) \right] \\
&&\times \exp \left( i\emph{S}_{J}\right)
\end{eqnarray*}
This gives the result 
\begin{eqnarray}
\left\langle x\left( t\right) x\left( t^{\prime }\right) \right\rangle
&=&x_{0}^{2}+\exp \left[ \left( a-3bx_{0}^{2}\right) t\right] D
\label{xtxtp} \\
&&\times \left\{ \frac{1}{2\left( -a+3bx_{0}^{2}\right) }\left( \exp \left[
2\left( -a+3bx_{0}^{2}\right) t\right] -1\right) \Theta \left( t^{\prime
}-t\right) \right.  \notag \\
&&\left. +\frac{1}{2\left( -a+3bx_{0}^{2}\right) }\left( \exp \left[ 2\left(
-a+3bx_{0}^{2}\right) t^{\prime }\right] -1\right) \Theta \left( t-t^{\prime
}\right) \right\}  \notag \\
&&\times \exp \left[ \left( a-3bx_{0}^{2}\right) t^{\prime }\right]  \notag
\end{eqnarray}
In the absence of the potential\ $a\rightarrow 0$ and $b\rightarrow 0$ ,
this is simply 
\begin{equation*}
\left\langle x\left( t\right) x\left( t^{\prime }\right) \right\rangle
=x_{0}^{2}+D\min \left( t,t^{\prime }\right)
\end{equation*}
\emph{i.e.} the diffusion around the position $x_{0}$. In the present case,
we note that $a-3bx_{0}^{2}\equiv V^{\prime }\left( x_{0}\right) =-U^{\prime
\prime }\left( x_{0}\right) $ with $U_{0}^{\prime \prime }\equiv U^{\prime
\prime }\left( x_{0}\right) >0$. Fixing the parameters, we take $t>t^{\prime
}$ and obtain 
\begin{equation*}
\left\langle x\left( t\right) x\left( t^{\prime }\right) \right\rangle
=x_{0}^{2}+D\exp \left( -U_{0}^{\prime \prime }t\right) \frac{1}{%
2U_{0}^{\prime \prime }}\left[ \exp \left( 2U_{0}^{\prime \prime }t^{\prime
}\right) -1\right] \exp \left( -U_{0}^{\prime \prime }t^{\prime }\right)
\end{equation*}
For $t=t^{\prime }$ (the dispersion) we obtain (with $\Theta \rightarrow 1/2$%
) 
\begin{equation}
\left\langle x\left( t\right) ^{2}\right\rangle =x_{0}^{2}+\left( \frac{D}{%
2U_{0}^{\prime \prime }}\right) \left[ 1-\exp \left( -2U_{0}^{\prime \prime
}t\right) \right]  \label{x2canoi}
\end{equation}
\begin{figure}[!htp]
\centering
\includegraphics[width=10cm]{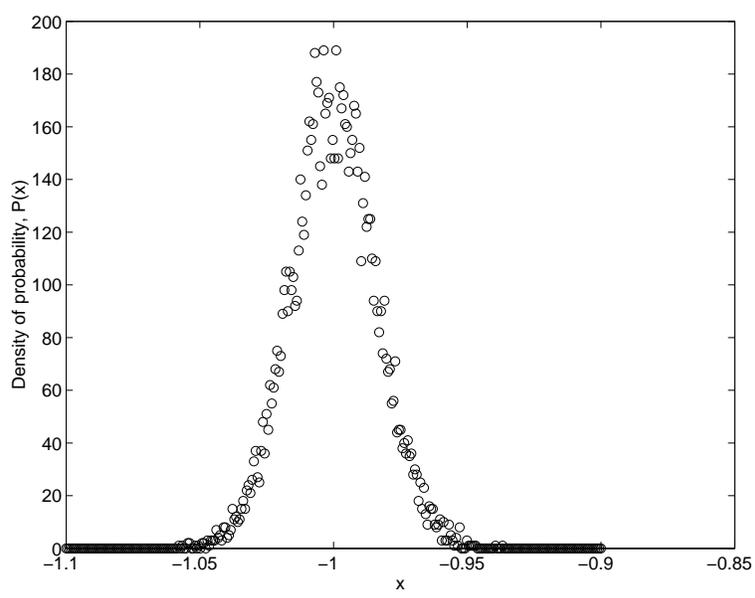}
\caption{The probability distribution in the harmonic region around the left
equilibrium stable point.}
\label{probab}
\end{figure}
\begin{figure}[!htp]
\centering
\includegraphics[width=10cm]{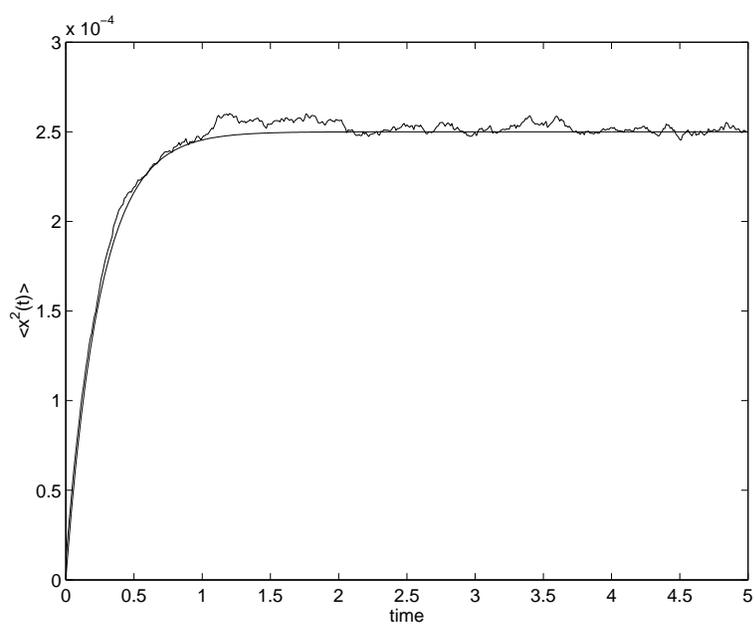}
\caption{Comparison between the analytical formula Eq.(\protect\ref{x2canoi}%
) (continuous line) and numerical integration of the stochastic equation.}
\label{x2th}
\end{figure}

It is straightforward to calculate the higher order statistics for this
process, since the functional derivatives can easily be done. We have to
remember that this derivation was based on the approximation consisting in
taking the equilibrium position in the Euler-Lagrange equation for $k\left(
t\right) $.

\subsection{Assembling the instanton-anti-instanton solutions}

\subsubsection{The formulation based on Onsager-Machlup action}

However, since $W\left( 0\right) -W\left( b\right) $ and $W\left( a\right)
-W\left( b\right) $ are small quantities and we can suppose that the
instantons, even if they are not exact solutions of the equations of motion,
can give a contribution to the action. The instantons connects the points of
maximum not of $-W\left( x\right) $ (because they are not equal) but of a
different potential, a \emph{corrected} $-W\left( x\right) $ to order one in 
$D$ that has degenerate maxima. This potential will be called $W^{\left(
0\right) }\left( x\right) $ and it will be considered in the calculation of
the contribution of the instantons and antiinstantons. 
\begin{eqnarray*}
W^{\left( 0\right) }\left( x\right) &=&W\left( x\right) -\delta W\left(
x\right) \\
\delta W\left( x\right) &=&0\;,\;x<x_{m} \\
&=&W_{0}-W_{b}\;,\;x>x_{m}
\end{eqnarray*}
where $x_{m}$ is the point corresponding to the minimum situated between the
two maximas.

It is introduced the family of trajectories $x_{I}\left( \tau -t_{0}\right) $
which leaves $x=b$ at time $-\infty $ and reach $x=0$ at time $\infty $.
They have all the same energy 
\begin{equation*}
E=-W_{b}
\end{equation*}
and the classical action $S_{b0}$.

We now consider the travel from $x=b$ to $x=0$ made by an instanton $%
x_{I}\left( \tau -t_{0}\right) $ with the center located at time $t_{0}$ ;
next the return made by an antiinstanton which is actually an instanton $%
x_{I}\left( t_{1}-\tau \right) $ starting from $x=0$ and going to $x=b$ with
the center located at time $t_{0}$. With these two instantons we create a
single classical solution 
\begin{equation*}
x_{IA}\left( \tau ;t_{0},t_{1}\right) =\left\{ 
\begin{array}{cc}
x_{I}\left( \tau -t_{0}\right) & \tau <\frac{1}{2}\left( t_{0}+t_{1}\right)
\\ 
x_{I}\left( t_{1}-\tau \right) & \tau >\frac{1}{2}\left( t_{0}+t_{1}\right)
\end{array}
\right.
\end{equation*}
The contribution to the action of this \emph{assambled} solution is 
\begin{eqnarray*}
S_{IA}\left( t;t_{0},t_{1}\right) &\simeq &\left( W_{0}-W_{b}\right) \left(
t_{1}-t_{0}\right) +W_{b}t+2S_{b0} \\
&&-2\int_{\left( t_{1}+t_{0}\right) /2}^{\infty }d\tau \frac{\left[ \overset{%
\cdot }{x}_{I}\left( \tau -t_{0}\right) \right] ^{2}}{2}
\end{eqnarray*}
The contributions in this formula comes from the potential energy and the
kinetic energy along the trajectory. We should remember that the action is
the integral on time of the density of Lagrangian, where there is the
kinetic energy term and minus the potential energy. 
\begin{equation*}
S=\int dt\left\{ \frac{1}{2}\overset{\cdot }{x}^{2}-\left[ -W\left( x\right) 
\right] \right\}
\end{equation*}

If the particle would have remained in $x=b$ imobile for all time $\left(
-t/2,t/2\right) $ then the contribution from the potential would have been 
\begin{equation*}
W\left( b\right) t
\end{equation*}

The instanton spends $\left( t_{1}-t_{0}\right) $ time in the point $0$
before returning to $b$. Then it accumulates the action equal with the
difference in potential between $b$ and $0$ multiplied with this time
interval. 
\begin{equation*}
\left[ \left( -W_{b}\right) -\left( -W_{0}\right) \right] \left(
t_{1}-t_{0}\right) =\left( W_{0}-W_{b}\right) \left( t_{1}-t_{0}\right)
\end{equation*}
(since $\left( -W_{b}\right) >\left( -W_{0}\right) $).

\bigskip

Define the kinetic energy and the energy 
\begin{eqnarray*}
K &=&\int_{t_{0}}^{t}\frac{1}{2}\overset{\cdot }{x}^{2}\left( t\right) dt\;
\\
E &=&\frac{1}{2}\overset{\cdot }{x}^{2}\left( t\right) +\left[ -W\left(
x\right) \right] \\
\overset{\cdot }{x}\left( t\right) &=&\sqrt{2\left( E+W\left( x\right)
\right) }
\end{eqnarray*}
then 
\begin{eqnarray*}
K &=&\int_{t_{0}}^{t}\left( E+W\left( x\right) \right) dt=\int_{t_{0}}^{t}%
\frac{E+W\left( x\right) }{\overset{\cdot }{x}\left( t\right) }dx \\
&=&\frac{1}{2}\int_{x_{0}}^{x}\sqrt{2\left( E+W\left( x\right) \right) }dx
\end{eqnarray*}
In this formula we have to replace the expression of the trajectory $%
x_{IA}\left( t\right) $ and integrate.

In our case the energy has the value of the initial position. Here the
velocity is zero and the potential is $-W\left( b\right) $%
\begin{equation*}
E|_{x=b}=-W\left( b\right)
\end{equation*}
and the potential is actually the current value of the zeroth -order
potential 
\begin{eqnarray*}
E+W\left( x\right) &=&\left[ -W\left( b\right) \right] -\left[ -W^{\left(
0\right) }\left( x\right) \right] \\
&=&W^{\left( 0\right) }\left( x\right) -W\left( b\right)
\end{eqnarray*}
If we want to calculate the integral of the \emph{kinetic energy} along the
trajectory, we have to consider separately the intervals where the kinetic
energy is strongly determined by the velocity, \emph{i.e.} the region where
the instanton transition occurs, from the rest of the trajectory where, the
velocity being practically zero, the potential is a better description and
can be easily approximated. The approximation will relay on the fact that
the particle is practically imobile in $b$ or in $0$, after the transition
has been made. So we will use both expressions for the kinetic energy 
\begin{equation*}
K|_{x=b}^{x=0}=\int_{0}^{\frac{t_{1}+t_{0}}{2}}dt\frac{1}{2}\overset{\cdot }{%
x}^{2}\left( t\right) +\int_{b}^{0}dx\left[ W^{\left( 0\right) }\left(
x\right) -W_{b}\right] ^{1/2}
\end{equation*}
The terms must be considered twice and the interval of integration can be
extended for the region of transition, since in any case it is very small 
\begin{equation*}
2\int_{b}^{0}dx\left[ W^{\left( 0\right) }\left( x\right) -W_{b}\right]
^{1/2}-2\int_{\frac{t_{1}+t_{0}}{2}}^{\infty }d\tau \frac{1}{2}\overset{%
\cdot }{x}^{2}\left( \tau -t_{0}\right)
\end{equation*}

The term $S_{b0}$ is 
\begin{equation*}
S_{b0}=\int_{b}^{0}dx\left[ W^{\left( 0\right) }\left( x\right) -W_{b}\right]
^{1/2}
\end{equation*}

Using an approximation for the form of the instanton, it results 
\begin{eqnarray*}
S_{IA}\left( t;t_{0},t_{1}\right) &=&\left( W_{0}-W_{b}\right) \left(
t_{1}-t_{0}\right) +W_{b}t+2S_{b0} \\
&&-\frac{x_{m}^{2}\left( W_{0}^{\prime \prime }\right) ^{1/2}}{\sqrt{2}}\exp %
\left[ \left( 2W_{0}^{\prime \prime }\right) ^{1/2}\left( 2\Delta
_{m0}-t_{1}+t_{0}\right) \right]
\end{eqnarray*}
where 
\begin{equation*}
W_{0}^{\prime \prime }=\left. \frac{d^{2}W_{0}}{dx^{2}}\right| _{x=0}
\end{equation*}
and 
\begin{equation*}
\Delta _{m0}=\frac{1}{2}\int_{x_{m}}^{0}dx\left\{ \frac{1}{\left[ W^{\left(
0\right) }\left( x\right) -W_{b}\right] ^{1/2}}-\frac{1}{\left[
W_{h}^{\left( 0\right) }\left( x\right) -W_{b}\right] ^{1/2}}\right\}
\end{equation*}

The local harmonic approximation to $W^{\left( 0\right) }$ is $W_{h}^{\left(
0\right) }$%
\begin{equation*}
W_{h}^{\left( 0\right) }=\left\{ 
\begin{array}{cc}
W_{b}+\frac{W_{0}^{\prime \prime }}{2}x^{2} & x>x_{m} \\ 
W_{b}+\frac{W_{b}^{\prime \prime }}{2}\left( x-b\right) ^{2} & x<x_{m}
\end{array}
\right.
\end{equation*}

\textbf{NOTE} Since the trajectory $x_{IA}\left( \tau ;t_{0},t_{1}\right) $
is not an \emph{exact} solution of the equation of motion in the potential $%
W $, the expansion of the action $S$ will not be limitted to the zeroth and
the second order terms. It will also contain a firts order term, $\left.
\left( \delta S/\delta x\right) \right| _{x=x_{c}\left( \tau \right) }$ . It
can be shown that this contribution is negligible in the order $O\left(
D\right) $.

\bigskip

The calculation of the contribution to the functional integral from the
second order expansion around $x_{IA}$ is done as usual by finding the
eigenvalues of the determinant of the corresponding operator. As before, the
product of the eigenvalues should not include the first eigenvalue since
this is connected with the translational symmetry of the instanton solution.
This time there will be two eigenvalues, one for $t_{0}$ (the transition
performed by the $b\rightarrow 0$ instanton) and the second for $t_{1}$ (the
transition prtformed by the antiinstanton, or the transition $0\rightarrow b$%
). Another way of expressing this invariance to the two time translations is
to say that the pair of instantons has not a determined central moment and,
in addition, there is an \emph{internal} degree of freedom of the \emph{%
breathing} solution, which actually is this pair instanton-antiinstanton.

To take into account these modes, whose eigenvalues are zero, we need to
integrate in the functional integral, over the two times, $t_{0}$ and $t_{1}$%
.

The measure of integration for the two translational symmetries $t_{0}$ and $%
t_{1}$ is 
\begin{equation*}
\frac{S_{b0}}{2\pi D}
\end{equation*}
This quantity is the Jacobian of the change of variables in the functional
integration over the fluctuations around the instanton solution. The
fluctuation that corresponds to the lowest (almost zero) eigenvalue is
replaced in the measure of integration with the differential of the time
variable representing the moment of transition. Then it results this
Jacobian.

It can be shown that in the approximation given by exponentially small
terms, the contribution to the path integral of the small fluctuations
around the classical instanton-antiinstanton $x_{IA}$ solution is the \emph{%
product} of the fluctuation terms around the instanton and antiinstanton
separately. 
\begin{eqnarray*}
K^{\left( 1\right) }\left( b,\frac{t}{2};b,-\frac{t}{2}\right)
&=&\int_{-t/2}^{t/2}dt_{0}\int_{t_{0}}^{t/2}dt_{1}\left( \frac{S_{bo}}{2\pi D%
}\right) \\
&&\times \left( \frac{\lambda _{0}}{4\pi D\psi }\right) _{I}^{1/2}\left( 
\frac{\lambda _{0}}{4\pi D\psi }\right) _{A}^{1/2} \\
&&\times \left[ \frac{2\pi D}{\left( 2W_{0}^{\prime \prime }\right) ^{1/2}}%
\right] ^{1/2}\exp \left[ -\frac{S_{IA}\left( t;t_{0},t_{1}\right) }{D}%
\right]
\end{eqnarray*}
where $\left( \lambda _{0}\right) _{I,A}$ represent the lowest eigenvalue of
the operator arising from the second order expansion of the action, defined
on the time intervals 
\begin{eqnarray*}
x_{c}\left( \tau \right) &\equiv &x_{I}\left( \tau -t_{0}\right) \;,\;-\frac{%
t}{2}<\tau <\frac{t_{0}+t_{1}}{2} \\
x_{c}\left( \tau \right) &\equiv &x_{I}\left( t_{1}-\tau \right) \;,\;\frac{%
t_{0}+t_{1}}{2}<\tau <\frac{t}{2}
\end{eqnarray*}
The notation $\psi _{I}$ and respectively $\psi _{A}$ represents 
\begin{equation*}
\psi _{I}\;\text{value at }\frac{t_{0}+t_{1}}{2}\text{ of the eigenfunction }%
\psi \text{ that starts at }t_{i}=-\frac{t}{2}
\end{equation*}
\begin{equation*}
\psi _{A}\;\text{value at }\frac{t}{2}\text{ of the eigenfunction }\psi 
\text{ that starts at }t_{i}=\frac{t_{0}+t_{1}}{2}
\end{equation*}
This corresponds to the formula of Coleman which replaces the infinite
product of eigenvalues with $\left( \frac{\lambda _{0}}{4\pi D\psi }\right)
^{1/2}$ where $\psi $ is calculated at the end of the interval of time,
where $\psi $ verifies the boundary conditions (\ref{psibo}).

The result is 
\begin{eqnarray}
&&K^{\left( 1\right) }\left( b,\frac{t}{2};b,-\frac{t}{2}\right)
\label{K(1)raw} \\
&=&\left[ \frac{2\pi D}{\left( 2W_{0}^{\prime \prime }\right) ^{1/2}}\right]
^{1/2}\frac{\left( x_{m}-b\right) \left| x_{m}\right| W_{0}^{\prime \prime
}W_{b}^{\prime \prime }}{2\pi ^{2}D^{2}}\exp \left[ -\frac{\varepsilon
_{0}^{\left( b\right) }t}{D}\right]  \notag \\
&&\times \exp \left[ -\frac{2S_{b0}}{D}+\Delta _{bm}\left( 2W_{b}^{\prime
\prime }\right) ^{1/2}+\Delta _{bm}\left( 2W_{0}^{\prime \prime }\right)
^{1/2}\right]  \notag \\
&&\hspace*{-1cm}\times \int_{-t/2}^{t/2}dt_{0}\int_{t_{0}}^{t/2}dt_{1}\exp
\left\{ -\frac{\varepsilon _{0}^{\left( 0\right) }-\varepsilon _{0}^{\left(
b\right) }}{D}\left( t_{1}-t_{0}\right) +\frac{C}{D}\exp \left[ -\left(
2W_{0}^{\prime \prime }\right) ^{1/2}\left( t_{1}-t_{0}\right) \right]
\right\}  \notag
\end{eqnarray}
where 
\begin{equation*}
\varepsilon _{0}^{\left( b\right) }=W_{i}+D\left( \frac{W_{i}^{\prime \prime
}}{2}\right) ^{1/2}
\end{equation*}
is the lowest eigenvalue of the Schrodinger equation associated with the
Fokker-Planck diffusion equation in the local harmonic approximation of the
potential $W$ in the well $i$. 
\begin{equation}
C=x_{m}^{2}\left( \frac{W_{0}^{\prime \prime }}{2}\right) ^{1/2}\exp \left[
2\left( 2W_{0}^{\prime \prime }\right) ^{1/2}\Delta _{m0}\right]
\label{constC}
\end{equation}
and 
\begin{equation*}
\Delta _{ij}=\frac{1}{2}\int_{x_{i}}^{x_{j}}dx\left\{ \frac{1}{\left[
W^{\left( 0\right) }\left( x\right) -W_{b}\right] ^{1/2}}-\frac{1}{\left[
W_{h}^{\left( 0\right) }\left( x\right) -W_{b}\right] ^{1/2}}\right\}
\end{equation*}

There is a very important problem with this formula: the coefficient $C$ is
positive and the contribution of this part in the $t_{1}$ integration comes
from time intervals 
\begin{equation*}
t_{1}-t_{0}\ll \frac{1}{\left( 2W_{0}^{\prime \prime }\right) ^{1/2}}\sim
\Delta t
\end{equation*}
Since this is a very small time interval, it results that the contributions
are due to states where the instanton and the antiinstanton are very close
one of the other, which is unphysical. It will be necessary to calculate in
a particular way this part, introducing acontour in the complex $t_{1}-t_{0}$
plane, with an excursion on the imaginary axis.

\bigskip

\textbf{The calculation of the propagator }$K^{\left( 1\right) }\left( b,%
\frac{t}{2};b,-\frac{t}{2}\right) $. Here $\left( 1\right) $ means that only
one pair of instanton and anti-instanton is considered.

We particularize the formula above using the expression for $W\left(
x\right) $ given in terms of $U\left( x\right) $ . Now we have 
\begin{eqnarray*}
W_{b} &=&-\frac{DU_{b}^{\prime \prime }}{2} \\
W_{0} &=&-\frac{DU_{0}^{\prime \prime }}{2} \\
W_{b} &=&-\frac{U_{b}^{\prime \prime 2}}{2} \\
W_{0} &=&-\frac{U_{0}^{\prime \prime 2}}{2}
\end{eqnarray*}
The expression of $S_{b0}$ is 
\begin{equation*}
S_{b0}=\frac{U_{0}-U_{b}}{2}+\frac{D}{2}\ln \left| \frac{U_{b}^{\prime
\prime }\left( x_{m}-b\right) }{U_{0}^{\prime \prime }x_{m}}\right| +\frac{D%
}{2}\left( U_{b}^{\prime \prime }\Delta _{bm}+U_{0}^{\prime \prime }\Delta
_{m0}\right)
\end{equation*}
\begin{eqnarray*}
K^{\left( 1\right) }\left( b,\frac{t}{2};b,-\frac{t}{2}\right) &=&\frac{%
U_{b}^{\prime \prime }\left| U_{0}^{\prime \prime }\right| x_{m}^{2}}{2}%
\left( \frac{\left| U_{0}^{\prime \prime }\right| }{2\pi D}\right)
^{3/2}I\left( t\right) \\
&&\times \exp \left( -\frac{U_{0}-U_{b}}{D}+2\Delta _{m0}\left|
U_{0}^{\prime \prime }\right| \right)
\end{eqnarray*}
where 
\begin{equation*}
I\left( t\right) =\int_{-t/2}^{t/2}dt_{0}\int_{\emph{C}}dz\exp \left[
-\left| U_{0}^{\prime \prime }\right| z+\frac{C}{D}\exp \left( -\left|
U_{0}^{\prime \prime }\right| z\right) \right]
\end{equation*}

\subsubsection{The formulation using dual functions}

We will use the functional approach in the setting that has been developed
by us to the calculation of the probability of transition from one minimum
to the same minimum with an intermediate stay at the unstable maximum point
(symmetric potential).

The initial equation is 
\begin{equation*}
\overset{\cdot }{x}\left( t\right) =V\left( x\right) +\xi \left( t\right)
\end{equation*}
The action functional with external current added is 
\begin{equation*}
\emph{S}=i\int_{-T}^{T}dt\emph{L}
\end{equation*}
\begin{equation*}
\emph{L}=-k\left( t\right) \overset{\cdot }{x}\left( t\right) +k\left(
t\right) V\left( x\right) +iDk^{2}\left( t\right) +J_{1}\left( t\right)
x\left( t\right)
\end{equation*}
The equations of motion are 
\begin{eqnarray*}
\overset{\cdot }{x}-V\left( x\right) -2iDk &=&0 \\
\overset{\cdot }{k}-k\left( \frac{dV}{dx}\right) &=&-J_{1}
\end{eqnarray*}
We have to solve these equations, and replace the solutions $x\left(
t\right) $ and $k\left( t\right) $ in the action functional. Then the
functional derivatives to the external current $J_{1}\left( t\right) $ will
give us the correlations for the stochastic variable $x\left( t\right) $. We
can also calculate the probability that the particle, starting from one
point at a certain time will be found at another point at other time. This
will be done below.

The first step is to obtain an analytical solution to the Euler-Lagrange
equation. The method to solve these equation is essentially a successive
approximation, as we have done above, for the diffusive motion around a
stable position in the potential (harmonic region). We know that the
classical trajectory must be of the type of a transition between the initial
point, taken here as the left minimum of the potential and the final point,
the unstable maximum of the potential. From there, the particle will return
to the left minimum by an inverse transition. We have to calculate
simultaneously $x\left( t\right) $ and $k\left( t\right) $, but we have
sufficient information to find an approximation for $x\left( t\right) $, by
neglecting the effect of diffusion (the last term in the differential
equation for $x\left( t\right) $). 
\begin{equation*}
\overset{\cdot }{x}=ax-bx^{3}
\end{equation*}
This gives 
\begin{equation*}
x\left( \tau \right) =\pm \sqrt{\frac{a}{b}}\frac{1}{\left\{ 1+\exp \left[
-2a(\tau -t_{0})\right] \right\} ^{1/2}}
\end{equation*}
This solution is a transition between either of the minima $\pm \sqrt{a/b}$
and $0$. It shows the same characteristics as found numerically or by
integrating the elliptic form of the equation, in the case of
Onasger-Machlup action. The particle spends long time in the initial and
final points and makes a fast transition between them at an arbitrary time $%
t_{0}$. The width of transition is small compared to the rest of
quasi-imobile stays in the two points, especially if $a$ is large. Then we
will make an approximation, taking the solution as 
\begin{equation*}
x_{\eta }^{\left( 1\right) }\left( \tau \right) =-\sqrt{\frac{a}{b}}\Theta
\left( t_{0}-\tau \right) +0\times \Theta \left( \tau -t_{0}\right)
\end{equation*}
where $-\sqrt{a/b}$ and $0$ are the initial and final positions, the indice $%
\eta $ means that this is the calssical solution (extremum of the action)
and $\left( 1\right) $ means the first part of the full trajectory, which
will also include the inverse transition, from $0$ to $-\sqrt{a/b}$. The
following structure of the total trajectory is examined: The total time
interval is between $-T$ and $T$ (later the parameter $T$ will be identified
with $-t/2$ for comparison with the results from the literature). The
current time variable is $\tau $ and in the present notations, $t$ is any
moment of time in the interval $\left( -T,T\right) $. At time $t_{0}$ the
particle makes a jump \ to the position $x=0$ and remains there until $\tau
=t_{1}$. At $t_{1}$ it performs a jump to the position $-\sqrt{a/b}$, where
it remains for the rest of time, until $T$.

\bigskip

With this approximative solution of the Euler-Lagrange equations (since we
have neglected the term with $D$ in the equation of $x\left( \tau \right) $)
we return to the equation for $k\left( \tau \right) $. We first calculate 
\begin{equation*}
\left. \frac{dV}{dx}\right| _{x_{\eta }^{\left( 1\right) }\left( \tau
\right) }=\left( -2a\right) \Theta \left( t_{0}-\tau \right) +a\Theta \left(
\tau -t_{0}\right)
\end{equation*}
We need the integration of this quantity in the inverse direction starting
from the end of the motion toward the initial time 
\begin{equation*}
\int_{T}^{t}d\tau \left( \frac{dV}{dx}\right) _{x_{\eta }^{\left( 1\right)
}\left( \tau \right) }=\left[ -a\left( T-t\right) \right] \Theta \left(
t-t_{0}\right) +\left[ -aT+3at_{0}-2at\right] \Theta \left( t_{0}-t\right)
\end{equation*}

We now introduce the second part of the motion: at time $\tau =t_{1}$ the
particle makes the inverse transition 
\begin{equation*}
x_{\eta }^{\left( 2\right) }\left( \tau \right) =0\times \Theta \left(
t_{1}-\tau \right) +\left( -\sqrt{\frac{a}{b}}\right) \Theta \left( \tau
-t_{1}\right)
\end{equation*}
with the similar quantities.

After explaining the steps of the calculation, we change to work with the
full process, assembling the two transitions and the static parts into a
single trajectory 
\begin{eqnarray*}
x_{\eta }\left( \tau \right) &=&x_{\eta }^{\left( 1\right) }\left( \tau
\right) +x_{\eta }^{\left( 2\right) }\left( \tau \right) \\
&=&\left( -\sqrt{\frac{a}{b}}\right) \Theta \left( t_{0}-\tau \right)
+\left( -\sqrt{\frac{a}{b}}\right) \Theta \left( \tau -t_{1}\right)
\end{eqnarray*}
\begin{equation*}
\left. \frac{dV}{dx}\right| _{x_{\eta }\left( \tau \right) }=\left(
-2a\right) \Theta \left( t_{0}-\tau \right) +a\Theta \left( t_{1}-\tau
\right) \Theta \left( \tau -t_{0}\right) +\left( -2a\right) \Theta \left(
\tau -t_{1}\right)
\end{equation*}
and 
\begin{equation*}
\int_{T}^{t}d\tau \left( \frac{dV}{dx}\right) _{x_{\eta }^{\left( 1\right)
}\left( \tau \right) }=W\left( t\right)
\end{equation*}
We have introduce the notation 
\begin{eqnarray}
W\left( t\right) &\equiv &\left( 2a\right) \left( T-t\right) \Theta \left(
t-t_{1}\right)  \label{expw} \\
&&+\left( 2aT-3at_{1}+at\right) \Theta \left( t-t_{0}\right) \Theta \left(
t_{1}-t\right)  \notag \\
&&+\left( 2aT-3at_{1}+3at_{0}-2at\right) \Theta \left( t_{0}-t\right)  \notag
\end{eqnarray}
Then the solution of the equation for the dual variable is 
\begin{equation*}
k_{\eta }\left( \tau \right) =\exp \left[ -W\left( \tau \right) \right]
\left\{ k_{T}+\int_{T}^{\tau }dt^{\prime }\left[ -J_{1}\left( t^{\prime
}\right) \right] \exp \left[ W\left( t^{\prime }\right) \right] \right\}
\end{equation*}
According to the procedure explained before we will need to express the
solutions as bilinear combinations of currents, so we identify 
\begin{equation}
\Delta _{21}\left( t,t^{\prime }\right) =\Theta \left( t^{\prime }-t\right)
\exp \left[ -W\left( t\right) \right] \exp \left[ W\left( t^{\prime }\right) %
\right]  \label{del21new}
\end{equation}
and the solution can be rewritten 
\begin{equation}
k_{\eta }\left( \tau \right) =k_{T}\exp \left[ -W\left( \tau \right) \right]
+\int_{-T}^{T}dt^{\prime }\Delta _{21}\left( t,t^{\prime }\right)
J_{1}\left( t^{\prime }\right)  \label{kdel21}
\end{equation}

Using these first approximations for the extremizing path $\left( x_{\eta
}\left( \tau \right) ,k_{\eta }\left( \tau \right) \right) $ we return to
the Euler Lagrange equations and expend the variable $x$ as 
\begin{equation}
x\left( \tau \right) =x_{\eta }\left( \tau \right) +\delta x\left( \tau
\right)  \label{xexpa}
\end{equation}
whose equation is 
\begin{equation*}
\overset{\cdot }{\delta x}\left( \tau \right) =2iDk_{\eta }\left( \tau
\right) +\left( \frac{dV}{dx}\right) _{x_{\eta }\left( \tau \right) }\delta
x\left( \tau \right)
\end{equation*}
and the solution 
\begin{equation*}
\delta x\left( t\right) =\exp \left[ \int_{0}^{t}d\tau \left( \frac{dV}{dx}%
\right) _{x_{\eta }\left( \tau \right) }\right] \left\{
B_{0}+\int_{0}^{t}dt^{\prime }2iDk_{\eta }\left( t^{\prime }\right) \exp %
\left[ -\int_{0}^{t^{\prime }}dt^{\prime \prime }\left( \frac{dV}{dx}\right)
_{x_{\eta }\left( t^{\prime \prime }\right) }\right] \right\}
\end{equation*}
Here $B_{0}$ is a constant to be determined by the condition that $\delta x$
vanishes at the final point. We note that here all integrations are
performed forward in time. We also notice that this expression will contain
the current $J_{1}$ and we will introduce the propagator $\Delta _{11}$.
Performing the detailed calculations we obtain 
\begin{eqnarray}
\delta x\left( t\right) &=&B_{0}\exp \left[ W\left( t\right) \right] +\exp %
\left[ W\left( t\right) \right] \int_{-T}^{T}dt^{\prime }\left(
2iDk_{T}\right) \Theta \left( t-t^{\prime }\right) \exp \left[ -W\left(
t^{\prime }\right) \right]  \label{delxp} \\
&&+\int_{-T}^{T}dt^{\prime }\Delta _{11}\left( t,t^{\prime }\right)
J_{1}\left( t^{\prime }\right)  \notag
\end{eqnarray}
The first two terms depend on constants of integrations and the propagator
is 
\begin{equation}
\Delta _{11}\left( t,t^{\prime }\right) =\exp \left[ W\left( t\right) \right]
\int_{-T}^{T}dt^{\prime \prime }2iD\Theta \left( t-t^{\prime \prime }\right)
\Delta _{21}\left( t^{\prime \prime },t^{\prime }\right) \exp \left[
-W\left( t^{\prime \prime }\right) \right]  \label{del11}
\end{equation}

Using the solutions (\ref{xexpa}) and (\ref{kdel21}) we can calculate the
action along this trajectory. 
\begin{eqnarray*}
\emph{S} &=&i\int_{-T}^{T}dt\left( -k\overset{\cdot }{x}+kV+iDk^{2}+J_{1}x%
\right) \\
&=&i\int_{-T}^{T}dt\left[ \frac{1}{2}\left( -k\overset{\cdot }{x}%
+kV+2iDk^{2}\right) +\frac{1}{2}\left( -k\overset{\cdot }{x}+kV\right)
+J_{1}x\right] \\
&=&i\int_{-T}^{T}dt\left( \frac{1}{2}x\overset{\cdot }{k}+\frac{1}{2}%
kV+J_{1}x\right) \\
&=&i\int_{-T}^{T}dt\left[ \frac{1}{2}x\left( -J_{1}+k\frac{dV}{dx}\right) +%
\frac{1}{2}kV+J_{1}x\right] \\
&=&i\int_{-T}^{T}dt\left\{ \frac{1}{2}J_{1}x+\frac{1}{2}k\left[ V-\left( 
\frac{dV}{dx}\right) x\right] \right\}
\end{eqnarray*}
[It becomes obvious that when the potential is linear (which means that the
Lagrangean is quadratic the two functional variables $x\left( t\right) $ and 
$k\left( t\right) $) the potential does not contribute to the action along
the extremal path].

Using the solutions we have 
\begin{eqnarray*}
\emph{S} &=&i\int_{-T}^{T}dt\frac{1}{2}J_{1}\left( t\right) \left[ x_{\eta
}\left( t\right) +\delta x\left( t\right) \right] \\
&&+i\int_{-T}^{T}dt\frac{1}{2}\left[ V\left( x_{\eta }\right) -\left( \frac{%
dV}{dx}\right) _{x_{\eta }}x_{\eta }\right] i\int_{-T}^{T}dt^{\prime }\Delta
_{21}\left( t,t^{\prime }\right) J_{1}\left( t^{\prime }\right)
\end{eqnarray*}
It will become clear later that the parts containing constants are not
significative for the final answer, the probability. We will focus on the
terms containing explicitely the current $J_{1}$ since the statistical
properties are determined by functional derivatives to this parameter. The
action is 
\begin{eqnarray*}
\emph{S} &=&i\int_{-T}^{T}dt\left\{ \frac{1}{2}J_{1}\left( t\right) x_{\eta
}\left( t\right) \right. \\
&&+\frac{1}{2}J_{1}\left( t\right) \int_{-T}^{T}dt^{\prime }\Delta
_{11}\left( t,t^{\prime }\right) J_{1}\left( t^{\prime }\right) \\
&&\left. +\frac{1}{2}\left[ V\left( x_{\eta }\left( t\right) \right) -\left( 
\frac{dV}{dx}\right) _{x_{\eta }\left( t\right) }x_{\eta }\left( t\right) %
\right] \int_{-T}^{T}dt^{\prime }\Delta _{21}\left( t,t^{\prime }\right)
J_{1}\left( t^{\prime }\right) \right\}
\end{eqnarray*}

The generating functional of the correlations (at any order) is 
\begin{equation}
\emph{Z}\left[ J_{1}\right] =\exp \left\{ \emph{S}\left[ J_{1}\right]
\right\}  \label{zands}
\end{equation}
and we can calculate any quantity by simply parforming functional
derivatives and finally taking $J_{1}=0$.

Instead of that and in order to validate our procedure, we will calculate
the probability for the process: a particle in the initial position $x=-%
\sqrt{a/b}$ at time $-T$ can be found at the final position $x=x_{c}$ at the
time $t=t_{c}$. (Later we will particularize to $x_{c}=-\sqrt{a/b}$ and $%
t_{c}=T$). The calculation of this probability $P$ can be done in the
functional approach in the following way.

We have defined the statistical ensemble of possible particle trajectories
starting at $x=-\sqrt{a/b}$ at $t=-T$ and reaching an arbitrary point $x$ at
time $t$. In the Martin-Siggia-Rose-Jensen approach it is derived the
generating functional as a functional integration over this statistical
ensemble of a weight measure expressed as the exponential of the classical
action. Now we restrict the statistical ensemble by imposing that the
particle is found at time $t=t_{c}$ in the point $x=x_{c}$. By integration
this will give us precisely the probability required.

The condition can be introduced in the functional integration by a Dirac $%
\delta $ function which modifies the functional measure 
\begin{equation*}
\emph{D}\left[ x\left( \tau \right) \right] \rightarrow \emph{D}\left[
x\left( \tau \right) \right] \delta \left[ x\left( t_{c}\right) -x_{c}\right]
\end{equation*}
and we use the Fourier transform of the $\delta $ function 
\begin{eqnarray*}
P\left( x_{c},t_{c};-\sqrt{\frac{a}{b}},-T\right) &=&\int \emph{D}\left[
x\left( \tau \right) \right] \delta \left[ x\left( t_{c}\right) -x_{c}\right]
\exp \left( \emph{S}\right) \\
&=&\frac{1}{2\pi }\int d\lambda \exp \left( -i\lambda x_{c}\right) \int 
\emph{D}\left[ x\left( \tau \right) \right] \exp \left[ S+i\lambda x\left(
t_{c}\right) \right]
\end{eqnarray*}
It is convenient to write 
\begin{eqnarray*}
\emph{S}+i\lambda x\left( t_{c}\right) &=&i\int_{-T}^{T}d\tau \emph{L}\left[
x\left( t\right) ,k\left( t\right) \right] +i\int_{-T}^{T}d\tau \lambda
x\left( \tau \right) \delta \left( \tau -t_{c}\right) \\
&=&i\int_{-T}^{T}d\tau \left[ \emph{L}+\lambda x\left( \tau \right) \delta
\left( \tau -t_{c}\right) \right]
\end{eqnarray*}
We can see that the new term plays the same r\^{o}le as the external current 
$J_{1}\left( \tau \right) $ \ and this sugests to return to the Eq.(\ref
{zands}) and to perform the modification 
\begin{equation*}
J_{1}\left( \tau \right) \rightarrow J_{1}^{\prime }\left( \tau \right)
=J_{1}\left( \tau \right) +\lambda \delta \left( \tau -t_{c}\right)
\end{equation*}
obtaining 
\begin{equation*}
P\left( x_{c},t_{c};-\sqrt{\frac{a}{b}},-T\right) =\left. \frac{1}{2\pi }%
\int d\lambda \exp \left( -i\lambda x_{c}\right) \emph{Z}^{\left( \lambda
\right) }\left[ J_{1}\right] \right| _{J_{1}=0}
\end{equation*}

Using the previous results we have 
\begin{eqnarray}
\left. \emph{Z}^{\left( \lambda \right) }\left[ J_{1}\right] \right|
_{J_{1}=0} &=&\exp \left\{ i\frac{1}{2}\lambda x_{\eta }\left( t_{c}\right)
\right.  \label{zlamb1} \\
&&+i\frac{1}{2}\lambda ^{2}\Delta _{11}\left( t_{c},t_{c}\right)  \notag \\
&&\left. \left. +i\frac{1}{2}\left[ V-\left( \frac{dV}{dx}\right) x\right]
_{t=t_{c}}\lambda \Delta _{21}\left( t_{c},t_{c}\right) \right\} \right|
_{J_{1}=0}  \notag
\end{eqnarray}
This expression will have to be calculated at $t_{c}=T$ and $x_{c}=-\sqrt{a/b%
}$. 
\begin{eqnarray*}
V\left( -\sqrt{\frac{a}{b}}\right) &=&0 \\
\left. \frac{dV}{dx}\right| _{x=-\sqrt{\frac{a}{b}}} &=&0
\end{eqnarray*}
so the last term is zero. 
\begin{equation}
\left. \emph{Z}^{\left( \lambda \right) }\left[ J_{1}\right] \right|
_{J_{1}=0}=\left. \exp \left[ -i\frac{1}{2}\lambda \sqrt{\frac{a}{b}}+i\frac{%
1}{2}\lambda ^{2}\Delta _{11}\left( T,T\right) \right] \right| _{J_{1}=0}
\label{zdel}
\end{equation}
We need the propagator $\Delta _{21}$ in the expression for $\Delta _{11}$, 
\begin{equation*}
\Delta _{21}\left( t^{\prime },T\right) =\Theta \left( T-t^{\prime }\right)
\exp \left[ -W\left( t^{\prime }\right) \right] \exp \left[ W\left( T\right) %
\right]
\end{equation*}
\begin{equation*}
W\left( T\right) =0
\end{equation*}
Then the intration giving the propagator $\Delta _{11}$ is 
\begin{eqnarray*}
&&\int_{-T}^{T}dt^{\prime }2iD\Theta \left( T-t^{\prime }\right) \Theta
\left( T-t^{\prime }\right) \exp \left[ -W\left( t^{\prime }\right) \right]
\exp \left[ -W\left( t^{\prime }\right) \right] \\
&=&2iD\int_{-T}^{T}dt^{\prime }\exp \left[ -2W\left( t^{\prime }\right) %
\right] \\
&=&2iD\left\{ \exp \left[ -4aT+6at_{1}-6at_{0}\right] \int_{-T}^{t_{0}}dt%
\exp \left( 4at\right) \right. \\
&&+\exp \left[ -4aT+6at_{1}\right] \int_{t_{0}}^{t_{1}}dt\exp \left(
-2at\right) + \\
&&\left. +\exp \left[ -4aT\right] \int_{t_{1}}^{T}dt\exp \left( 4at\right)
\right\}
\end{eqnarray*}
We introduce the notation 
\begin{eqnarray*}
Y\left( t_{0},t_{1}\right) &\equiv &1+ \\
&&+3\exp \left[ -4a\left( T-t_{1}\right) \right] \left\{ \exp \left[
2a\left( t_{1}-t_{0}\right) \right] -1\right\} \\
&&-\exp \left[ -8aT\right] \exp \left[ 6a\left( t_{1}-t_{0}\right) \right]
\end{eqnarray*}
and we obtain the expression 
\begin{equation*}
\Delta _{11}\left( T,T\right) =\frac{iD}{2a}Y\left( t_{0},t_{1}\right)
\end{equation*}
Returning to (\ref{zdel}) 
\begin{equation*}
\emph{Z}^{\left( \lambda \right) }\left[ J_{1}=0\right] =\exp \left[ -i\frac{%
\lambda }{2}\sqrt{\frac{a}{b}}-\frac{\lambda ^{2}D}{4a}Y\right]
\end{equation*}
and the probability is 
\begin{eqnarray*}
P\left( t_{0},t_{1}\right) &=&\frac{1}{2\pi }\int_{-\infty }^{\infty
}d\lambda \exp \left[ i\lambda \sqrt{\frac{a}{b}}\right] \emph{Z}^{\left(
\lambda \right) }\left[ J_{1}=0\right] \\
&=&\frac{1}{2\pi }\int_{-\infty }^{\infty }d\lambda \exp \left[ \frac{1}{2}%
i\lambda \sqrt{\frac{a}{b}}-\frac{\lambda ^{2}D}{4a}Y\left(
t_{0},t_{1}\right) \right]
\end{eqnarray*}
We have made manifest the dependence of the probability on the arbitrary
times of jump $t_{0},t_{1}$. Integrating on $\lambda $%
\begin{equation*}
P\left( t_{0},t_{1}\right) =\sqrt{\frac{a}{\pi DY}}\exp \left( -\frac{a^{2}}{%
4bDY}\right)
\end{equation*}

This expression contains an arbitrary parameter, which is the duration of
stay in the intermediate position, \emph{i.e.} at the unstable extremum of
the potantial, $x=0$. Integrating over this duration, $t_{0}-t_{1}$ ,
adimensionalized with the unit $1/a$, for all possible values between $0$
and all the time interval, $2T$, will give the probability. In doing this
time integration we will make a simplification to consider that $T$ is a
large quantity, such that the exponential terms in $Y$ will vanish. Then $Y$
reduces to $1$ and the integration is trivial 
\begin{eqnarray*}
P &=&\int_{0}^{2T}d\left[ a\left( t_{1}-t_{0}\right) \right] P\left(
t_{0},t_{1}\right) \\
&=&2Ta\sqrt{\frac{a}{\pi D}}\exp \left( -\frac{a^{2}}{4bD}\right)
\end{eqnarray*}
This is our answer. To compare with the result of Caroli et al., we first
adopt their notation $T\equiv t/2$ and we write the potential as 
\begin{equation*}
V\left( x\right) =-U^{\prime }\left( x\right)
\end{equation*}
with 
\begin{equation*}
U\left( x\right) =-\frac{1}{2}ax^{2}+\frac{1}{4}bx^{4}
\end{equation*}
We note that 
\begin{equation*}
U\left( x=0\right) -U\left( -\sqrt{\frac{a}{b}}\right) =0+\frac{a^{2}}{4b}
\end{equation*}
which allows to write 
\begin{equation*}
P=t\frac{a^{3/2}}{\sqrt{\pi D}}\exp \left[ -\frac{U\left( 0\right) -U\left( -%
\sqrt{\frac{a}{b}}\right) }{D}\right]
\end{equation*}
which corresponds to the result of Caroli et al. for a single
instanton-anti-instanton pair, where 
\begin{equation*}
a^{3/2}\sim \left| U_{0}^{\prime \prime }\right| ^{1/2}U_{-\sqrt{a/b}%
}^{\prime \prime }
\end{equation*}

The only problem is the need to consider the reverse of sign due to the
virtual reflection at the position $x=0$. In the tratment of Caroli, it is
included by an integration in complex time plane, after the second jump, $%
t_{1}$.

\subsection{The trajectory reflection and the integration along the contour
in the complex time plane}

The problem arises at the examination of the Eq.(\ref{K(1)raw}) where it is
noticed that the constant $C$ multiplying the exponential term \emph{in the
exponential} in the integrations over $t_{0}$ and $t_{1}$ is positive (see
Eq.(\ref{constC})). Then the dependence of the integrand on $t_{1}-t_{0}$
makes that the most important contribution to the integration over the
variable $t_{1}$ to come from 
\begin{equation*}
t_{1}-t_{0}\ll \frac{1}{\left( 2W_{0}^{\prime \prime }\right) ^{1/2}}=\omega
_{0}^{-1}\sim \Delta t
\end{equation*}
Here the integrand is of the order 
\begin{equation*}
\sim \exp \left( \frac{C}{D}\right)
\end{equation*}
This would imply, in physical terms, that the most important configurations
contributing to the path integral would be pairs of instanton anti-instanton
very closely separated, \emph{i.e.} transitions with very short time of stay
in the point $0$. This is not correct since we expect that in large time
regime the trajectories assembled from instantons and anti-instantons with
arbitrary large separations (duration of residence in $0$ before returning
to $b$) should contribute to $K$.

This problem can be solved by extending the integration over the variable
representing the time separation between the transitions 
\begin{equation*}
z=t_{1}-t_{0}
\end{equation*}
in complex.

\textbf{NOTE}. The quantity 
\begin{equation*}
\left( 2W_{0}^{\prime \prime }\right) ^{1/2}=\omega _{0}
\end{equation*}
is the frequency of the harmonic oscillations in the parabolic approximation
of the potential.

\bigskip

The contour \emph{C} is defined as 
\begin{eqnarray*}
\func{Im}z &=&\frac{\pi }{\left| U_{0}^{\prime \prime }\right| } \\
0 &\leqslant &\func{Re}z\leqslant \frac{t}{2}-t_{0}
\end{eqnarray*}
The form of the trajectory $x_{IA}$ connecting the point $b$ to itself, $b$,
assembled from one instanton $x_{I}$ and one anti-instanton $x_{A}$ can now
be described as follows.

The trajectory starts in the point $b$ and follows the usual form of the
instanton solution, which means that for times 
\begin{equation*}
-\frac{t}{2}<\tau \lesssim t_{0}
\end{equation*}
it practically conicides with the fixed position $b$. At $t_{0}$ it makes
the transition to the point $0$, the duration of the transition (width of
the instanton) being $\Delta t$. Then it remains almost fixed at the point $%
0 $. The problem is that there is \emph{no} turning point at $x=0$ since
there the potential $-W\left( x\right) $ is smaller than at $x=b$. Then we
do not have the classical instanton which takes an infinite time to reach
exactly $x=0$, etc. Also there is no exact instanton solution which starts
from $x=0$ to go back to $x=b$. In the assambled solution $x_{I}+x_{A}$,
there is a discontinuity at $x=0$ in the derivative, the velocity simply is
forced (assumed) to change from almost zero positive (directed toward $x=0$)
to almost zero negative (directed toward $x=b$). What we do is to introduce
a turning point, which limits effectively the transition solution, but for
this we have to go to complex time, to compensate for the negative value of
the potential. Making time imaginary changes the sign of the velocity
squared (kinetic energy) and correspondingly can be seen as a change of sign
of the potential which now becomes positive and presents a wall from where
the solution can reflect. This is actually a local Euclidean-ization of the
theory, around the point $x=0$.

At time $\left( t_{0}+t_{1}\right) /2$ the trajectory starts moving along
the imaginary $z$ axis. It reaches an imaginary turning point at $\frac{i\pi 
}{\left| U_{0}^{\prime \prime }\right| }$ and returns to the point $\func{Re}%
z=x_{IA}\left( \frac{t_{0}+t_{1}}{2}\right) $, $\func{Im}z=0$.

This trajectory can be considered the limiting form of the trajectory of a
fictitious particle that traverses a part of potential (forbidden
classically) when this part goes to zero.

\bigskip

Then the expression is 
\begin{equation*}
K^{\left( 1\right) }\left( b,\frac{t}{2};b,-\frac{t}{2}\right) =-t\left( 
\frac{\left| U_{0}^{\prime \prime }\right| }{2\pi D}\right) ^{1/2}\frac{%
\left( U_{b}^{\prime \prime }\left| U_{0}^{\prime \prime }\right| \right)
^{1/2}}{2\pi }\exp \left( -\frac{U_{0}-U_{b}}{D}\right)
\end{equation*}
where the minus sign is due to the reflection at the turning point.

\subsection{Summation of terms from multiple transitions}

\textbf{Calculation of the transitions implying the points }$b$ and $a$

This is the term 
\begin{equation*}
K_{ba}^{\left( 2\right) }\left( b,-\frac{t}{2};a,\frac{t}{2}\right)
\end{equation*}
and corresponds to the transitions 
\begin{equation*}
b\rightarrow 0\rightarrow a\rightarrow 0\rightarrow b
\end{equation*}
Again we have to introduce a new potential replacing $W$. The real potential 
$W\left( x\right) $ as it results from the Langevin equation, depende on $D$
which is supposed to be small. Then the ``expansion'' in this small
parameter $D$ has as zerth order the new potential $W^{\left( 0\right)
}\left( x\right) $ that will be used in the computations below. This is 
\begin{equation*}
W^{\left( 0\right) }=W-\delta W
\end{equation*}
\begin{equation*}
\delta W\left( x\right) =\left\{ 
\begin{array}{cc}
0 & x<x_{m} \\ 
W_{0}-W_{b} & x_{m}<x<x_{p} \\ 
W_{a}-W_{b} & x>x_{p}
\end{array}
\right.
\end{equation*}
where the new point $x_{p}$ is in the region of the right-hand minimum of $%
-W\left( x\right) $. This point will correspond to the center of the
instanton connecting $0$ to $a$.

We introduce the new family of instanton solution 
\begin{equation*}
\widetilde{x}_{I}\left( \tau -t_{1}\right)
\end{equation*}
in the potential $-W^{\left( 0\right) }$ that leave $x=0$ at $-\infty $ and
reach $x=a$ at time $\infty $.

With this instanton and the previously defines ones, we assamble a function
that provide approximately the transitions conncetin $b$ to $b$ via $a$.
This, obviously, is not a classical solution of the equation of motion,
although it is very close to a solution. 
\begin{equation*}
x_{IIAA}\left( \tau ;t_{0},t_{1},t_{2},t_{3}\right) =\left\{ 
\begin{array}{cc}
x_{I}\left( \tau -t_{0}\right) & \tau <\frac{1}{2}\left( t_{0}+t_{1}\right)
\\ 
\widetilde{x}_{I}\left( \tau -t_{1}\right) & \frac{1}{2}\left(
t_{0}+t_{1}\right) <\tau <\frac{1}{2}\left( t_{1}+t_{2}\right) \\ 
\widetilde{x}_{I}\left( t_{2}-\tau \right) & \frac{1}{2}\left(
t_{1}+t_{2}\right) <\tau <\frac{1}{2}\left( t_{2}+t_{3}\right) \\ 
x_{I}\left( t_{3}-\tau \right) & \tau >\frac{t_{2}-t_{3}}{2}
\end{array}
\right.
\end{equation*}
These trajectories are rather artificially, being assambled form pieces of
true solutions. They have singularities: in the middel, at $\left(
t_{1}+t_{2}\right) /2$ the function is continuous but the derivative is
discontinuous; in the points where one instanton arrives and another must
continue (\emph{i.e.} $\left( t_{0}+t_{1}\right) /2$ and $\left(
t_{2}+t_{3}\right) /2$, both the function and the derivatives are
discontinuous.

this is corrected by extending the trajectories in complex time plane. One
defines the variables 
\begin{eqnarray*}
z_{1} &=&t_{1}-t_{0} \\
z_{2} &=&t_{2}-t_{1} \\
z_{3} &=&t_{3}-t_{2}
\end{eqnarray*}
and the three integration contours 
\begin{equation*}
\begin{array}{ccc}
\emph{C}_{1} & \func{Im}z_{1}=\frac{\pi }{\left| U_{0}^{\prime \prime
}\right| } & 0\leqslant \func{Re}z_{1}\leqslant \frac{t}{2}-t_{0} \\ 
\emph{C}_{2} & \func{Im}z_{2}=\frac{\pi }{U_{a}^{\prime \prime }} & 
0\leqslant \func{Re}z_{2}\leqslant \func{Re}\left( \frac{t}{2}-t_{1}\right)
\\ 
\emph{C}_{3} & \func{Im}z_{3}=\frac{\pi }{\left| U_{0}^{\prime \prime
}\right| } & 0\leqslant \func{Re}z_{3}\leqslant \frac{t}{2}-t_{2}
\end{array}
\end{equation*}
Together these contours must assure the continuity of the function $x_{IA}$
in the region $0$, from which the condition arises 
\begin{equation*}
x_{I}\left[ \func{Re}\left( \frac{t_{1}-t_{0}}{2}\right) \right] =-%
\widetilde{x}_{I}\left[ \func{Re}\left( \frac{t_{0}-t_{1}}{2}\right) \right]
\end{equation*}
which gives for large separations $t_{1}-t_{0}$%
\begin{equation*}
x_{p}\exp \left( \left| U_{0}^{\prime \prime }\right| \Delta _{0p}\right)
=\left| x_{m}\right| \exp \left( \left| U_{0}^{\prime \prime }\right| \Delta
_{m0}\right)
\end{equation*}

The exprsssion of $K$ at this moment is 
\begin{eqnarray*}
K_{ba}^{\left( 2\right) }\left( b,\frac{t}{2};b,-\frac{t}{2}\right)
&=&\left( \frac{2\pi D}{U_{a}^{\prime \prime }}\right) ^{1/2}\frac{%
U_{b}^{\prime \prime }U_{a}^{\prime \prime }U_{0}^{\prime \prime 2}}{4}%
x_{m}^{2}x_{p}^{2}\left( \frac{\left| U_{0}^{\prime \prime }\right| }{2\pi D}%
\right) ^{3} \\
&&\times \exp \left[ -\frac{2U_{0}-U_{a}-U_{b}}{D}+2\left| U_{0}^{\prime
\prime }\right| \left( \Delta _{m0}+\Delta _{0p}\right) \right] \\
&&\times \int_{-t/2}^{t/2}dt_{0}\int_{\emph{C}_{1}}dz_{1}\int_{\emph{C}%
_{2}}dz_{2}\int_{\emph{C}_{3}}dz_{3}\exp \left[ F\left( z_{1}\right)
+G\left( z_{2}\right) +F\left( z_{3}\right) \right]
\end{eqnarray*}
where 
\begin{eqnarray}
F\left( z\right) &=&\frac{C_{0}}{D}\exp \left( -\left| U_{0}^{\prime \prime
}\right| z\right)  \label{Fz} \\
&&-\left| U_{0}^{\prime \prime }\right| \func{Re}z  \notag
\end{eqnarray}
\begin{equation}
G\left( z\right) =\frac{C_{a}}{D}\exp \left( -U_{a}^{\prime \prime }z\right)
\label{Gz}
\end{equation}
and 
\begin{equation}
C_{0}=\frac{x_{m}^{2}\left| U_{0}^{\prime \prime }\right| }{2}\exp \left(
2\left| U_{0}^{\prime \prime }\right| \Delta _{m0}\right)  \label{co}
\end{equation}
\begin{equation}
C_{a}=\frac{\left( a-x_{p}\right) ^{2}U_{a}^{\prime \prime }}{2}\exp \left(
2U_{a}^{\prime \prime }\Delta _{pa}\right)  \label{ca}
\end{equation}

It is necessary to carry out the complex time integrals.

\textbf{The }$z_{3}$\textbf{\ integral}. Here we have 
\begin{equation*}
\int_{\emph{C}_{3}}dz_{3}\exp \left\{ \frac{C_{0}}{D}\exp \left( -\left|
U_{0}^{\prime \prime }\right| z\right) -\left| U_{0}^{\prime \prime }\right| 
\func{Re}z\right\}
\end{equation*}
along the real axis of $z$, we can substitute 
\begin{equation*}
u=\exp \left( -\left| U_{0}^{\prime \prime }\right| z\right)
\end{equation*}
and obtain 
\begin{equation*}
\int_{\func{Re}\emph{C}_{3}^{\prime }}\frac{-1}{\left| U_{0}^{\prime \prime
}\right| }du\exp \left( -\frac{C_{0}}{D}u\right) =-\frac{D}{\left|
U_{0}^{\prime \prime }\right| C_{0}}\int dw\exp \left( -w\right)
\end{equation*}
This simply results in the constant 
\begin{equation}
\frac{D}{C_{0}\left| U_{0}^{\prime \prime }\right| }  \label{rezc3}
\end{equation}

\textbf{The }$z_{2}$\textbf{\ integral}; The $z_{2}$ integral involves only
the exponential of $G\left( z_{2}\right) $ and the latter function does not
contain a term proportional with $\func{Re}z_{2}$ (comapre with (\ref{Fz})),
but only the exponential term. The function $G\left( z_{2}\right) $ is very
close to $1$ as soon as 
\begin{equation*}
\func{Re}z_{2}\gg t_{S}^{\left( a\right) }\simeq \frac{1}{U_{a}^{\prime
\prime }}\ln \left( \frac{C_{a}}{D}\right)
\end{equation*}
where $t_{S}^{\left( a\right) }$ is the Suzuki time for the region $a$%
\begin{equation*}
t_{S}^{\left( a\right) }\ll t\sim \tau _{K}
\end{equation*}
Then, with very good approximation 
\begin{equation*}
\int_{\emph{C}_{2}}dz_{2}\exp \left[ G\left( z_{2}\right) \right] =\int_{0}^{%
\frac{t}{2}-t_{1}}dz_{2}\exp \left[ G\left( z_{2}\right) \right] \simeq 
\frac{t}{2}
\end{equation*}
\textbf{The }$z_{1}$ \textbf{integral}. This is similar to the integral over 
$z_{3}$ and will imply again $\frac{D}{C_{0}\left| U_{0}^{\prime \prime
}\right| }$ as in (\ref{rezc3}).

Finally we note that only the constant $C_{0}$ relative to the intermediate
point $0$ appears, and the one for the farthest point of the trajectory, $a$%
, is absent.

There will be then a time integration over the moment of the first
transition, $t_{0}$, of an integrand that contains the time $t$ arising from
the integration on the contour $\emph{C}_{2}$, in the complex plane of the
variable $z_{2}=t_{2}-t_{1}$. This variable represents the duration between
the last transition to $a$ and the first transition from $a$, \emph{i.e.} is
the duration of stay in $a$. This time integration will produce a term with $%
t^{2}$.

Replacing in the expressions resulting from the integrations on $z_{1}$ and $%
z_{3}$ the constant $C_{0}$ from (\ref{co}) the factors $x_{m}$ and $x_{p}$
disappear.

The result is 
\begin{eqnarray*}
K_{ba}^{\left( 2\right) }\left( b,\frac{t}{2};b,-\frac{t}{2}\right) &=&\frac{%
t^{2}}{2!}\left( \frac{\left| U_{b}^{\prime \prime }\right| }{2\pi D}\right)
^{1/2}\frac{\left( \left| U_{0}^{\prime \prime }\right| U_{a}^{\prime \prime
}\right) ^{1/2}}{2\pi }\frac{\left( \left| U_{0}^{\prime \prime }\right|
U_{b}^{\prime \prime }\right) ^{1/2}}{2\pi } \\
&&\times \exp \left( -\frac{2U_{0}-U_{a}-U_{b}}{D}\right)
\end{eqnarray*}

\section{The probability}

The objective is to calculate $P\left( b,\frac{t}{2};b,-\frac{t}{2}\right) $
on the basis of the above results. We have to sum over any number of
independent pseudomolecules of the four species 
\begin{equation*}
\begin{array}{ccc}
\left( b0b\right) & \longleftrightarrow & -\alpha _{b} \\ 
\left( b0a\right) & \longleftrightarrow & \alpha _{b} \\ 
\left( a0b\right) & \longleftrightarrow & \alpha _{a} \\ 
\left( b0b\right) & \longleftrightarrow & -\alpha _{a}
\end{array}
\end{equation*}
where the factor which is associated to each pseudomolecule is 
\begin{equation*}
\alpha _{i}=\frac{\left( \left| U_{0}^{\prime \prime }\right| U_{a}^{\prime
\prime }\right) ^{1/2}}{2\pi }\exp \left( -\frac{U_{0}-U_{i}}{D}\right)
\;,\;i=a,b
\end{equation*}

A particular path containing $n$ $(b0b)$ and $m$ $\left( a0a\right) $ has
the form 
\begin{equation*}
\left( -1\right) ^{n+m}\frac{t^{n+m+2p}}{\left( n+m+2p\right) !}\alpha
_{b}^{n+p}\alpha _{a}^{p+m}\left( \frac{\left| U_{b}^{\prime \prime }\right| 
}{2\pi D}\right) ^{1/2}
\end{equation*}
It was taken into account that a path $\left( b\rightarrow b\right) $
contains the same number $p$ of $\left( b0a\right) $ and of $\left(
a0b\right) $.

\subsubsection{Transition and self-transition probabilities at the
equilibrium points}

The results from the previous calculations are listed below.

\textbf{The probability that a particle initially at }$\left( b,-\frac{t}{2}%
\right) $ \textbf{will be found at }$\left( b,\frac{t}{2}\right) $ 
\begin{equation*}
P\left( b,\frac{t}{2};b,-\frac{t}{2}\right) =\left( \frac{\left|
U_{b}^{\prime \prime }\right| }{2\pi D}\right) ^{1/2}\frac{1}{\alpha
_{a}+\alpha _{b}}\left\{ \alpha _{a}+\alpha _{b}\exp \left[ -t\left( \alpha
_{a}+\alpha _{b}\right) \right] \right\}
\end{equation*}

\textbf{The probability that a particle initially at }$\left( b,-\frac{t}{2}%
\right) $ \textbf{will be found at }$\left( a,\frac{t}{2}\right) $%
\begin{equation*}
P\left( a,\frac{t}{2};b,-\frac{t}{2}\right) =\left( \frac{U_{a}^{\prime
\prime }}{2\pi D}\right) ^{1/2}\frac{\alpha _{b}}{\alpha _{a}+\alpha _{b}}%
\left\{ 1-\exp \left[ -t\left( \alpha _{a}+\alpha _{b}\right) \right]
\right\}
\end{equation*}

\textbf{The probability that a particle initially at }$\left( a,-\frac{t}{2}%
\right) $ \textbf{will be found at }$\left( b,\frac{t}{2}\right) $%
\begin{equation*}
P\left( b,\frac{t}{2};a,-\frac{t}{2}\right) =\left( \frac{U_{b}^{\prime
\prime }}{2\pi D}\right) ^{1/2}\frac{\alpha _{a}}{\alpha _{a}+\alpha _{b}}%
\left\{ 1-\exp \left[ -t\left( \alpha _{a}+\alpha _{b}\right) \right]
\right\}
\end{equation*}

\textbf{The probability that a particle initially at }$\left( a,-\frac{t}{2}%
\right) $ \textbf{will be found at }$\left( a,\frac{t}{2}\right) $%
\begin{equation*}
P\left( a,\frac{t}{2};a,-\frac{t}{2}\right) =\left( \frac{\left|
U_{a}^{\prime \prime }\right| }{2\pi D}\right) ^{1/2}\frac{1}{\alpha
_{a}+\alpha _{b}}\left\{ \alpha _{b}+\alpha _{a}\exp \left[ -t\left( \alpha
_{a}+\alpha _{b}\right) \right] \right\}
\end{equation*}

\subsubsection{Transitions between arbitrary points}

\textbf{The long time limit of the probability density} 
\begin{equation*}
P\left( x,\frac{t}{2};x_{0},-\frac{t}{2}\right)
\end{equation*}
The formula has been derived for the case where the two positions $x$ and $%
x_{0}$ belong to the harmonic regions $\left( a\right) $ and $\left(
b\right) $. This is because asymptotically these regions will be populated.

Consider the case where 
\begin{eqnarray*}
x_{0} &\in &\left( b\right) \\
x &\in &\left( a\right)
\end{eqnarray*}
The first trajectory contributing to the path integral is the direct
connection between $b$ and $a$. The contribution to $K$ is 
\begin{equation*}
K^{direct}\left( x,\frac{t}{2};x_{0},-\frac{t}{2}\right) \sim \exp \left( -%
\frac{S_{b0}+S_{0a}}{D}\right)
\end{equation*}

Since the points of start and/or arrival $x_{0},x$ are different of $b,a$,
there is finite (\emph{i.e.} non exponentially small) slope of the solution,
and there is nomore degeneracy with respect to the translation of the center
of the instantons. then there will be not a proportionality with time in the
one-instanton term.

The terms connecting $x_{0}$ with $x$ with onely one intermediate step in
either $b$ or $a$ have comparable contributions to the action.

Now the term with two intermediate steps 
\begin{equation*}
x_{0}\rightarrow b\rightarrow a\rightarrow x
\end{equation*}
In the very long time regime, $K^{\left( 1\right) }$ can be factorized 
\begin{eqnarray*}
K^{\left( 1\right) }\left( x,\frac{t}{2};x_{0},-\frac{t}{2}\right)
&=&K_{harm}\left( x,\frac{t}{2};a,t^{\prime }\right) \\
&&\times \left( \frac{2\pi D}{U_{a}^{\prime \prime }}\right) ^{1/2}K^{\left(
1\right) }\left( a,t^{\prime };b,t^{\prime \prime }\right) \left( \frac{2\pi
D}{U_{b}^{\prime \prime }}\right) ^{1/2} \\
&&\times K_{harm}\left( b,t^{\prime \prime };x_{0},-\frac{t}{2}\right)
\end{eqnarray*}
This equation is independent of $t^{\prime }$ and $t^{\prime \prime }$ as
long as 
\begin{equation*}
t_{K}\gg \frac{t}{2}-t^{\prime }\;\text{and}\;t^{\prime \prime }-\left( -%
\frac{t}{2}\right) \;\gg t_{S}
\end{equation*}
Then 
\begin{equation*}
K_{harm}\left( x,\frac{t}{2};a,t^{\prime }\right) \simeq \left( \frac{%
U_{a}^{\prime \prime }}{2\pi D}\right) ^{1/2}\exp \left[ -\frac{U\left(
x\right) -U_{a}}{2D}\right]
\end{equation*}
There is also the approximate equality 
\begin{equation*}
K^{\left( 1\right) }\left( a,t^{\prime };b,t^{\prime \prime }\right) \approx
K^{\left( 1\right) }\left( a,\frac{t}{2};b,-\frac{t}{2}\right)
\end{equation*}
It results that all the contributions are included if in the expression of $%
K^{\left( 1\right) }\left( x,\frac{t}{2};x_{0},-\frac{t}{2}\right) $ we
replace the middle factor, which has become $K^{\left( 1\right) }\left( a,%
\frac{t}{2};b,-\frac{t}{2}\right) $ by the full $K\left( a,\frac{t}{2};b,-%
\frac{t}{2}\right) $. It is obtained 
\begin{eqnarray*}
P\left( x,\frac{t}{2};x_{0},-\frac{t}{2}\right) &=&\exp \left[ -\frac{%
U\left( x\right) -U_{a}}{2D}\right] P\left( a,\frac{t}{2};b,-\frac{t}{2}%
\right) \\
&=&\exp \left[ -\frac{U\left( x\right) -U_{a}}{2D}\right] \left( \frac{%
U_{a}^{\prime \prime }}{2\pi D}\right) ^{1/2}\frac{\alpha _{b}}{\alpha
_{a}+\alpha _{b}}\left\{ 1-\exp \left[ -t\left( \alpha _{a}+\alpha
_{b}\right) \right] \right\}
\end{eqnarray*}

\textbf{The case where both }$x$ \textbf{and }$x_{0}$ \textbf{belong to the
harmonic region }$\left( b\right) $

The same calculation shows that 
\begin{equation*}
P\left( x,\frac{t}{2};x_{0},-\frac{t}{2}\right) =\exp \left[ -\frac{U\left(
x\right) -U_{b}}{2D}\right] \left( \frac{U_{b}^{\prime \prime }}{2\pi D}%
\right) ^{1/2}\frac{1}{\alpha _{a}+\alpha _{b}}\left\{ \alpha _{a}+\alpha
_{b}\exp \left[ -t\left( \alpha _{a}+\alpha _{b}\right) \right] \right\}
\end{equation*}

\section{Results}

In the following we reproduce the graphs of the time-dependent probability distribution of a system governed by the basic Langevin equation. Various initialisations are considered, showing rapid redistribution of the density of presence of the system. The speed of redistribuition is, naturally, connected to the asymmetry of the potential. Each figure consists of a set of graphs: 
\begin{itemize}
\item the potential functions $V(x)$, $U(x)$ and $W(x)$
\item four probabilities of passage from and between the two equilibrium points $a$ and $b$;
\item the average value of the position of the system, as function of time, for the two most characteristic initializations
\end{itemize}
The two parameters $a$ and $D$ take different values, for illustration. 
(We apologize for the quality of the figures. Better but larger PS version can be downloaded from http://florin.spineanu.free.fr/sciarchive/topicalreview.ps )

\begin{figure}[htbp]
\centering
\begin{minipage}[b]{0.40\textwidth}
    \centering
    \includegraphics[height=11cm]{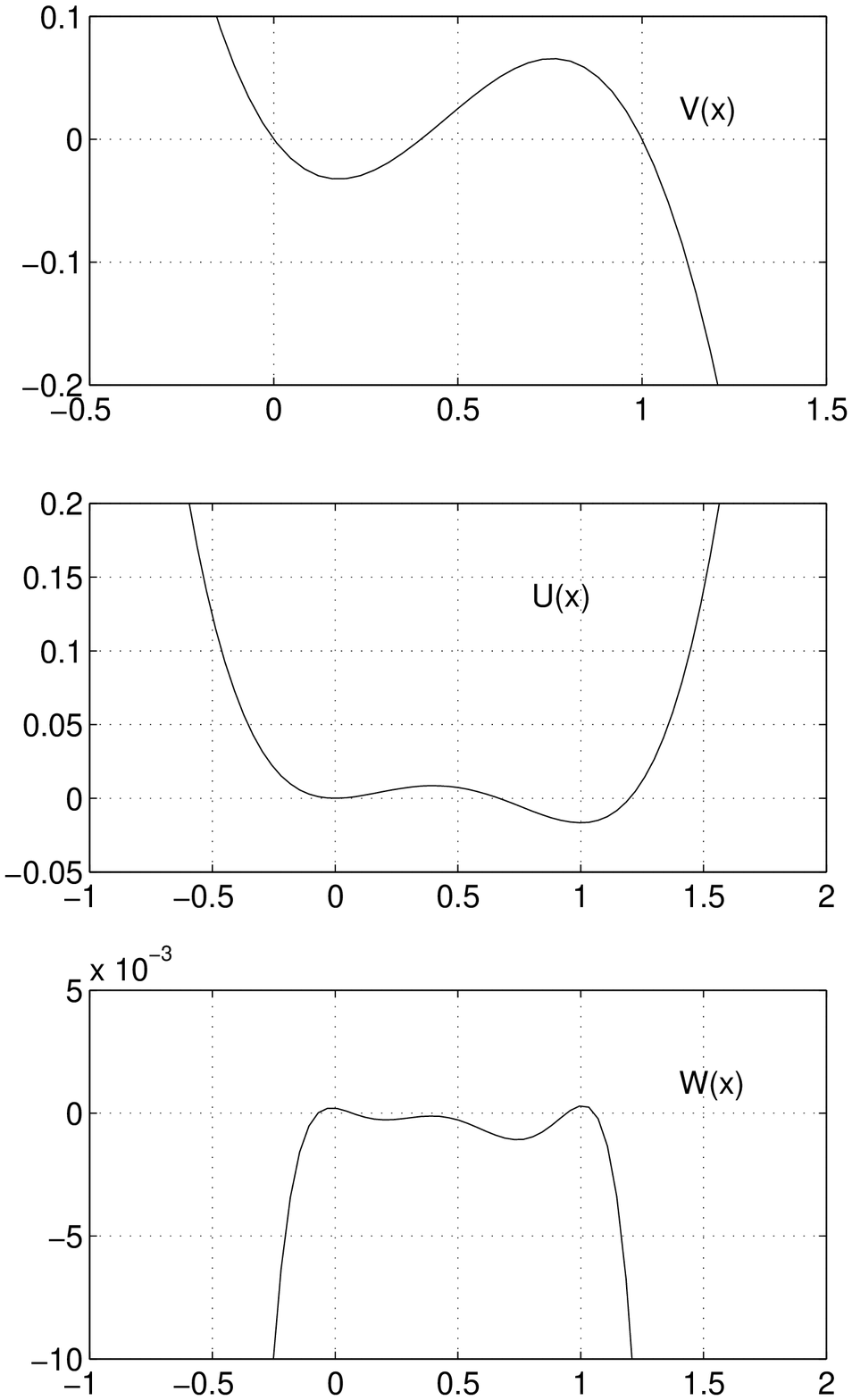}
   \end{minipage} \hspace{0.1\textwidth} 
\begin{minipage}[b]{0.40\textwidth}
    \centering
    \includegraphics[height=11.5cm]{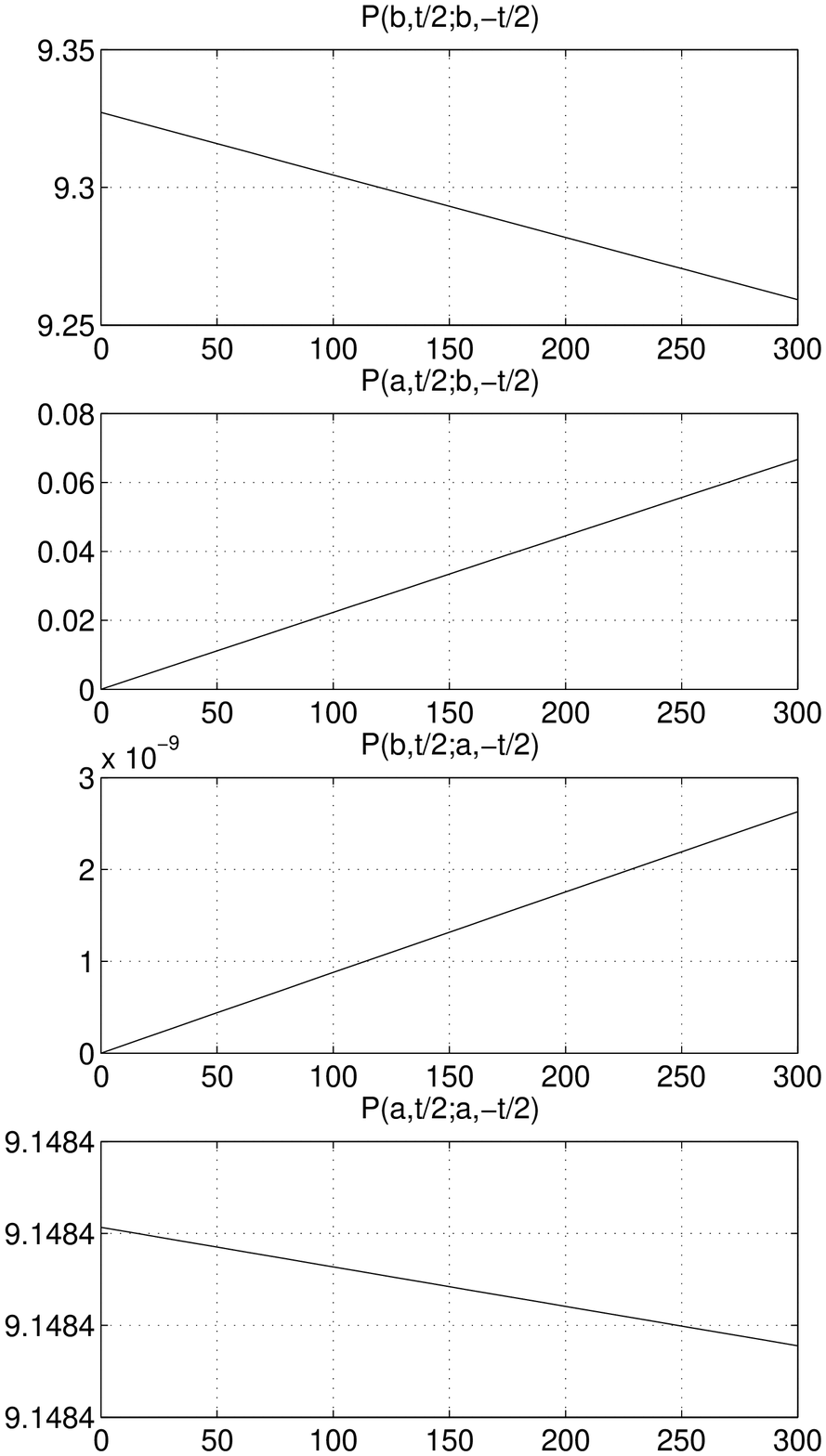}
   \end{minipage}
\par
\bigskip \centering
\begin{minipage}[b]{0.40\textwidth}
    \centering
    \includegraphics[height=5.cm]{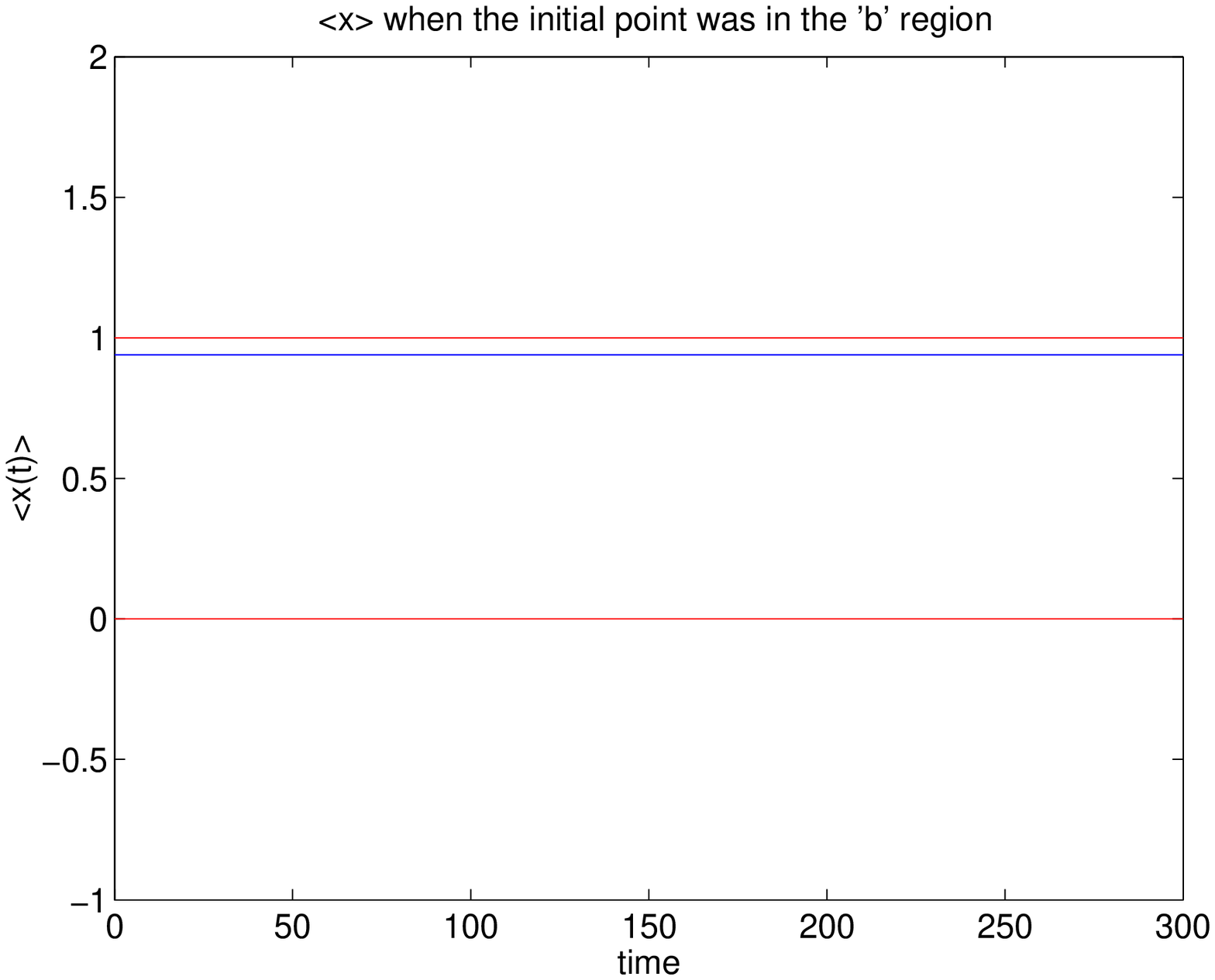}
   \end{minipage} \hspace{0.1\textwidth} 
\begin{minipage}[b]{0.40\textwidth}
    \centering
    \includegraphics[height=5.cm]{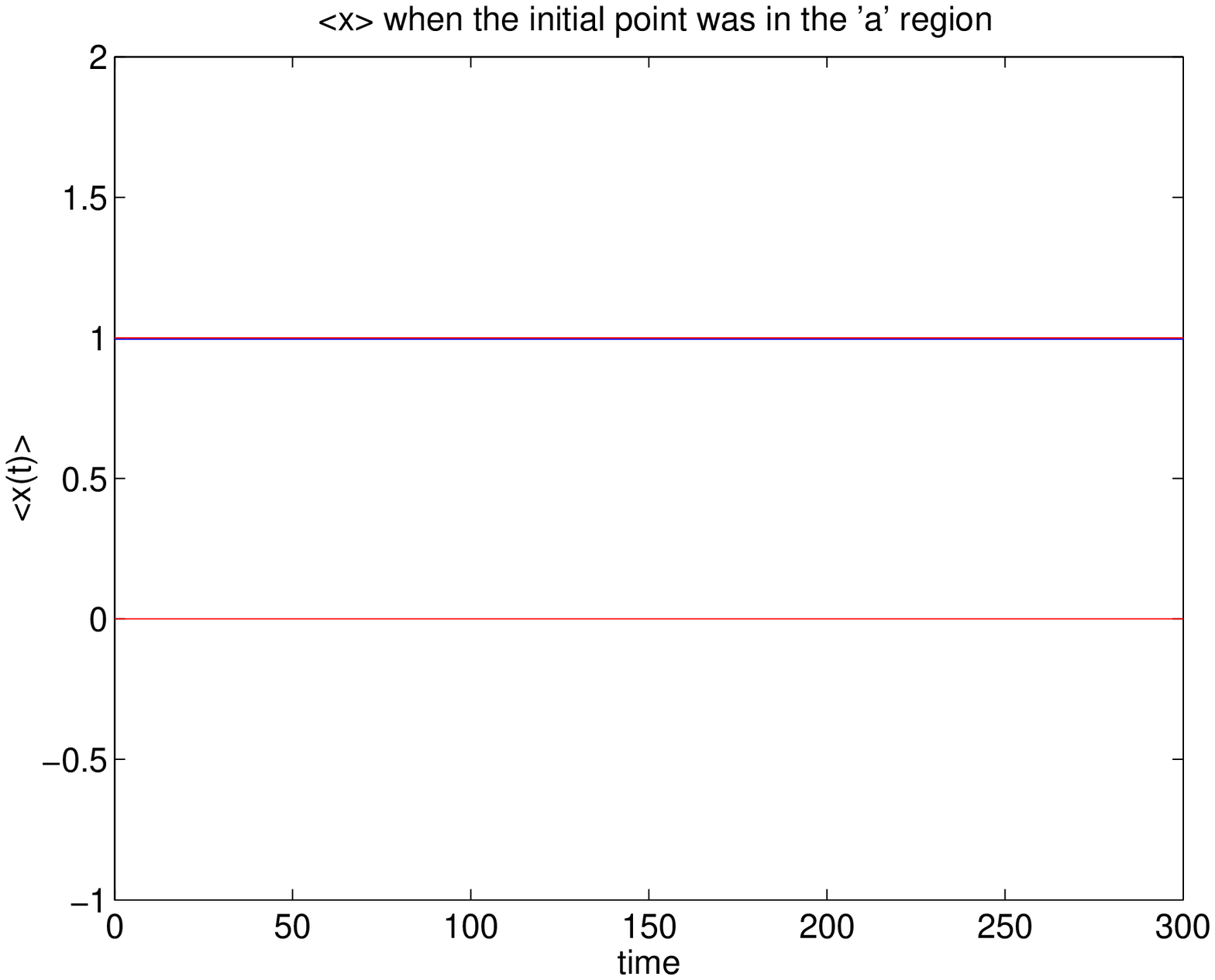}
   \end{minipage}
\end{figure}

\begin{figure}[htbp]
\centering
\includegraphics[height=9cm]{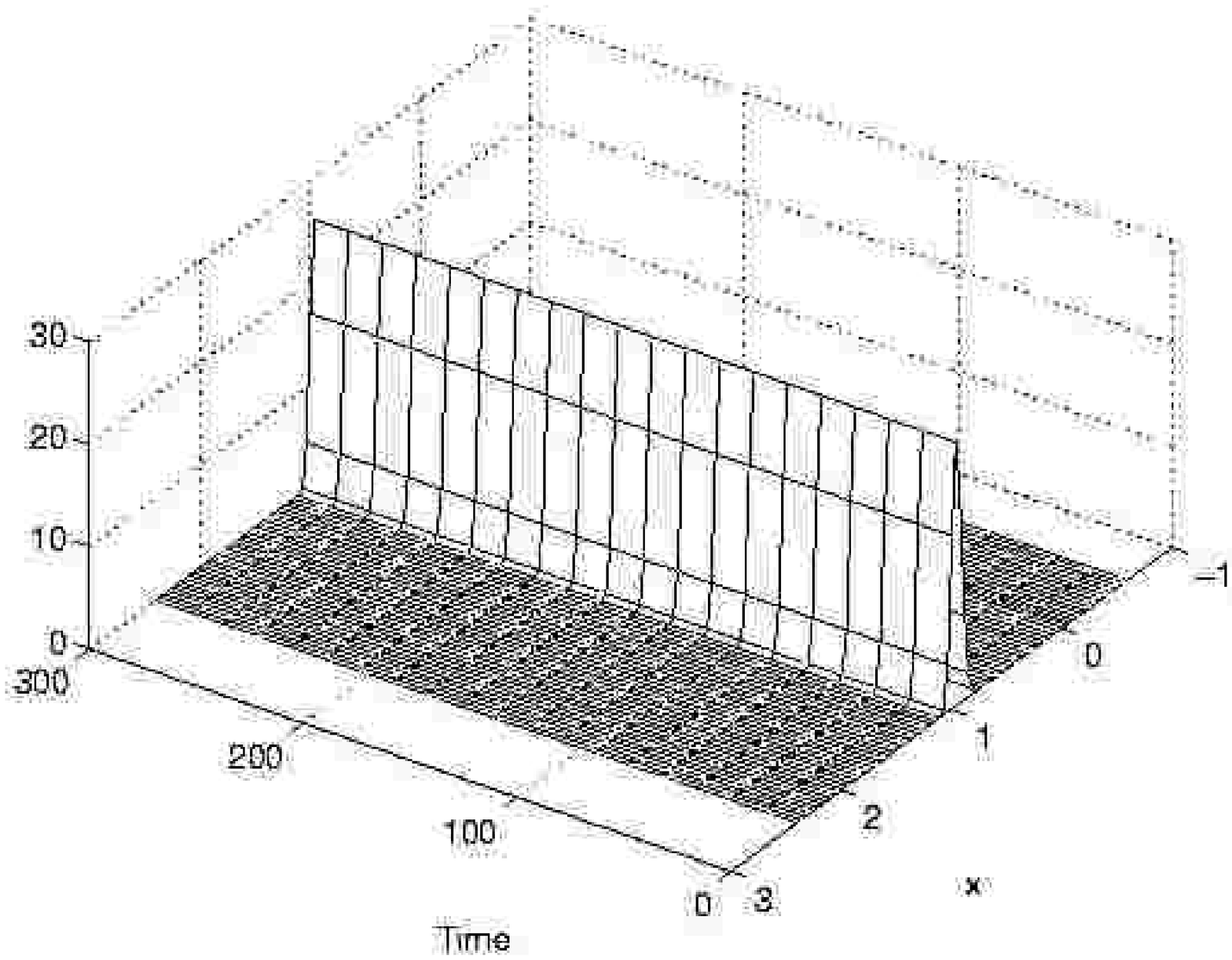}
\end{figure}
\begin{figure}[htbp]
\centering
\includegraphics[width=9cm]{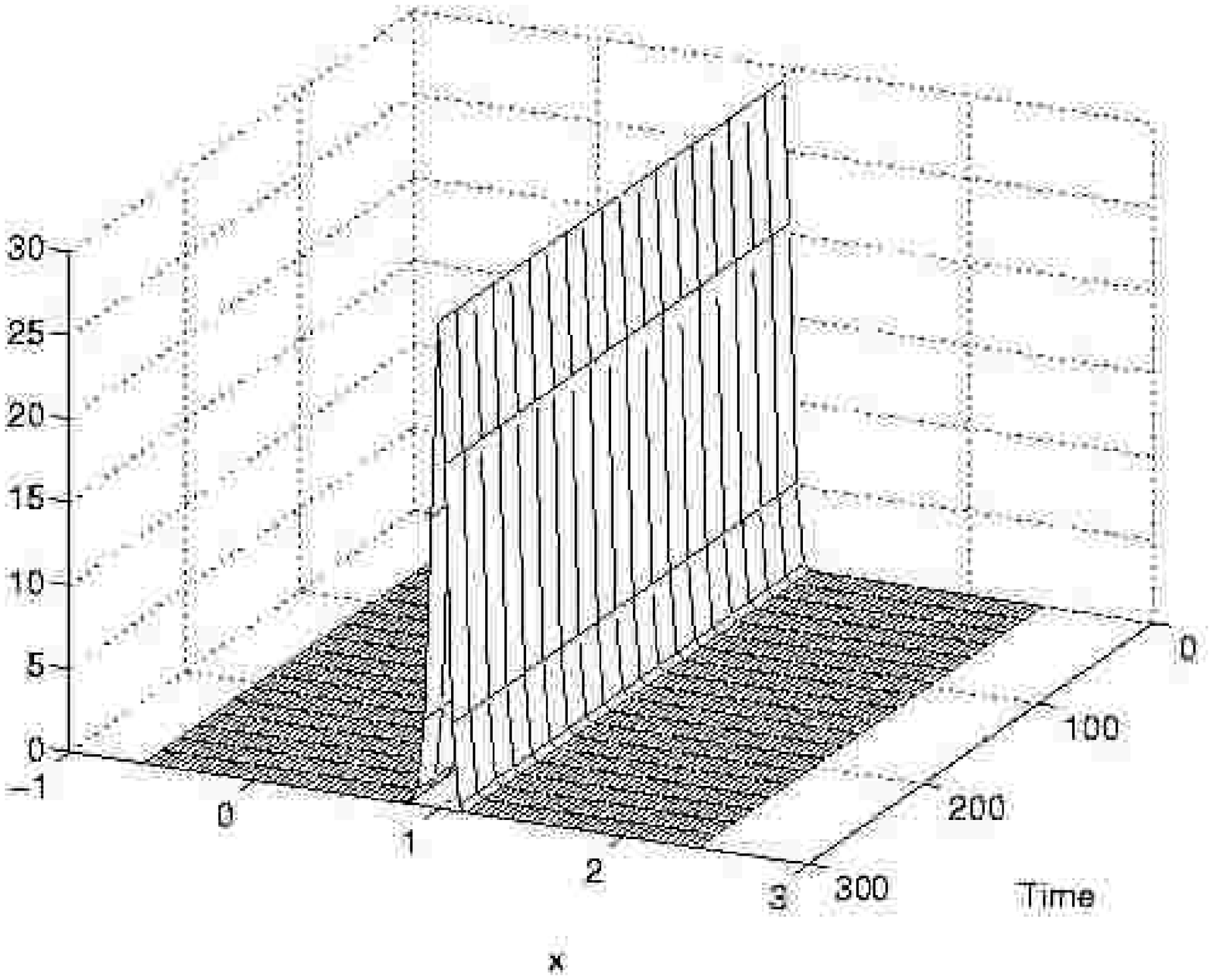}
\caption{Parameters: $a = 0.4$ and $D = 0.001$}
\label{Figa0.4D0.001_5}
\end{figure}

\clearpage

\begin{figure}[htb]
\centering
\begin{minipage}[b]{0.40\textwidth}
    \centering
    \includegraphics[height=11cm]{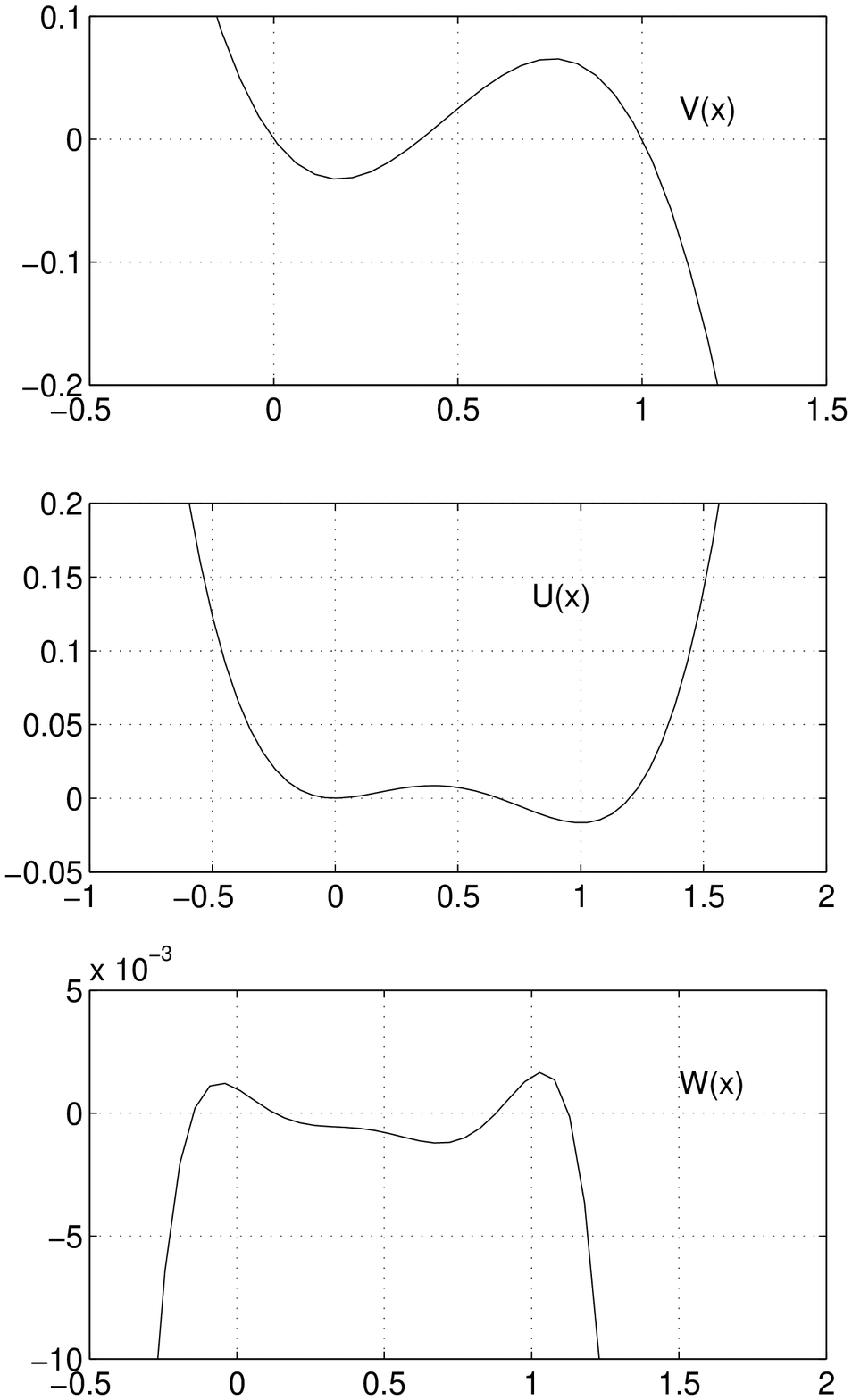}
   \end{minipage} \hspace{0.1\textwidth} 
\begin{minipage}[b]{0.40\textwidth}
    \centering
    \includegraphics[height=11.5cm]{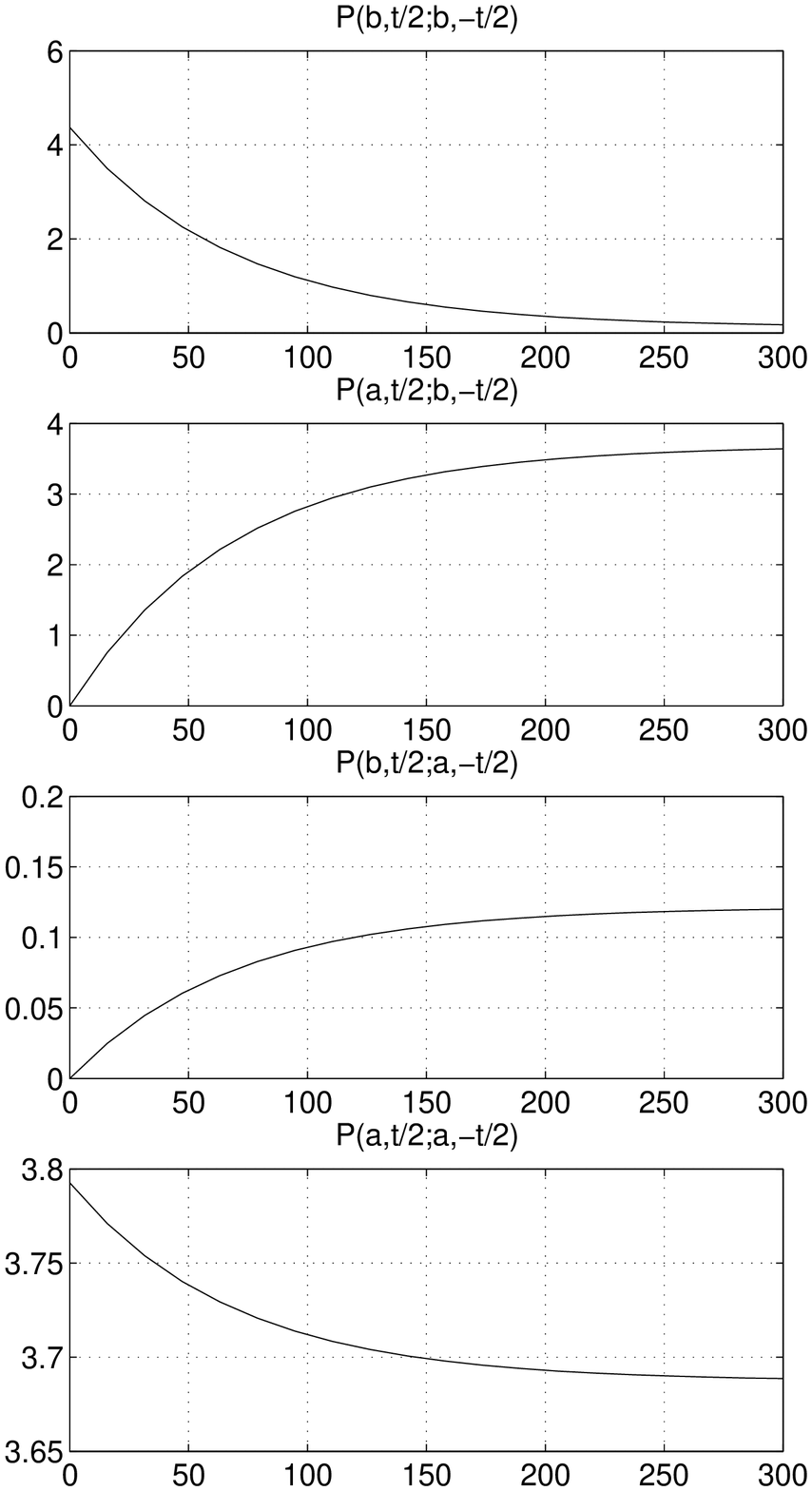}
   \end{minipage}
\par
\bigskip \centering
\begin{minipage}[b]{0.40\textwidth}
    \centering
    \includegraphics[height=5.cm]{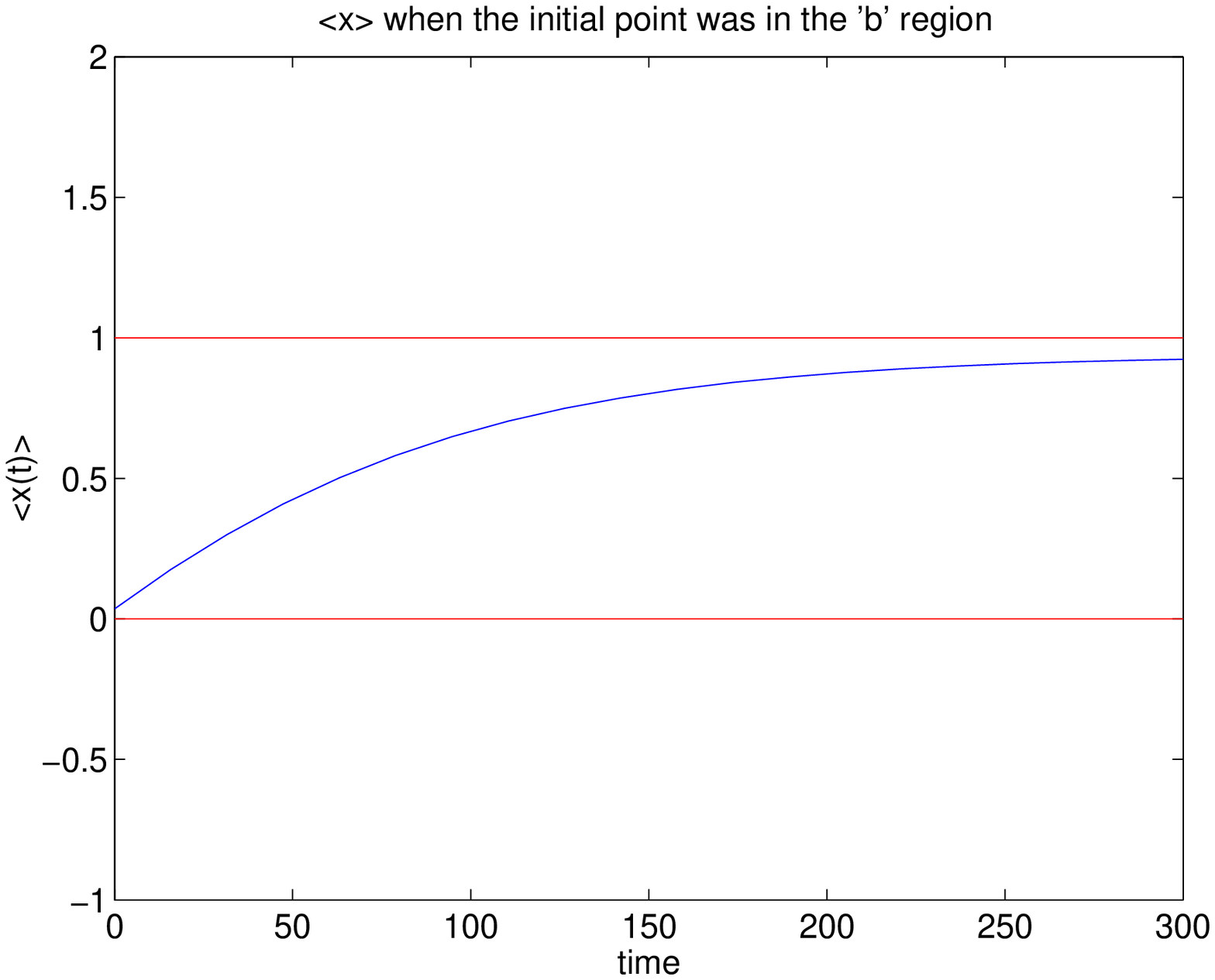}
   \end{minipage} \hspace{0.1\textwidth} 
\begin{minipage}[b]{0.40\textwidth}
    \centering
    \includegraphics[height=5.cm]{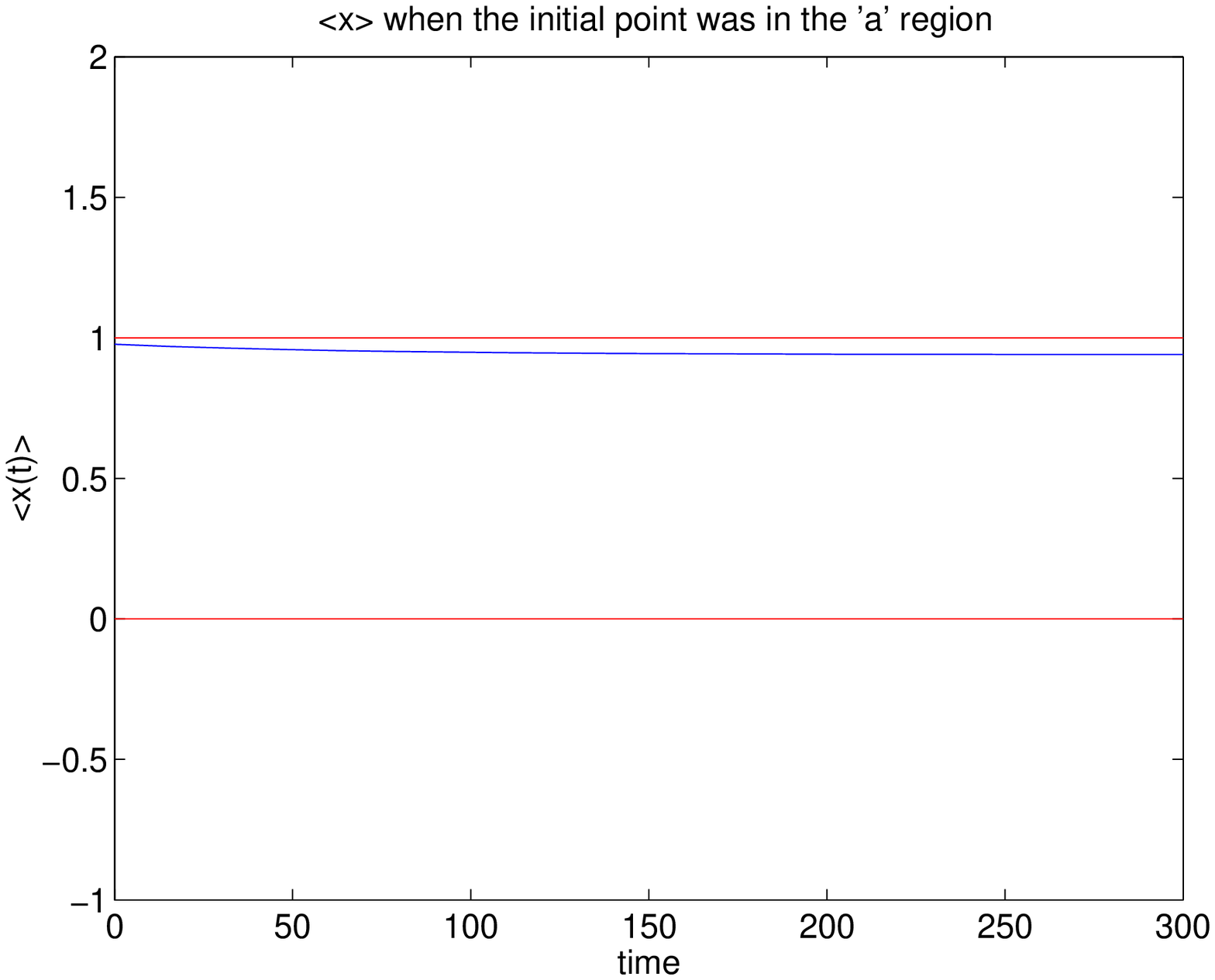}
   \end{minipage}
\end{figure}

\begin{figure}[htb]
\centering
\includegraphics[width=11cm]{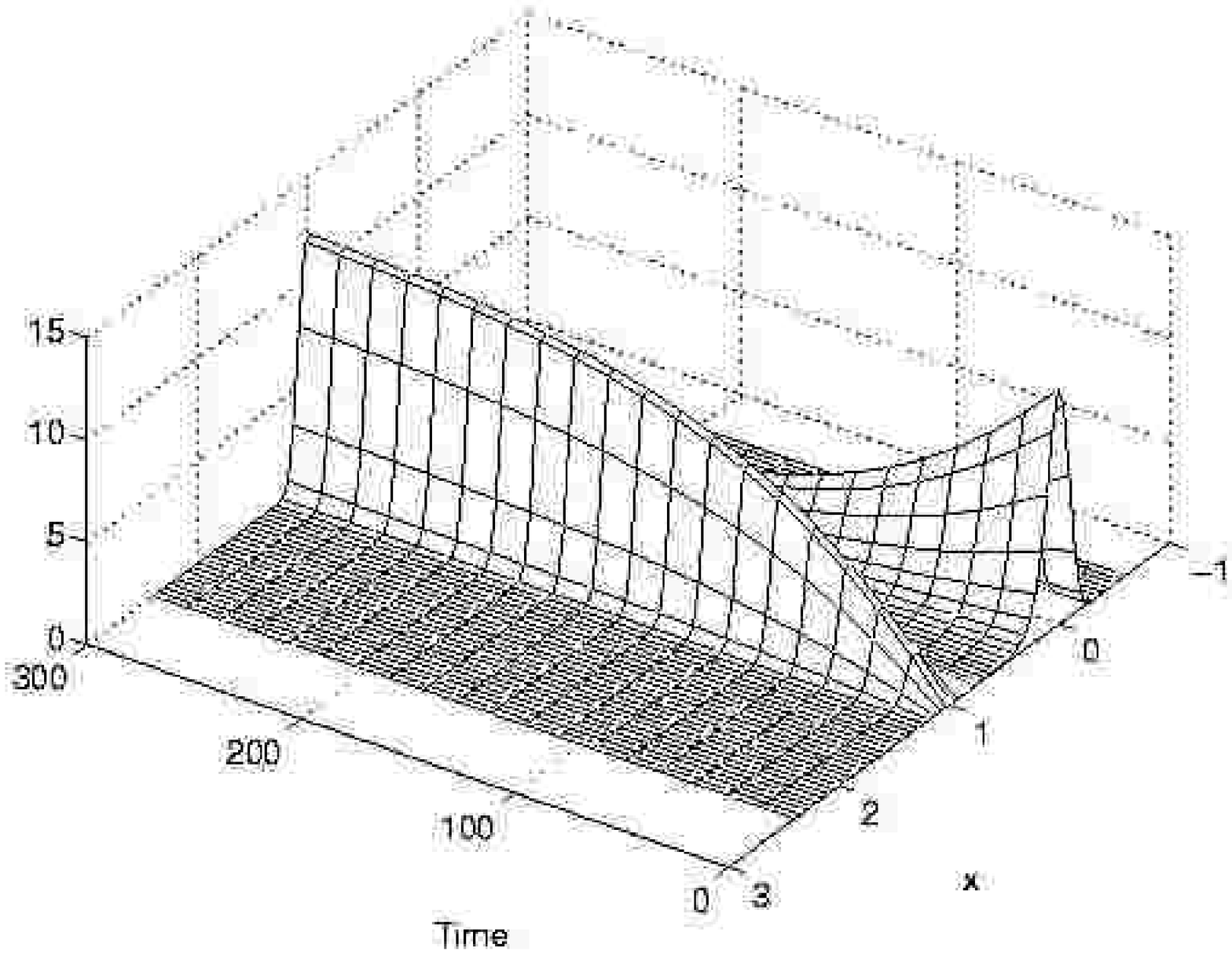}
\end{figure}
\begin{figure}[htbp]
\centering
\includegraphics[width=11cm]{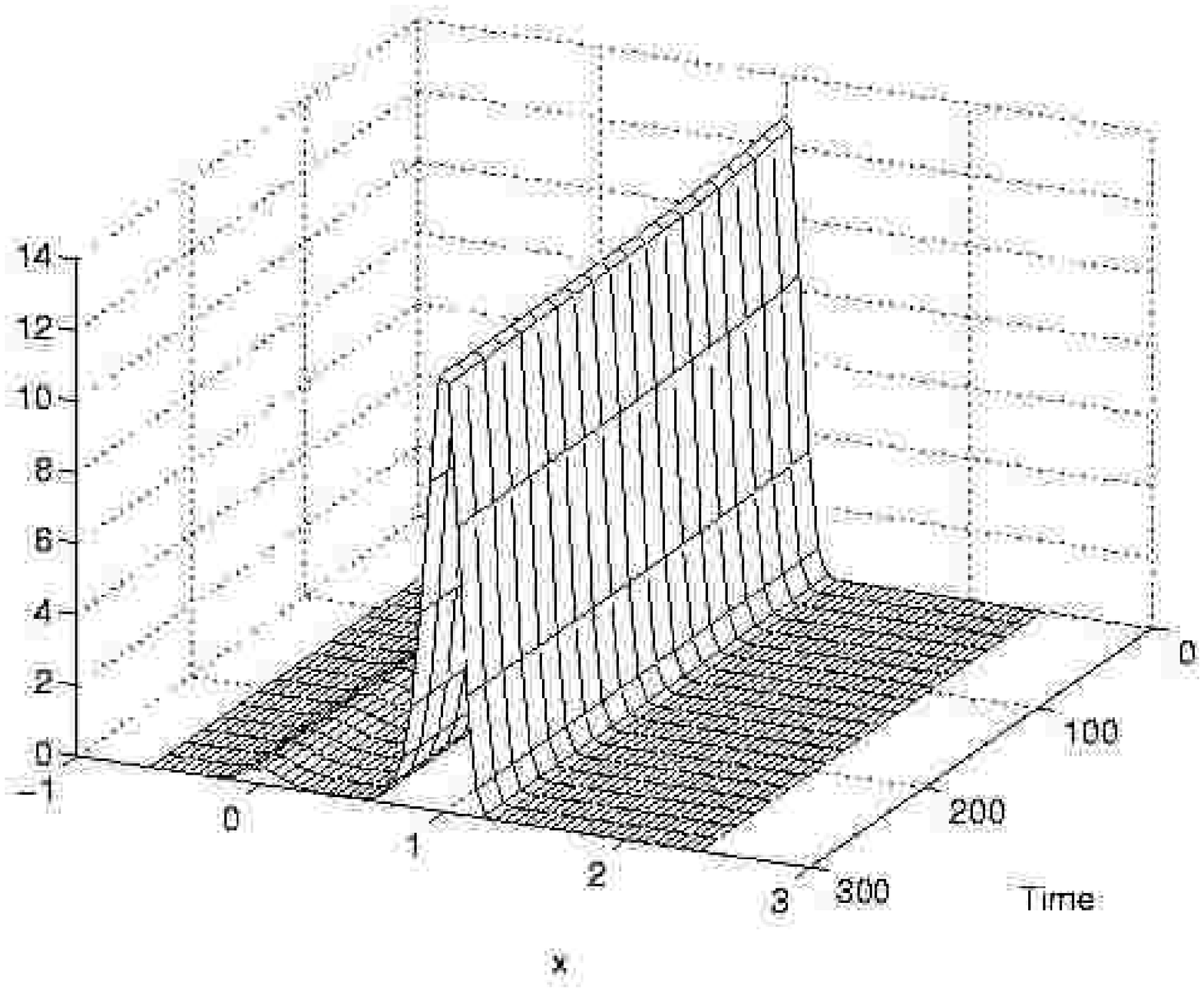}
\caption{Parameters: $a = 0.4$ and $D = 0.005$}
\label{Fig3}
\end{figure}
\clearpage

\begin{figure}[htb]
\centering
\begin{minipage}[b]{0.40\textwidth}
    \centering
    \includegraphics[height=11cm]{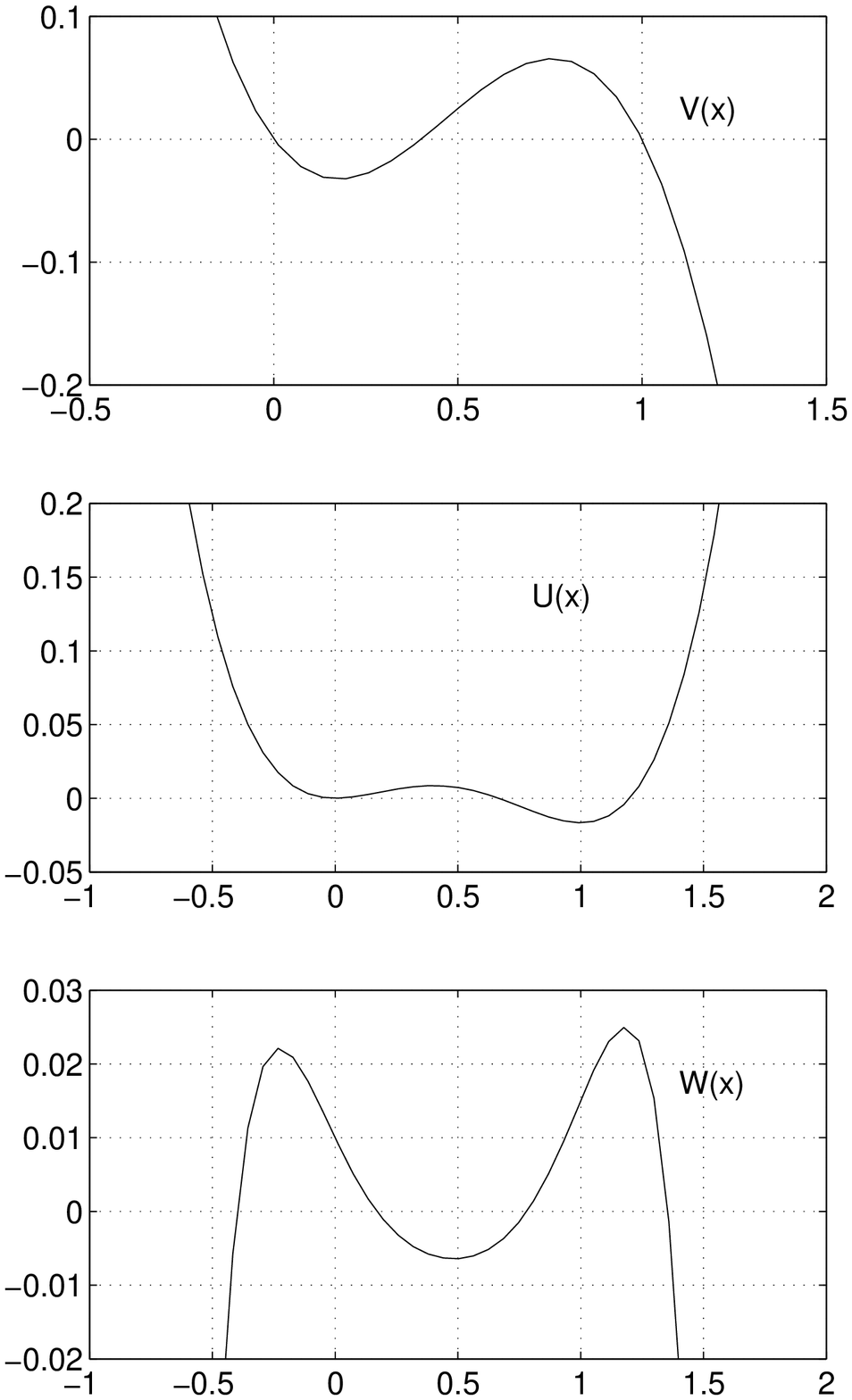}
   \end{minipage} \hspace{0.1\textwidth} 
\begin{minipage}[b]{0.40\textwidth}
    \centering
    \includegraphics[height=11.5cm]{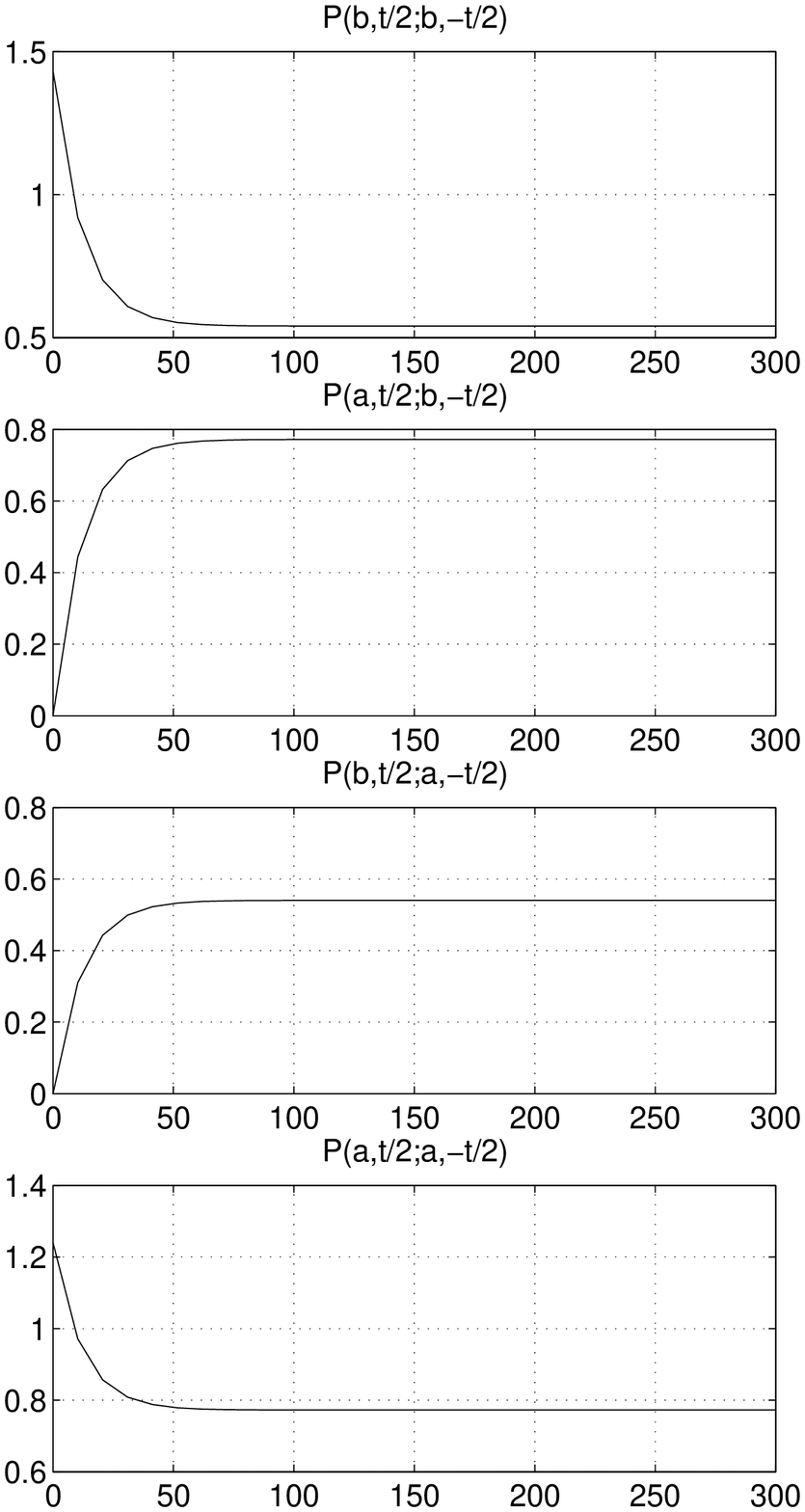}
   \end{minipage}
\par
\bigskip \centering
\begin{minipage}[b]{0.40\textwidth}
    \centering
    \includegraphics[height=5.cm]{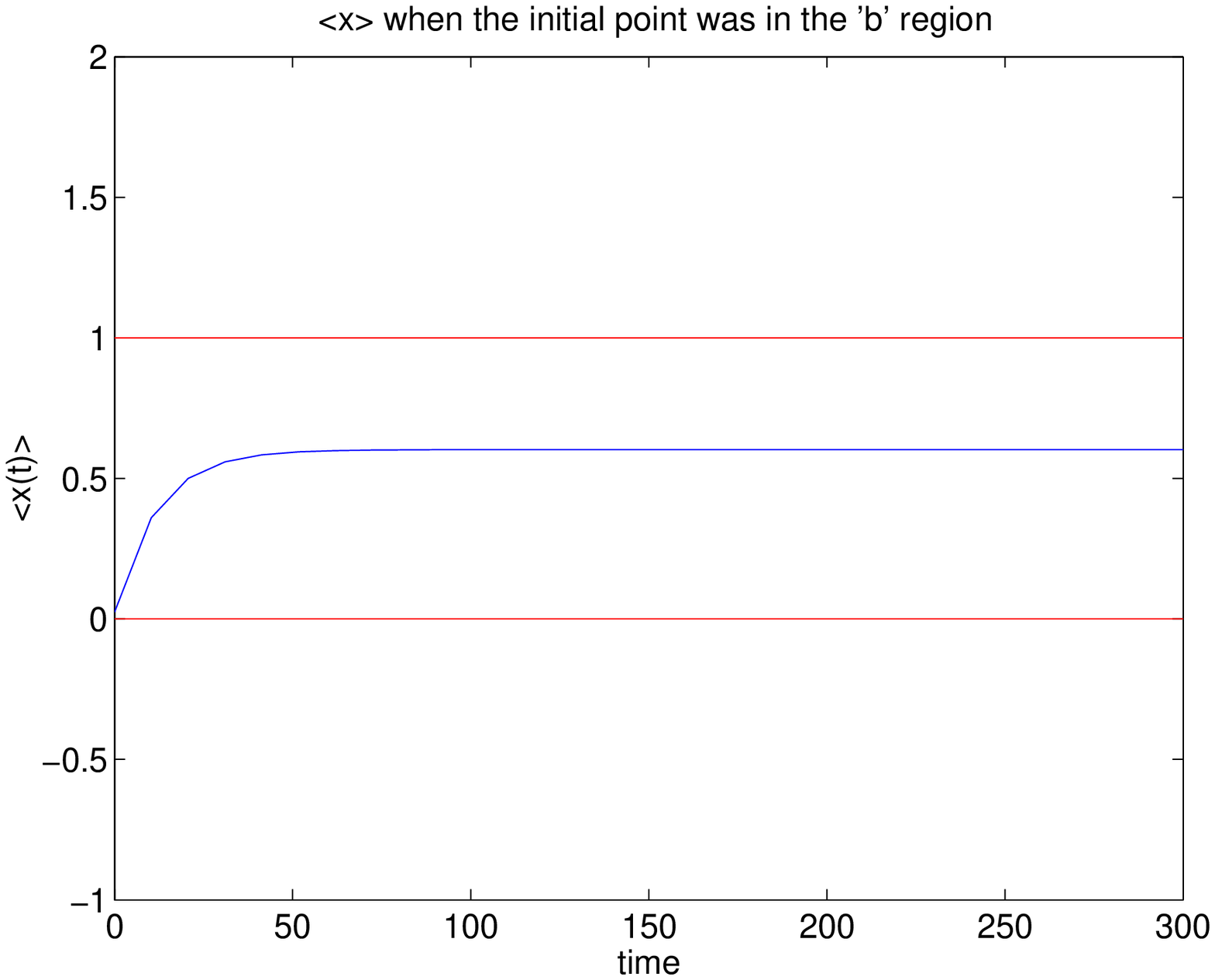}
   \end{minipage} \hspace{0.1\textwidth} 
\begin{minipage}[b]{0.40\textwidth}
    \centering
    \includegraphics[height=5.cm]{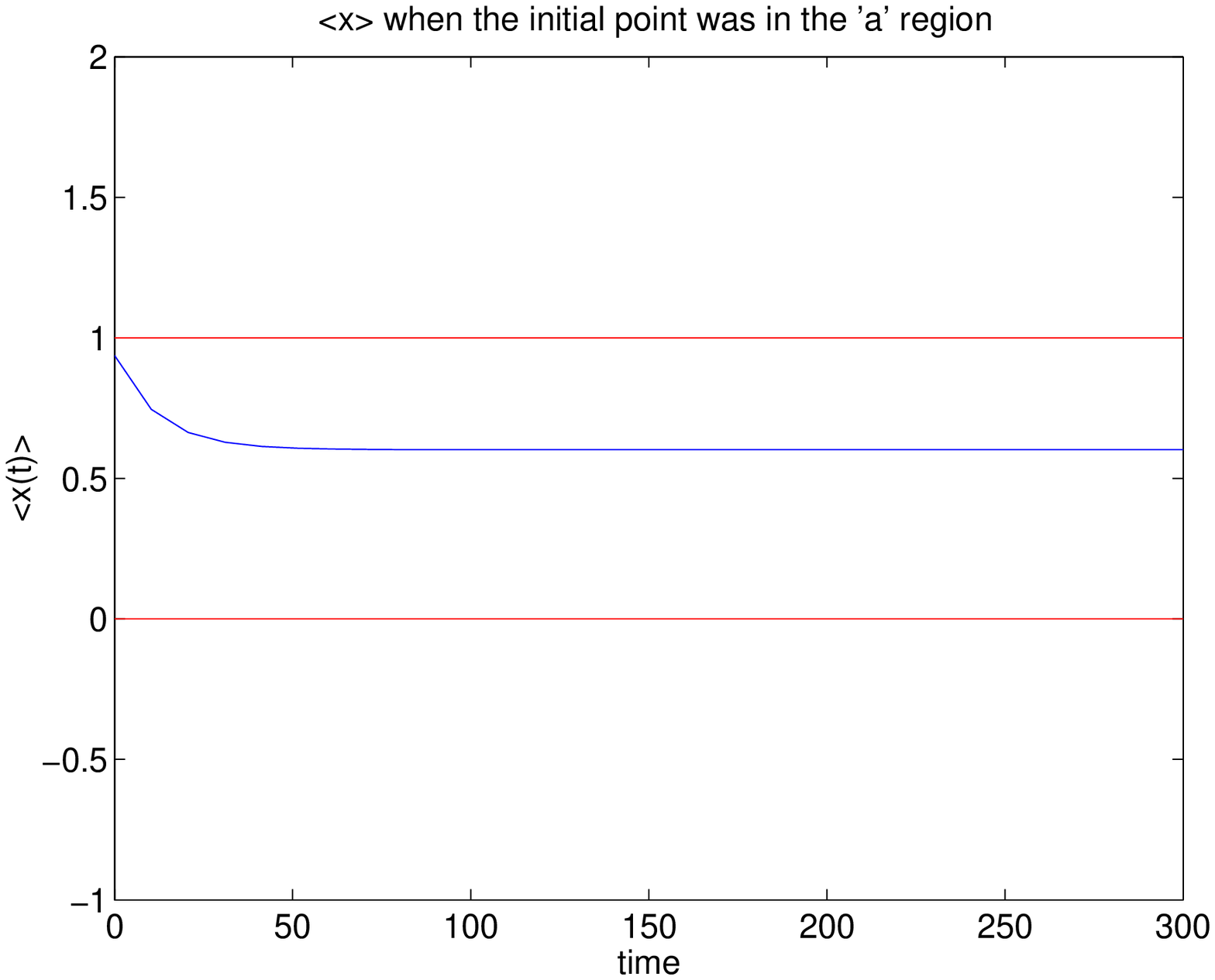}
   \end{minipage}
\end{figure}

\begin{figure}[htb]
\centering
\includegraphics[width=11cm]{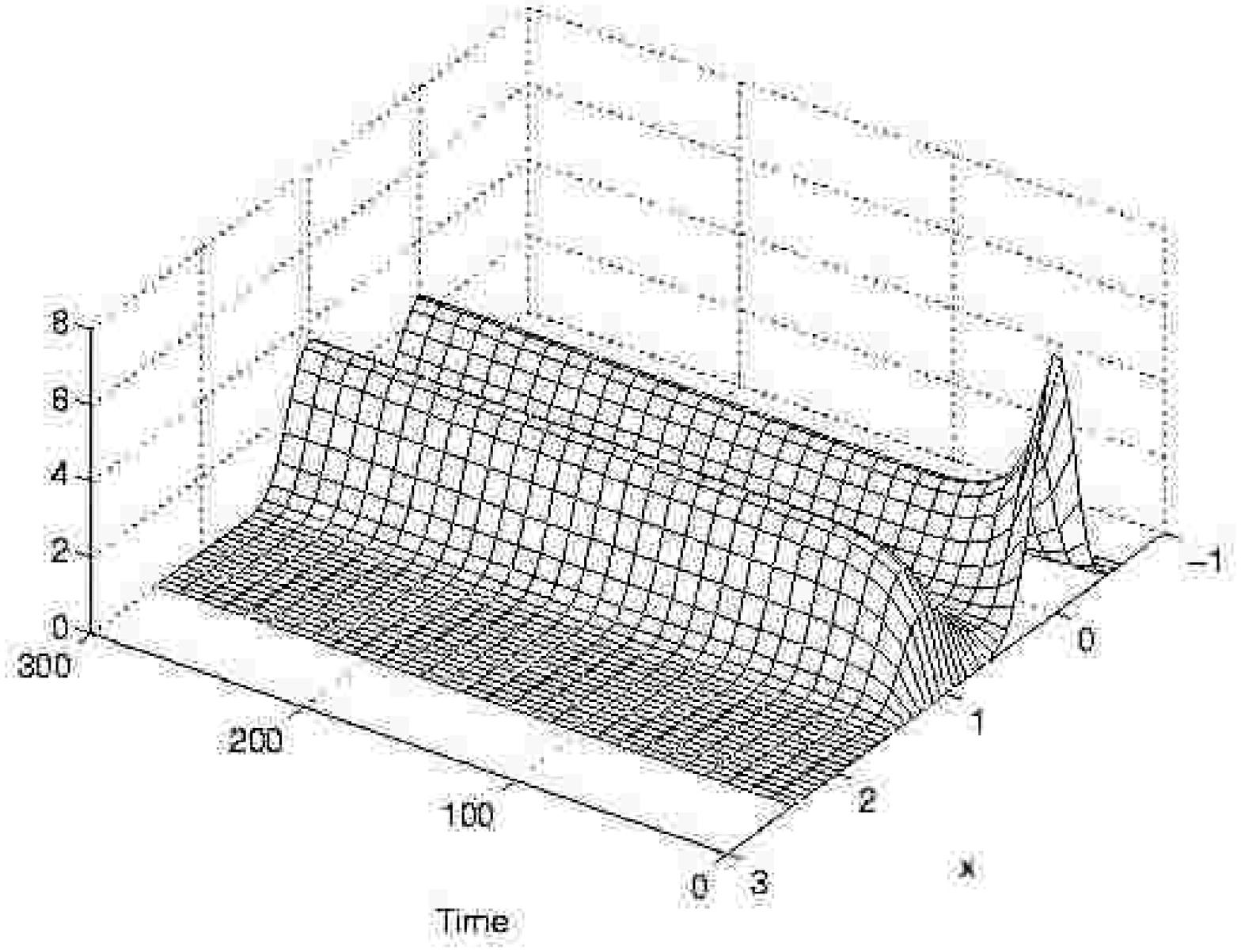}
\end{figure}
\begin{figure}[htbp]
\centering
\includegraphics[width=11cm]{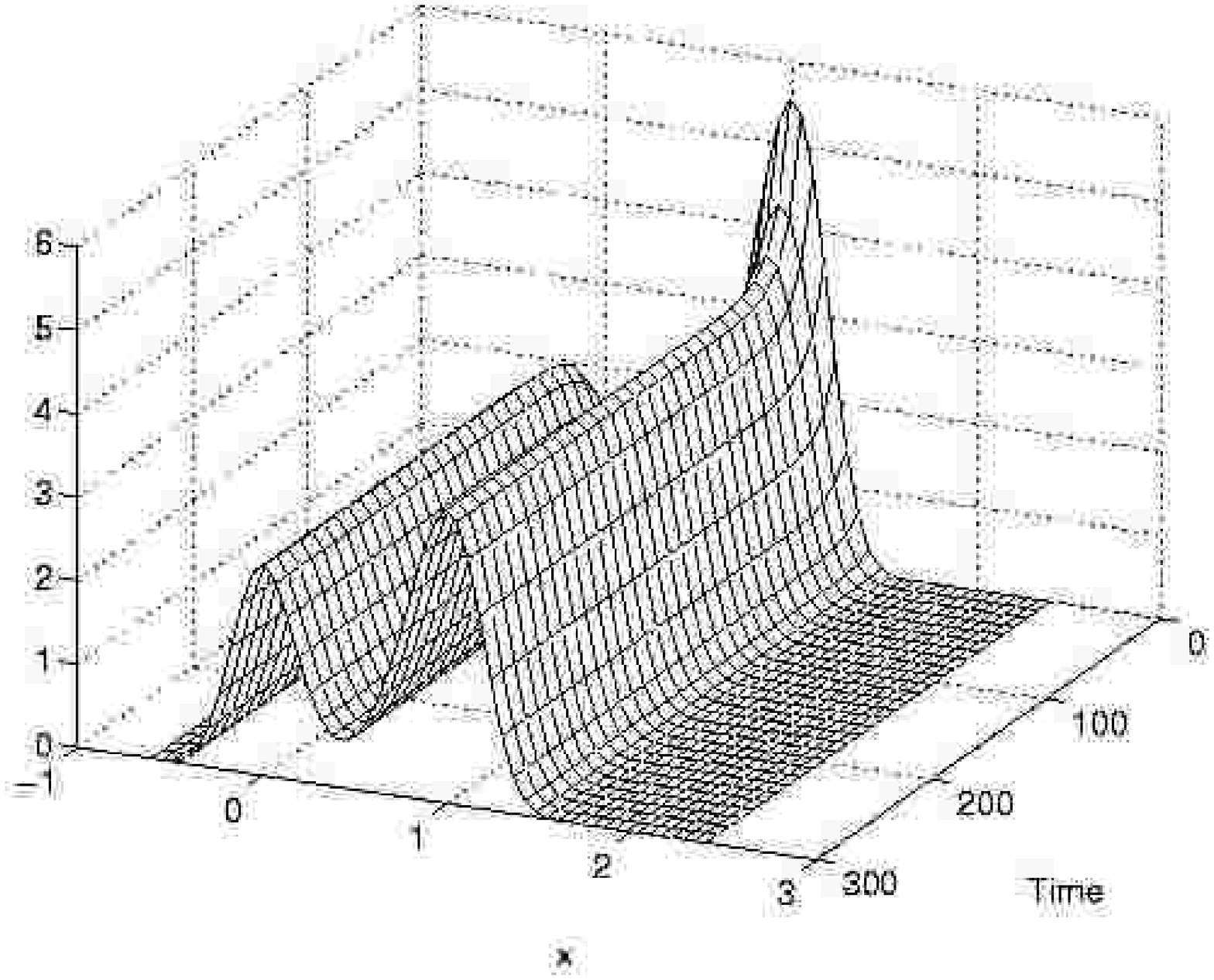}
\caption{Parameters: $a = 0.4$ and $D = 0.05$}
\label{Fig5}
\end{figure}

\clearpage

\begin{figure}[htb]
\centering
\begin{minipage}[b]{0.40\textwidth}
    \centering
    \includegraphics[height=11cm]{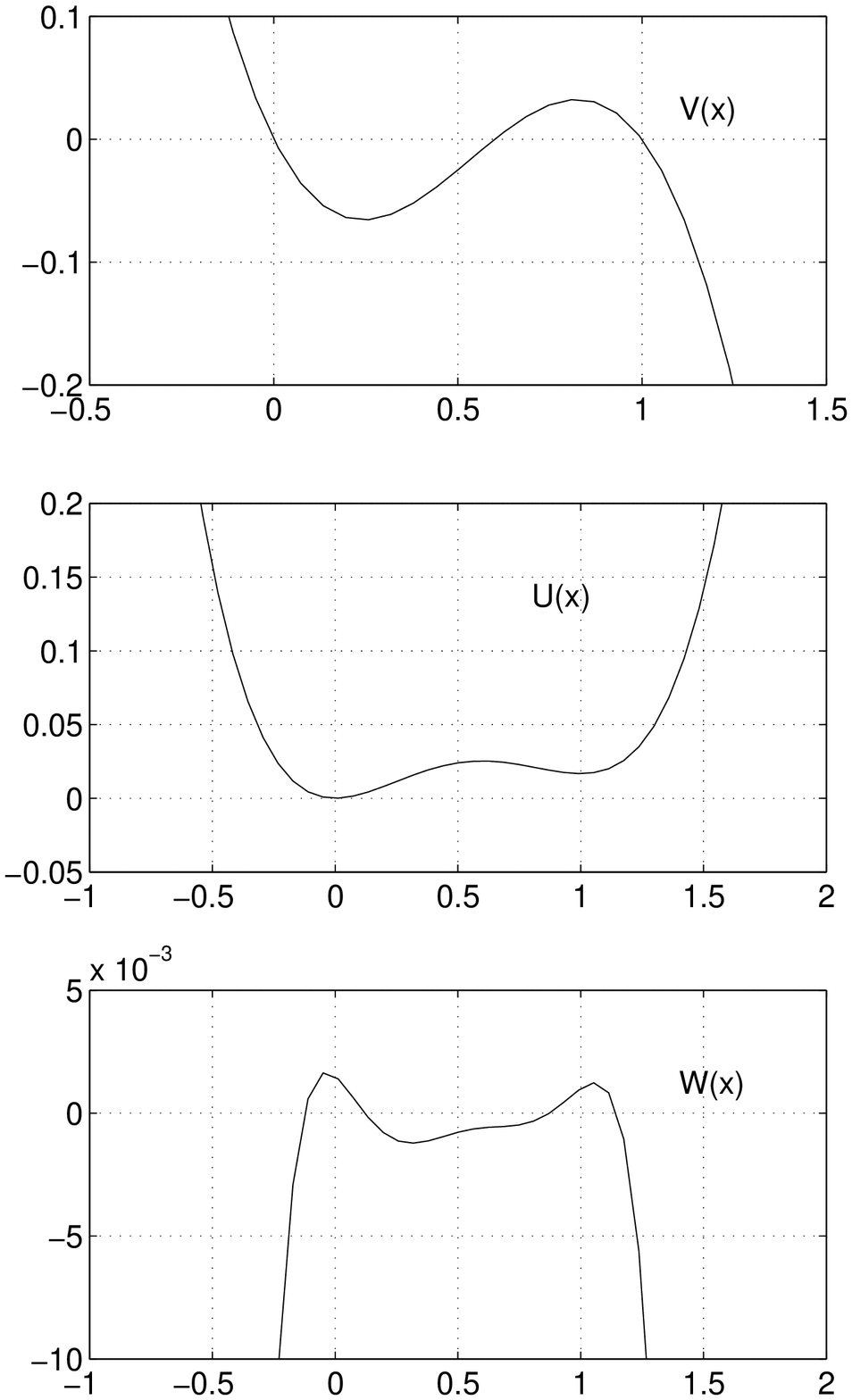}
   \end{minipage} \hspace{0.1\textwidth} 
\begin{minipage}[b]{0.40\textwidth}
    \centering
    \includegraphics[height=11.5cm]{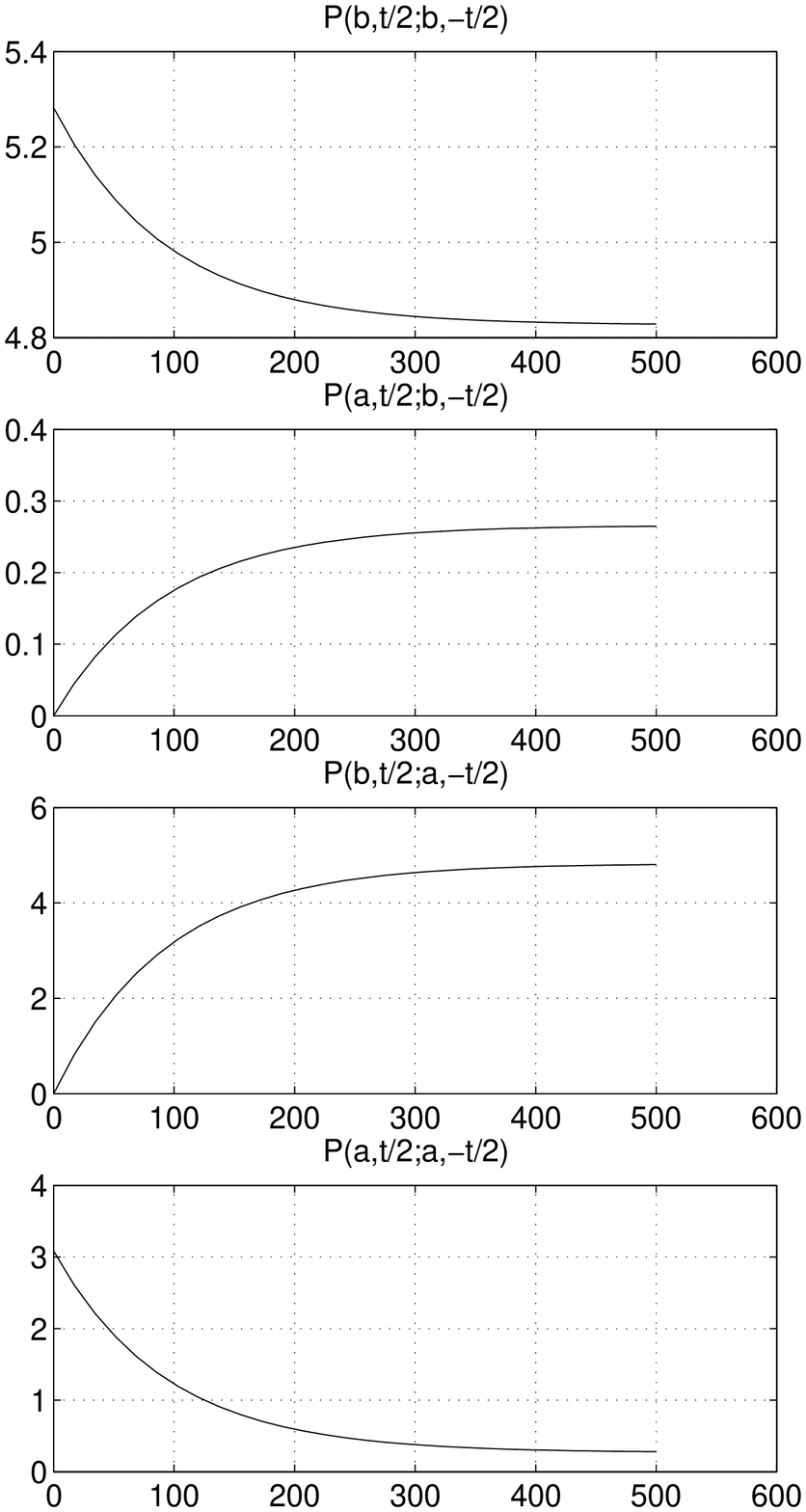}
   \end{minipage}
\par \bigskip \centering
\begin{minipage}[b]{0.40\textwidth}
    \centering
    \includegraphics[height=5.cm]{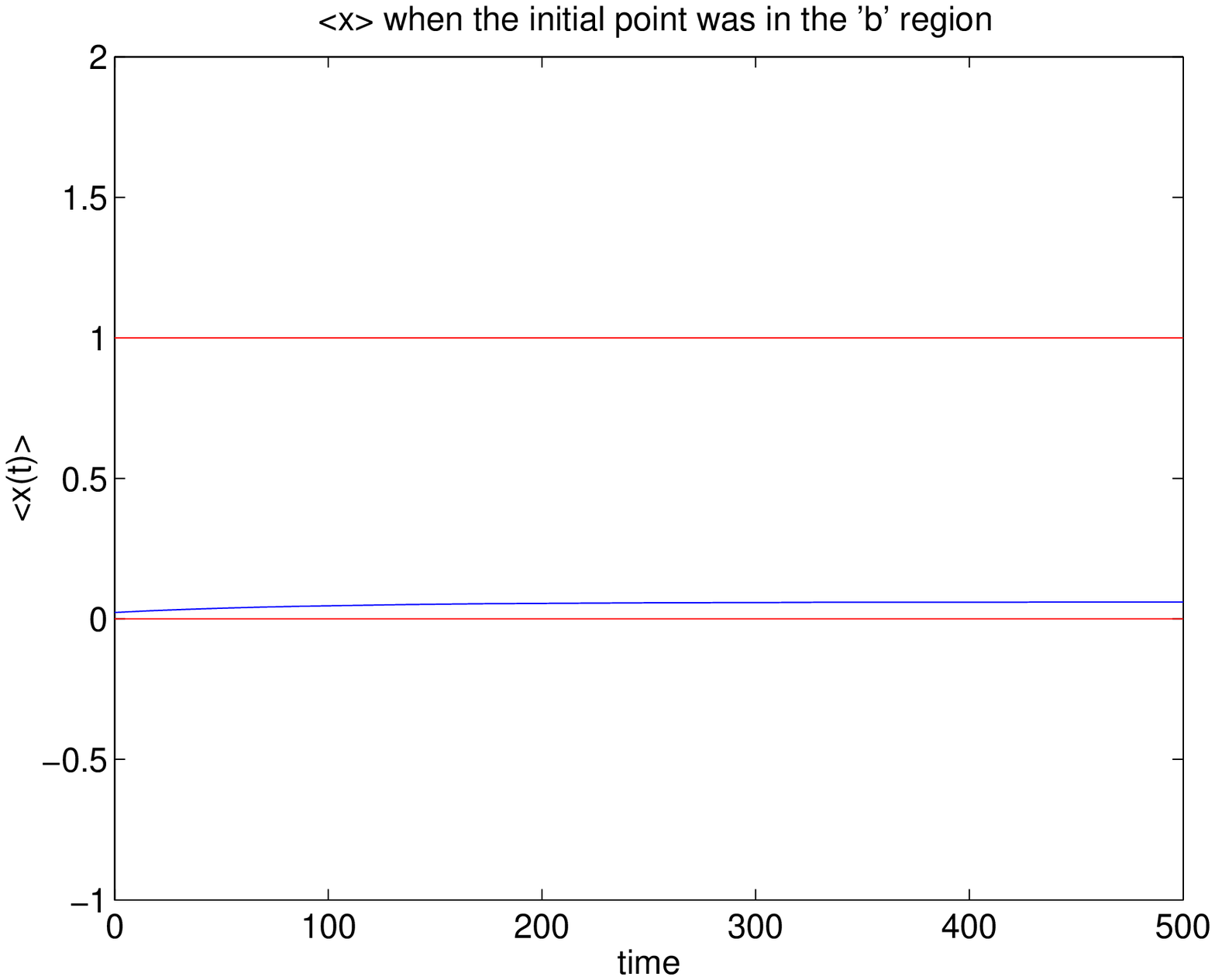}
   \end{minipage} \hspace{0.1\textwidth} 
\begin{minipage}[b]{0.40\textwidth}
    \centering
    \includegraphics[height=5.cm]{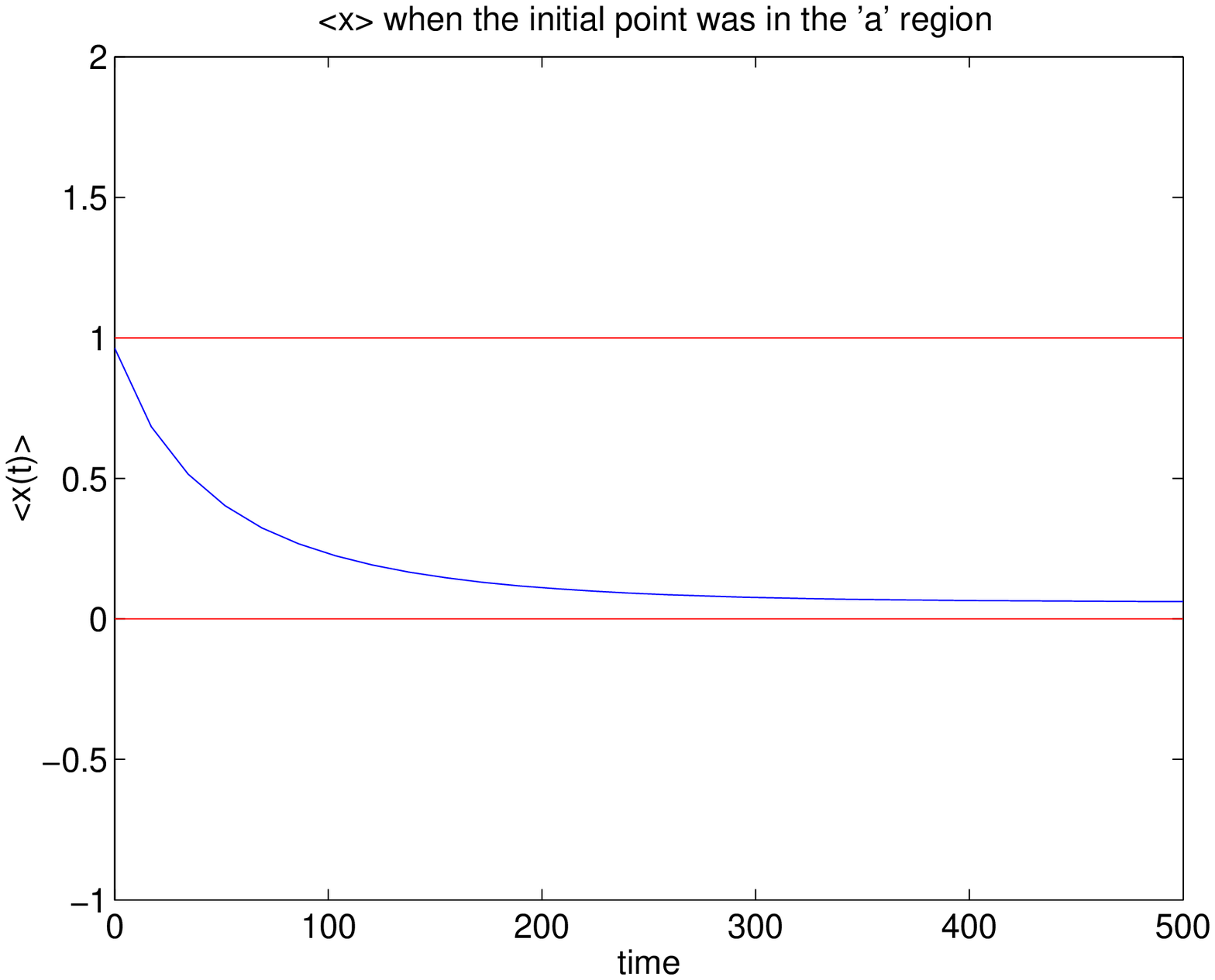}
   \end{minipage}
\end{figure}

\begin{figure}[htb]
\centering
\includegraphics[width=11cm]{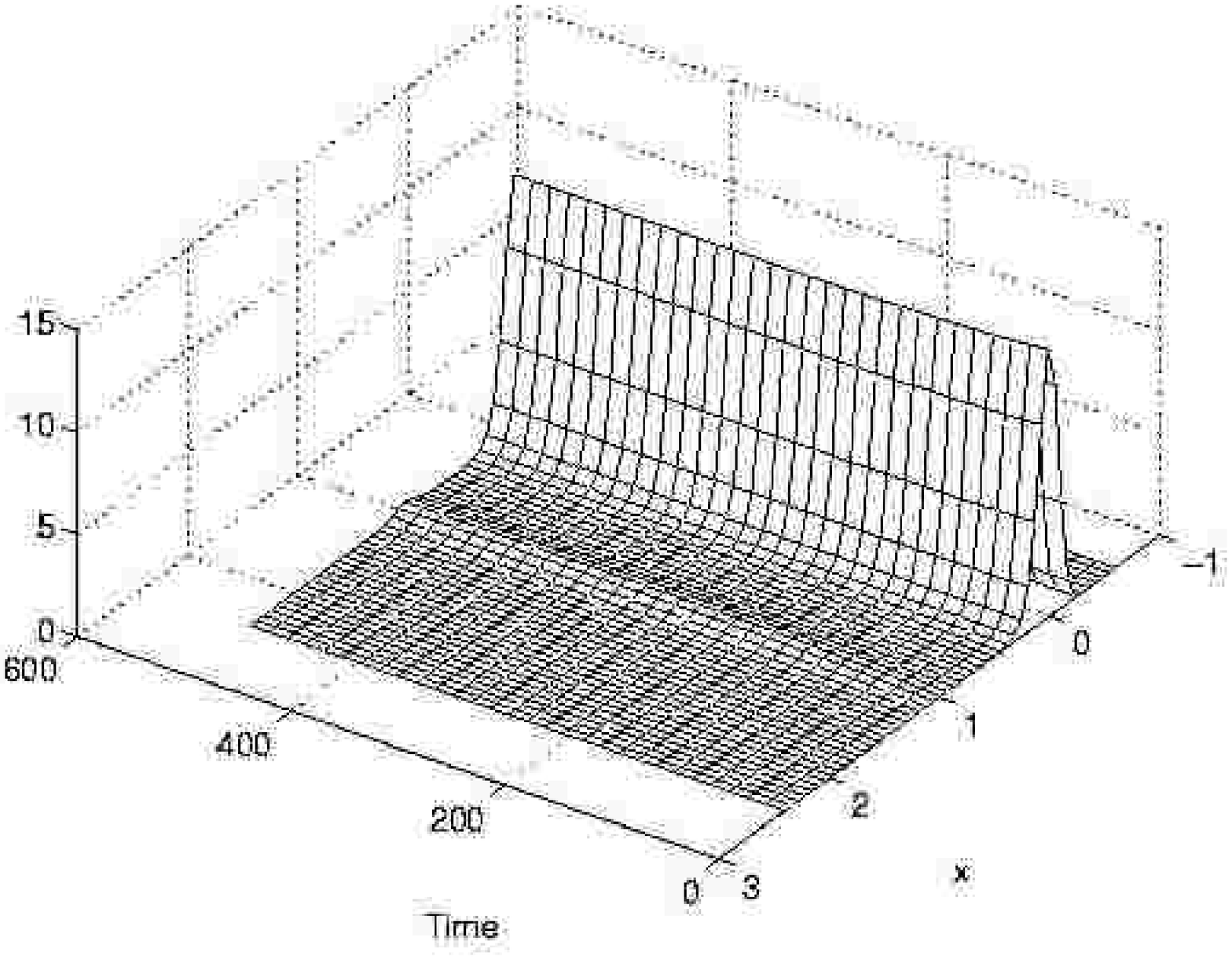}
\end{figure}
\begin{figure}[htbp]
\centering
\includegraphics[width=11cm]{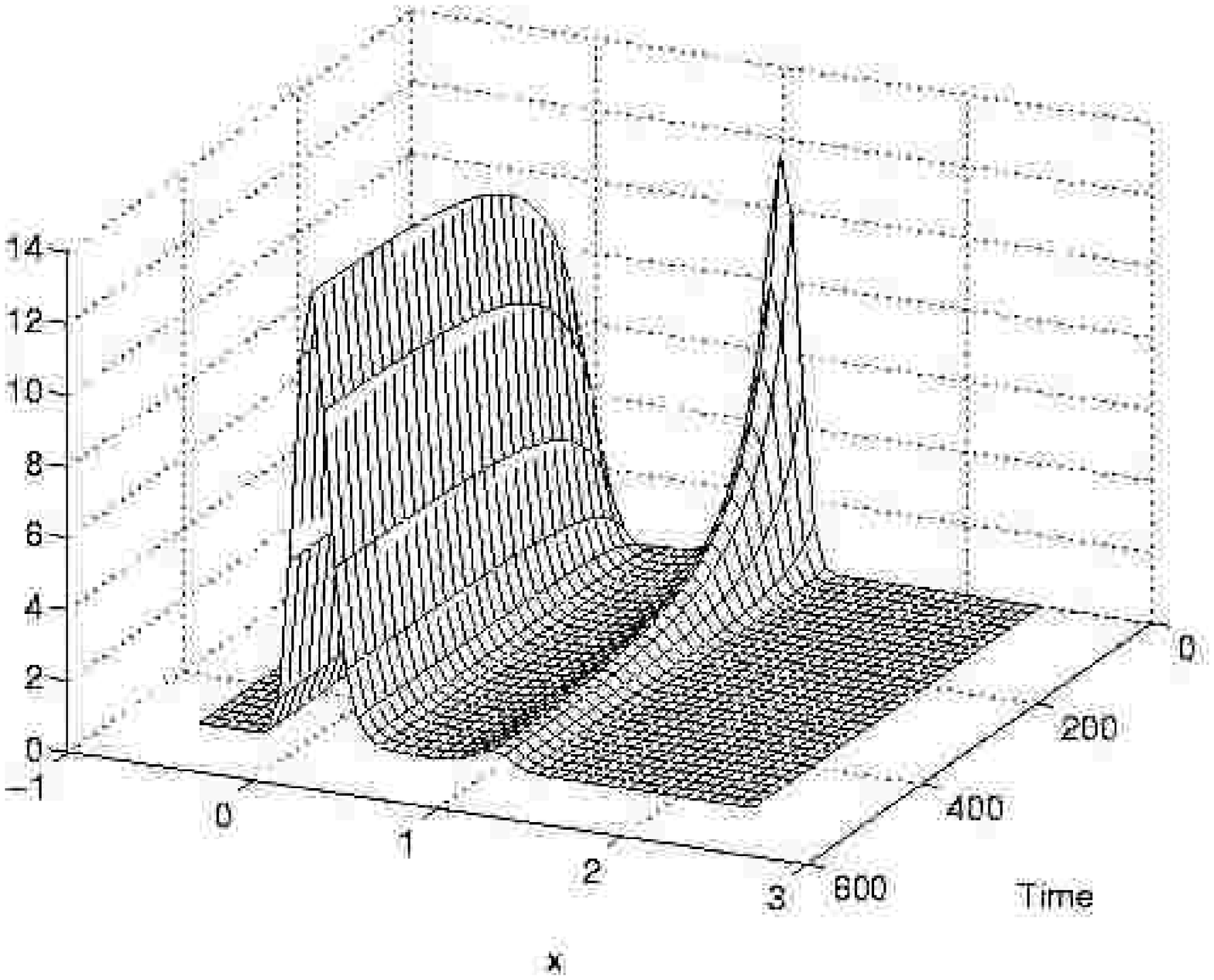}
\caption{Parameters: $a = 0.6$ and $D = 0.005$}
\label{Fig7}
\end{figure}

\clearpage

\begin{figure}[htb]
\centering
\begin{minipage}[b]{0.40\textwidth}
    \centering
    \includegraphics[height=11cm]{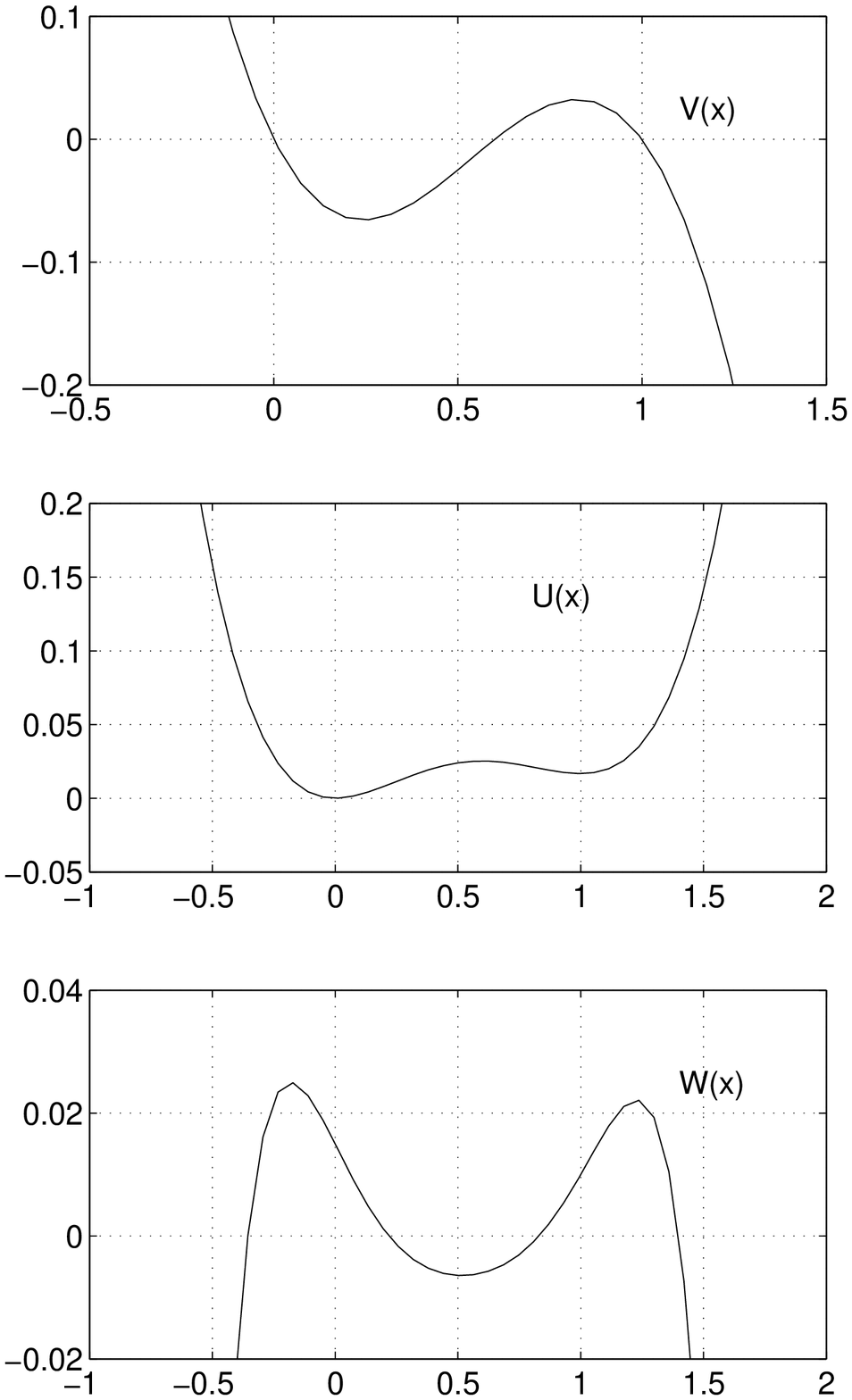}
   \end{minipage} \hspace{0.1\textwidth} 
\begin{minipage}[b]{0.40\textwidth}
    \centering
    \includegraphics[height=11.5cm]{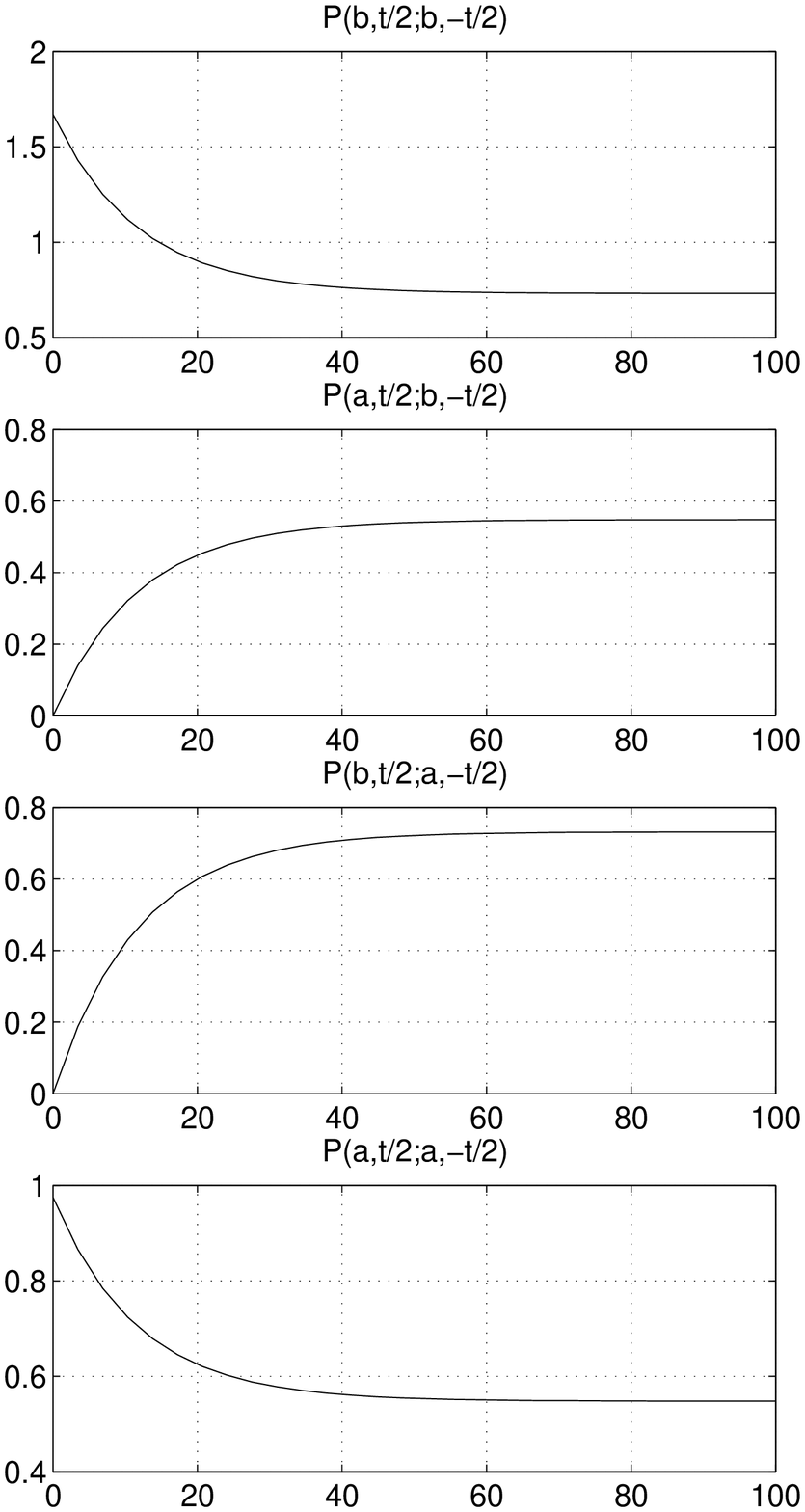}
   \end{minipage}
\par
\bigskip \centering
\begin{minipage}[b]{0.40\textwidth}
    \centering
    \includegraphics[height=5.cm]{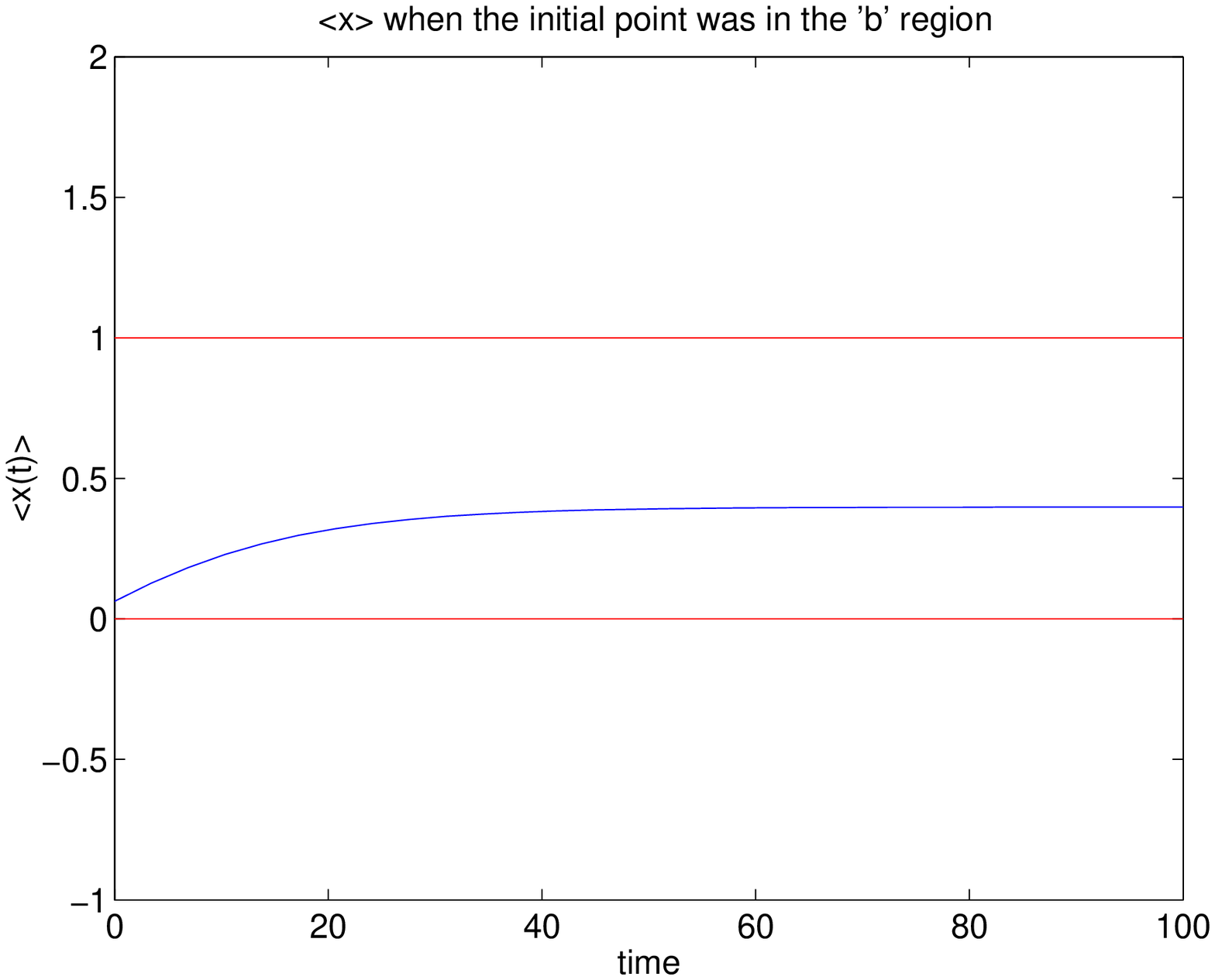}
   \end{minipage} \hspace{0.1\textwidth} 
\begin{minipage}[b]{0.40\textwidth}
    \centering
    \includegraphics[height=5.cm]{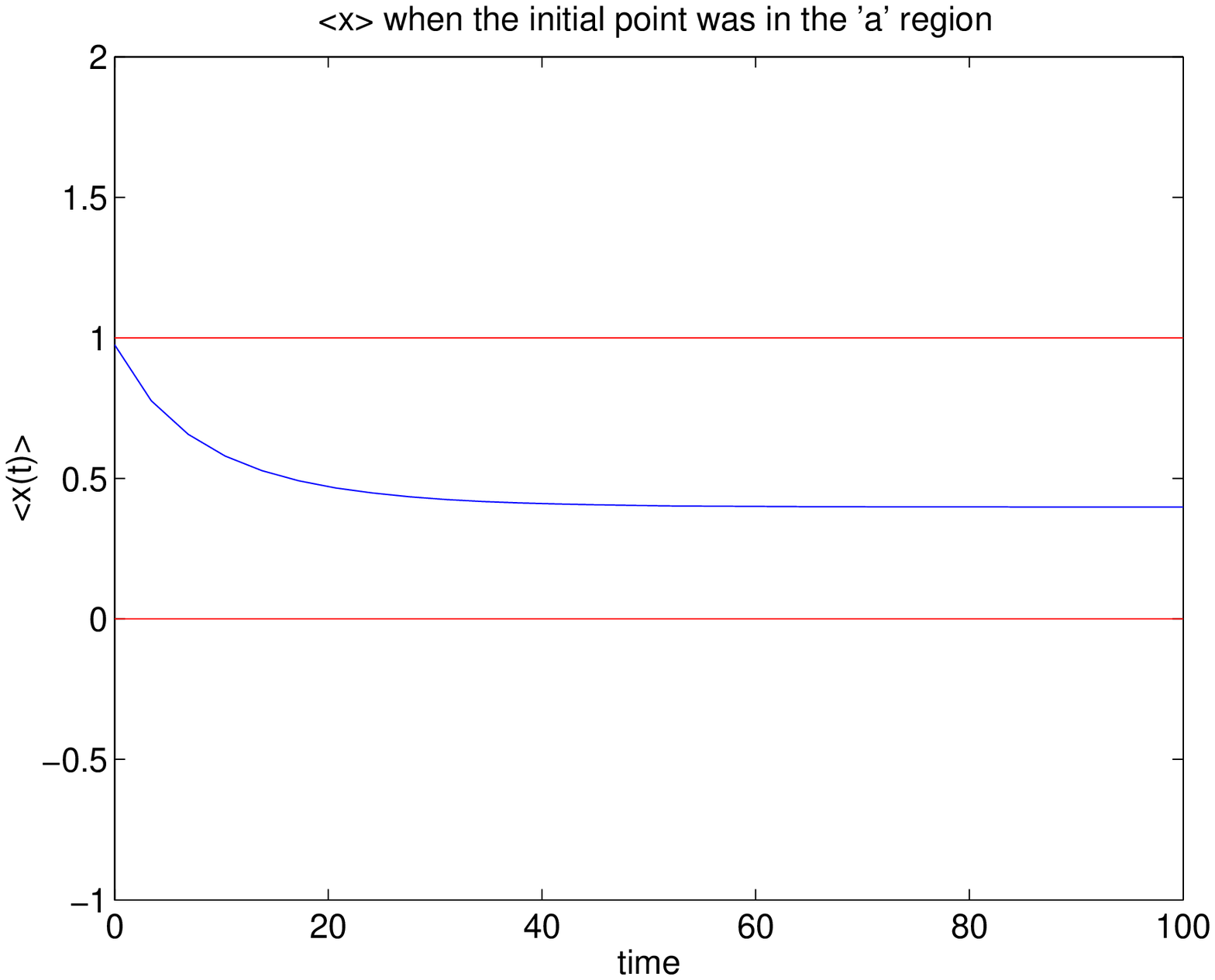}
   \end{minipage}
\end{figure}

\begin{figure}[htb]
\centering
\includegraphics[width=11cm]{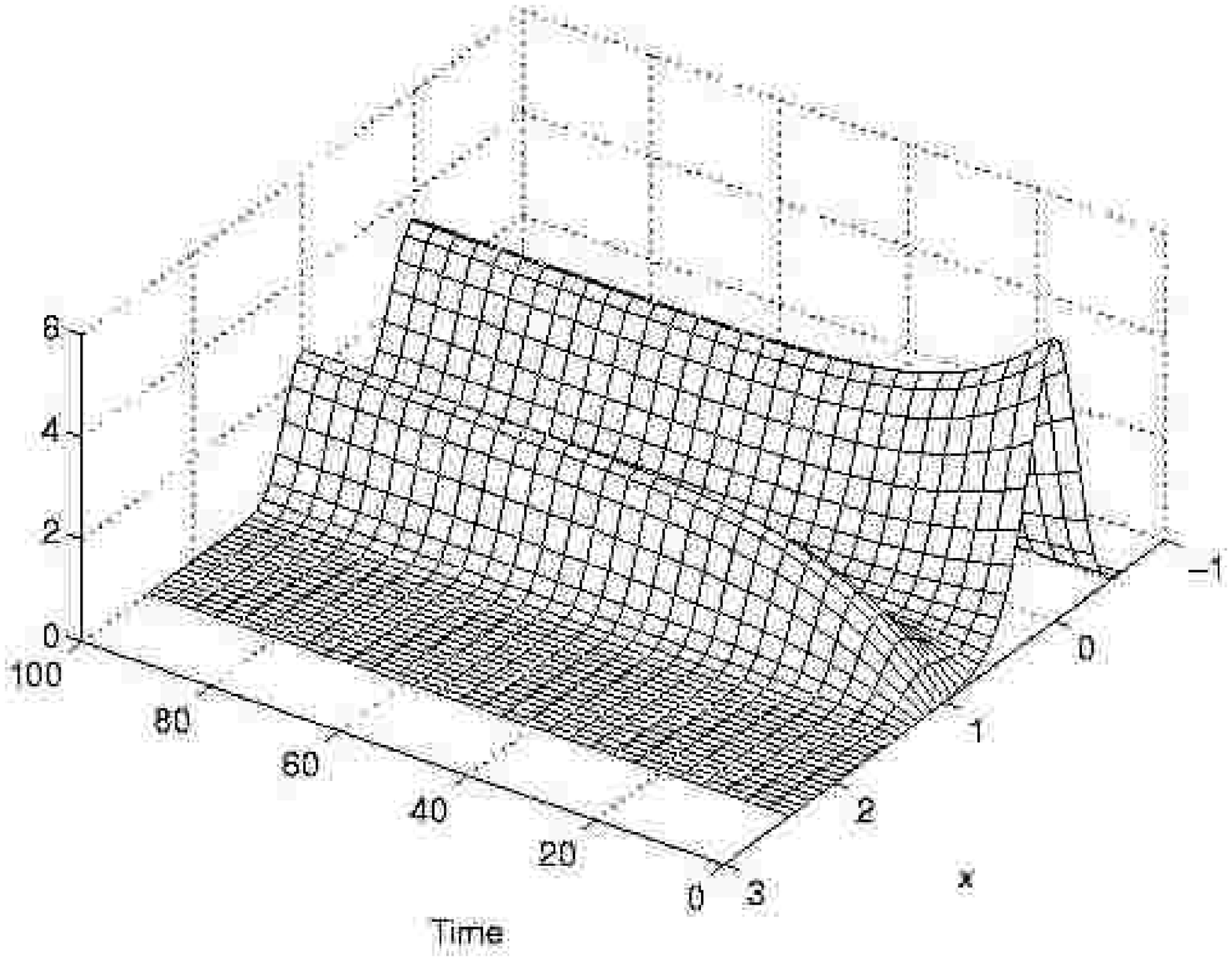}
\end{figure}
\begin{figure}[htbp]
\centering
\includegraphics[width=11cm]{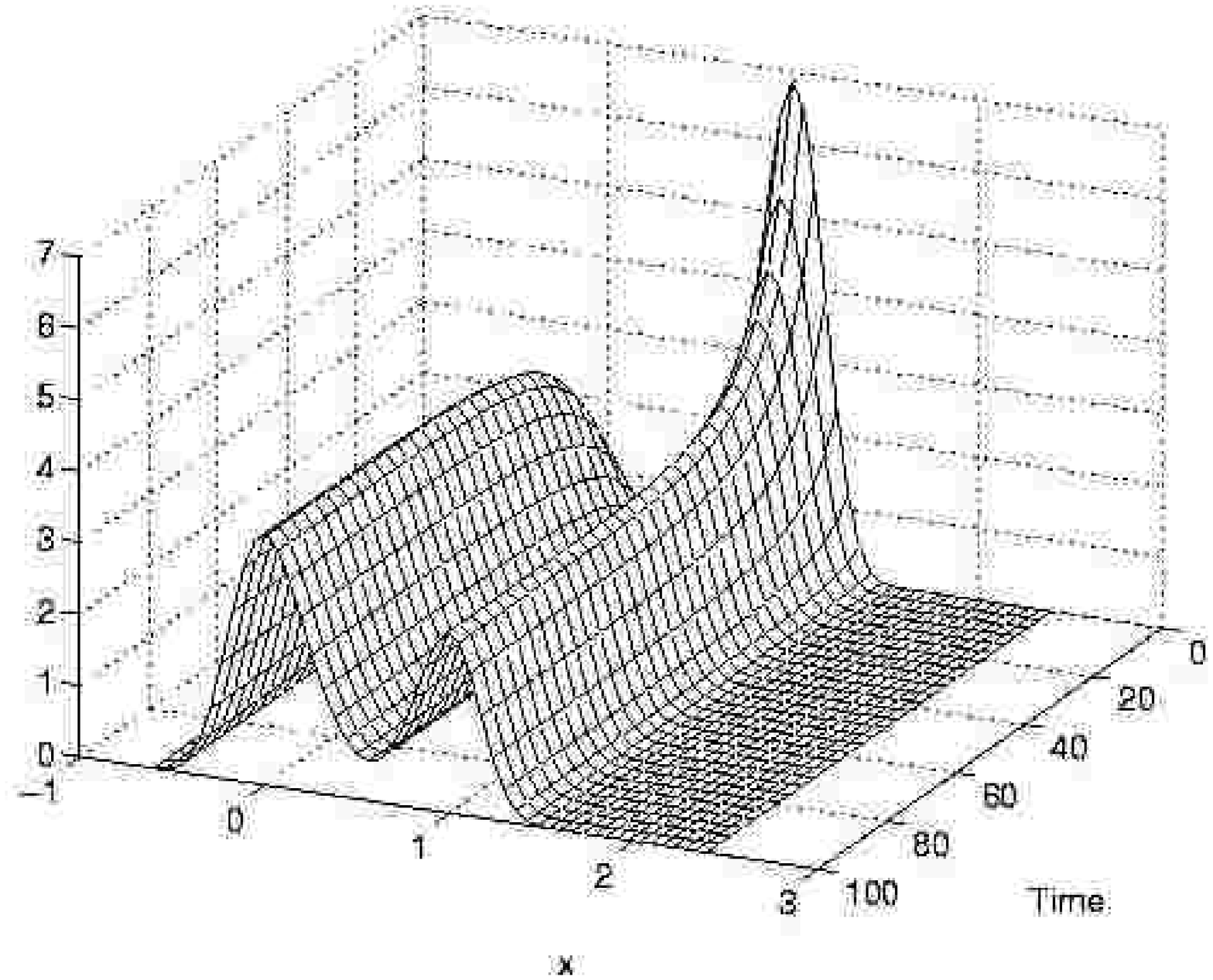}
\caption{Parameters: $a = 0.6$ and $D = 0.05$}
\label{Fig9}
\end{figure}

\clearpage

\begin{figure}[htb]
\centering
\begin{minipage}[b]{0.40\textwidth}
    \centering
    \includegraphics[height=11cm]{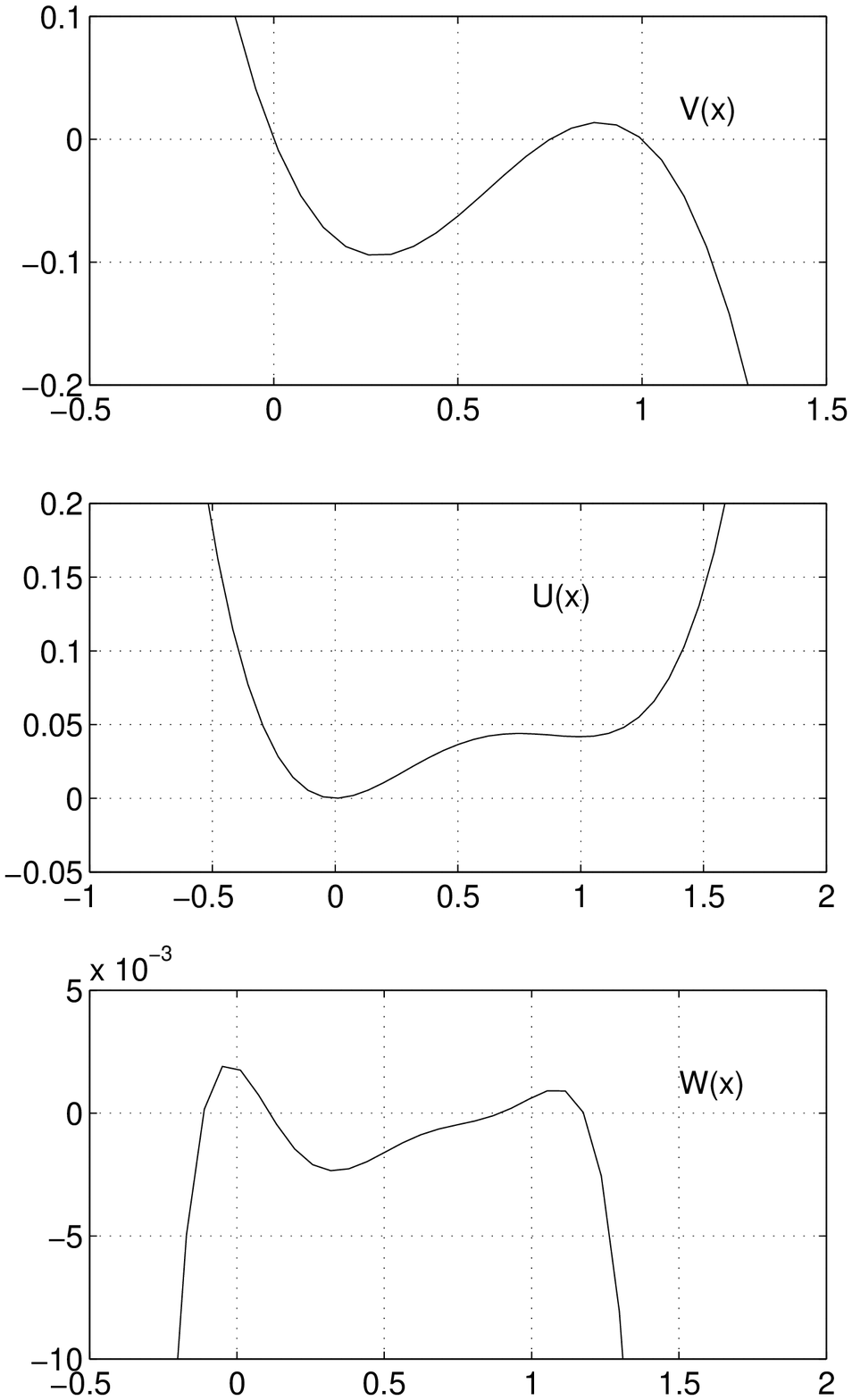}
   \end{minipage} \hspace{0.1\textwidth} 
\begin{minipage}[b]{0.40\textwidth}
    \centering
    \includegraphics[height=11.5cm]{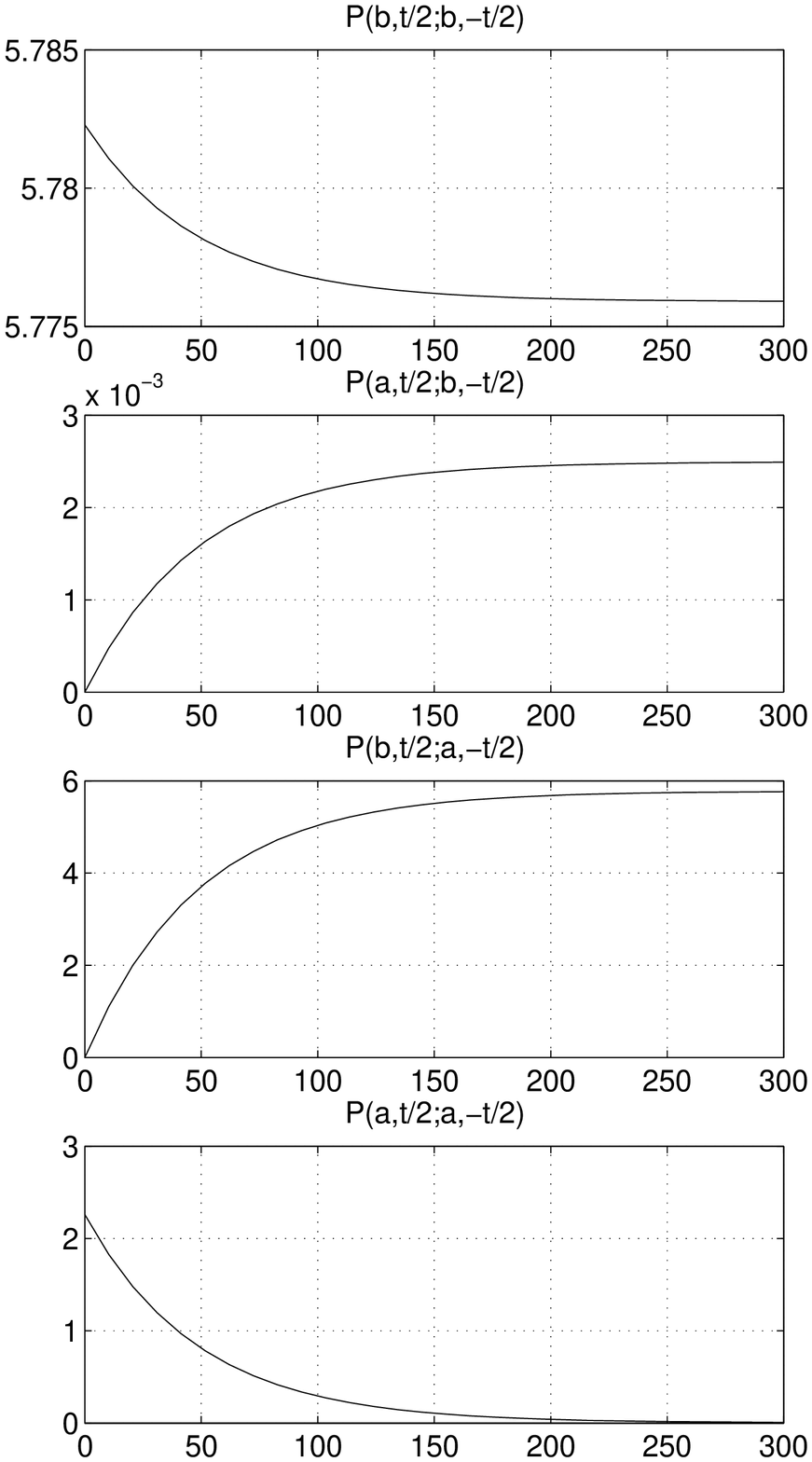}
   \end{minipage}
\par
\bigskip \centering
\begin{minipage}[b]{0.40\textwidth}
    \centering
    \includegraphics[height=5.cm]{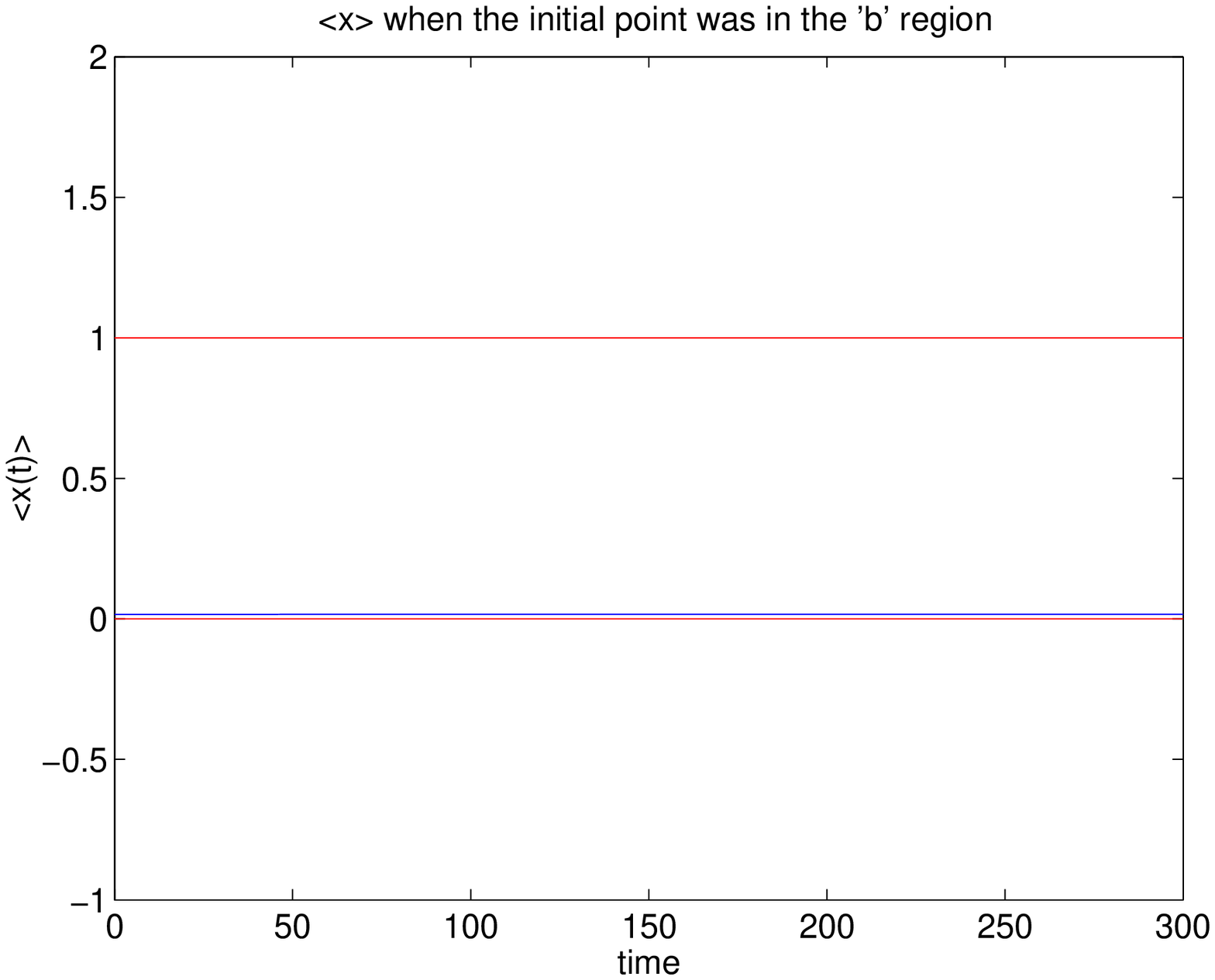}
   \end{minipage} \hspace{0.1\textwidth} 
\begin{minipage}[b]{0.40\textwidth}
    \centering
    \includegraphics[height=5.cm]{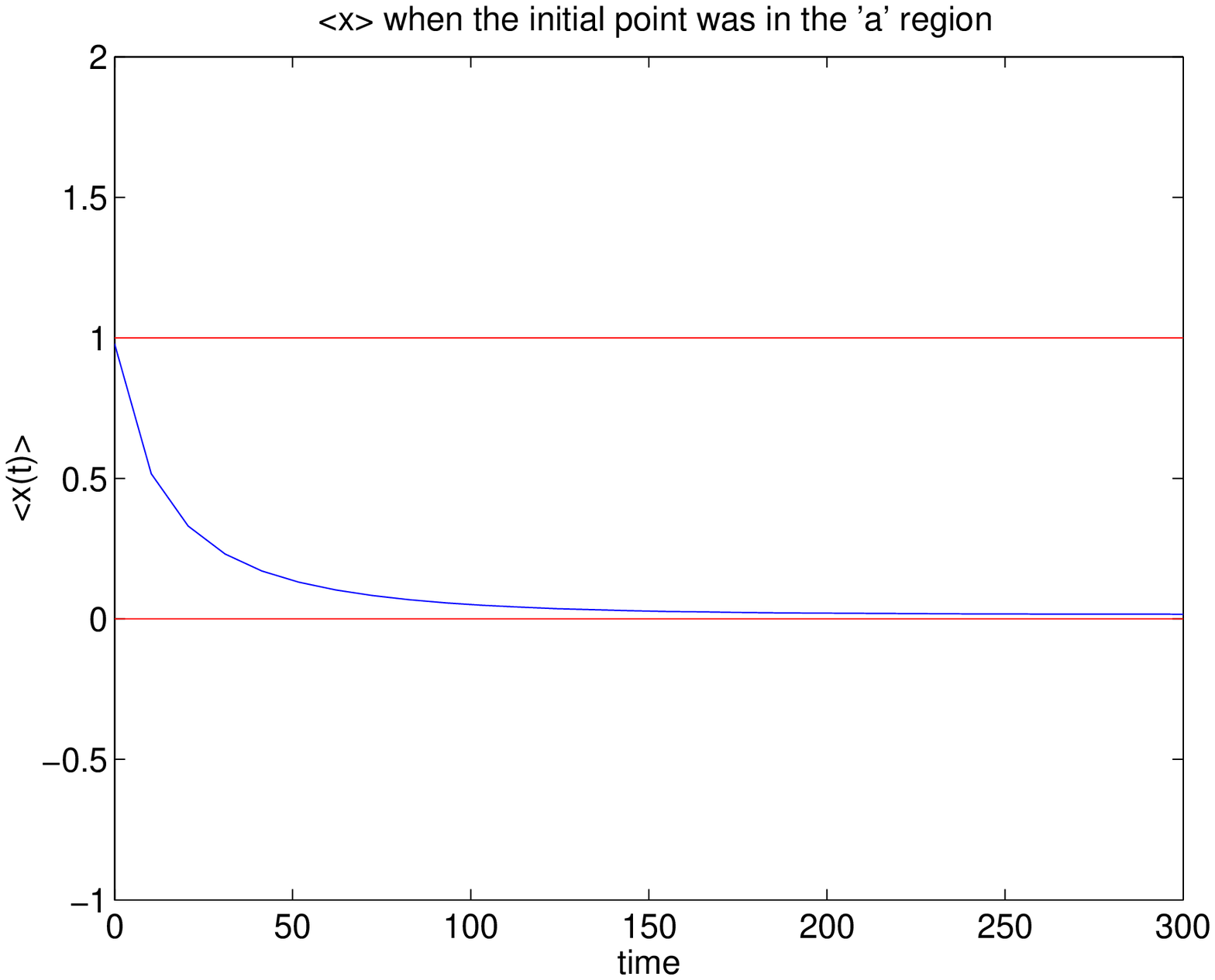}
   \end{minipage}
\end{figure}

\begin{figure}[htb]
\centering
\includegraphics[width=11cm]{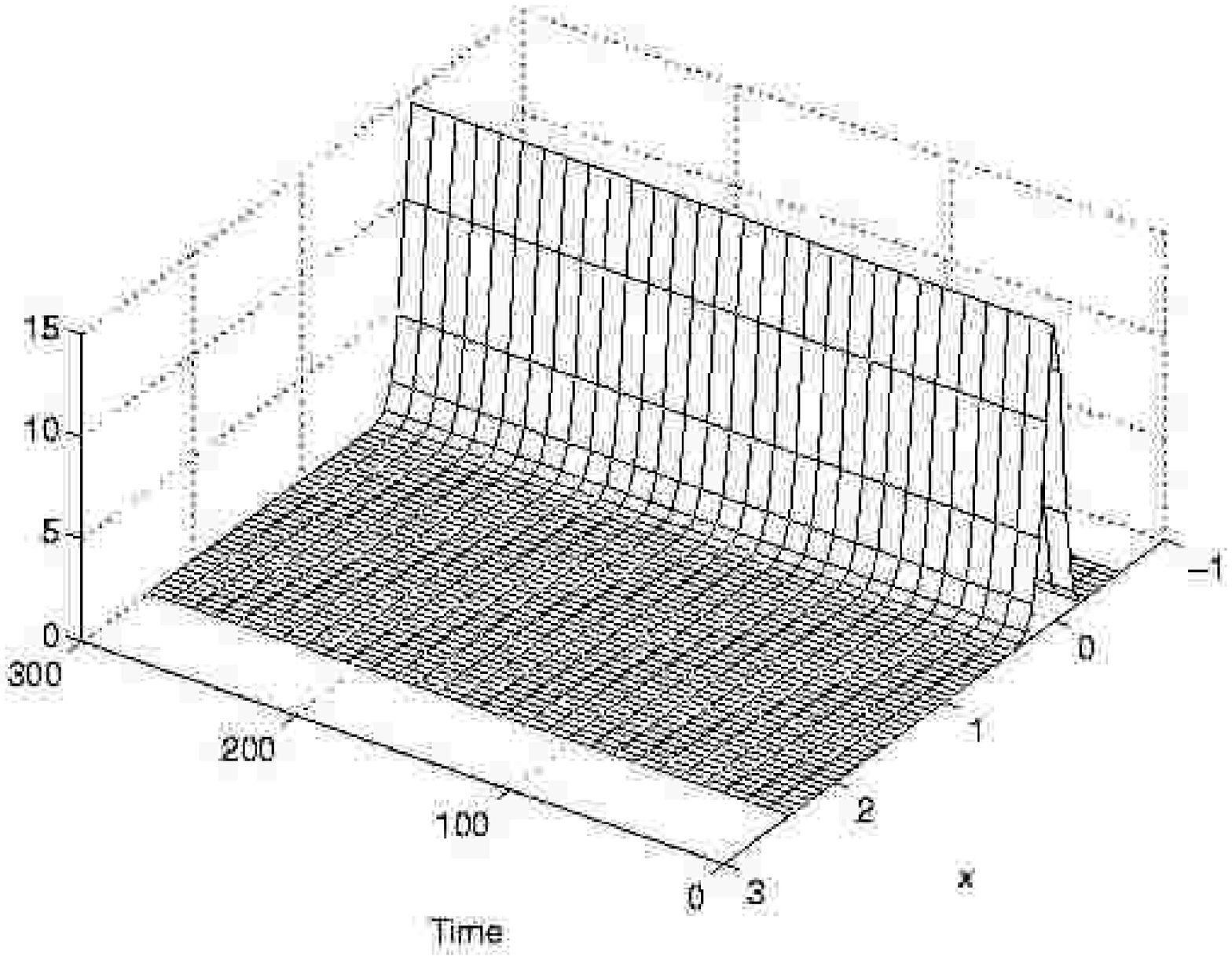}
\end{figure}
\begin{figure}[htbp]
\centering
\includegraphics[width=11cm]{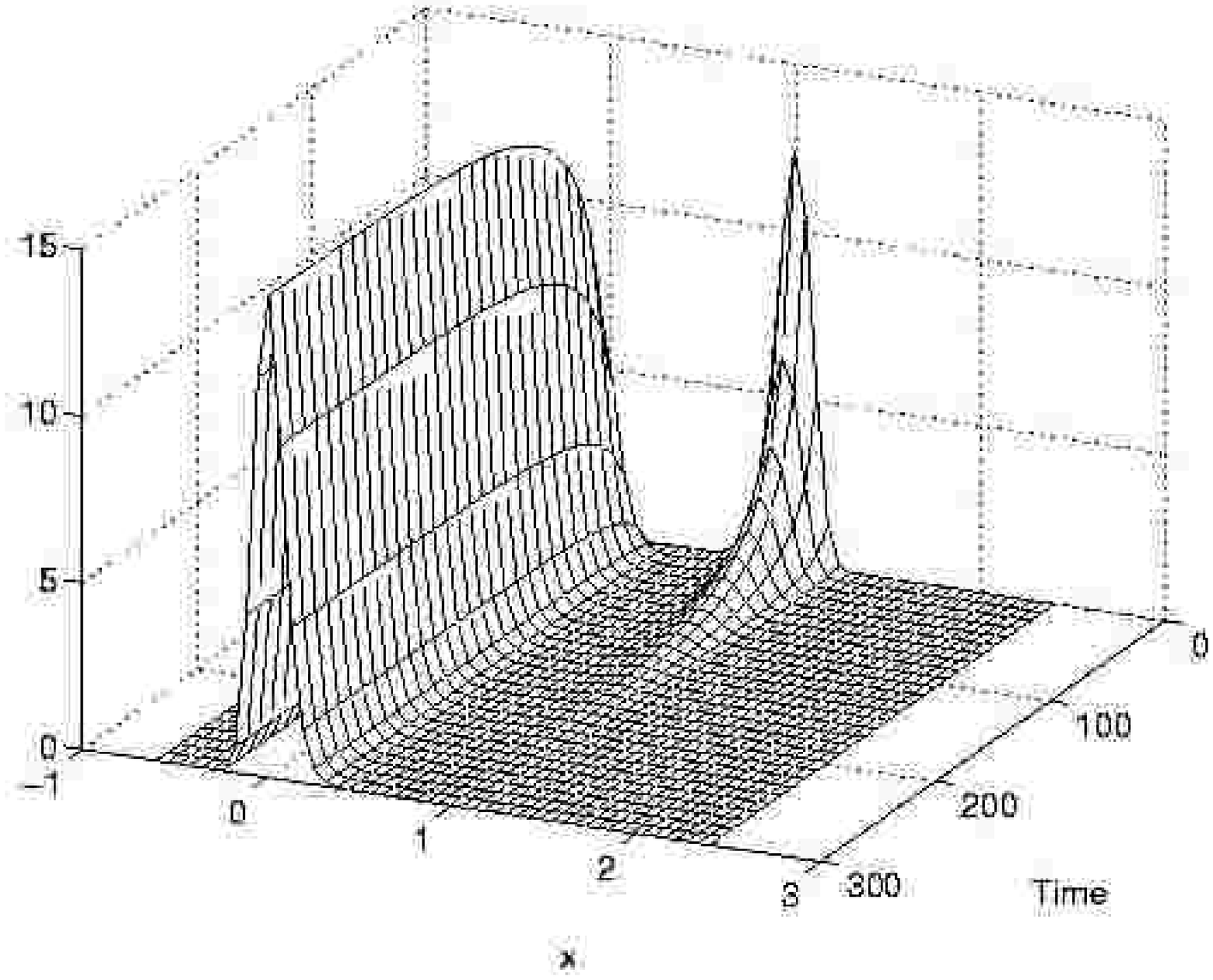}
\caption{Parameters: $a = 0.75$ and $D = 0.005$}
\label{Fig11}
\end{figure}

\clearpage

\begin{figure}[htb]
\centering
\begin{minipage}[b]{0.40\textwidth}
    \centering
    \includegraphics[height=11cm]{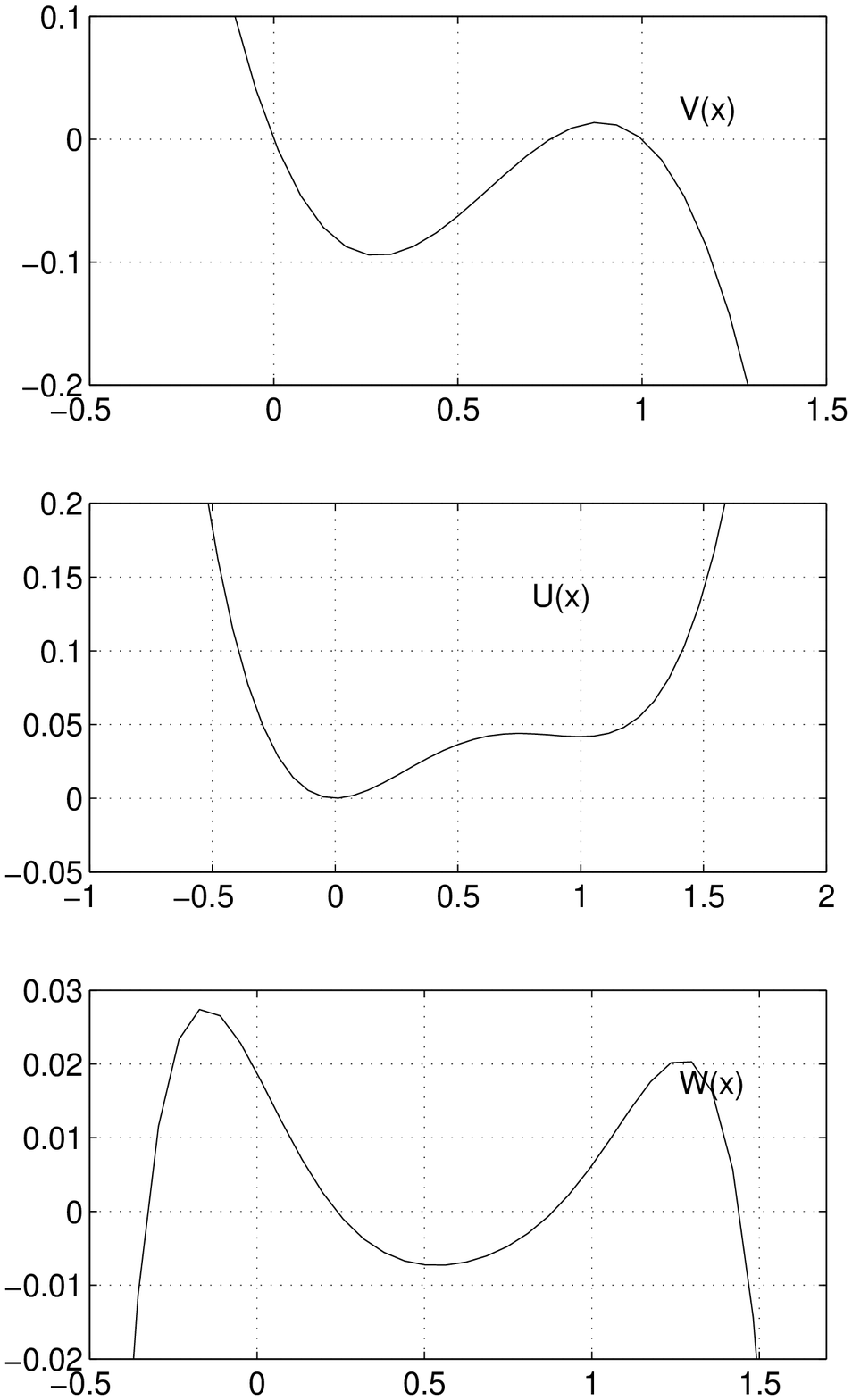}
   \end{minipage} \hspace{0.1\textwidth} 
\begin{minipage}[b]{0.40\textwidth}
    \centering
    \includegraphics[height=11.5cm]{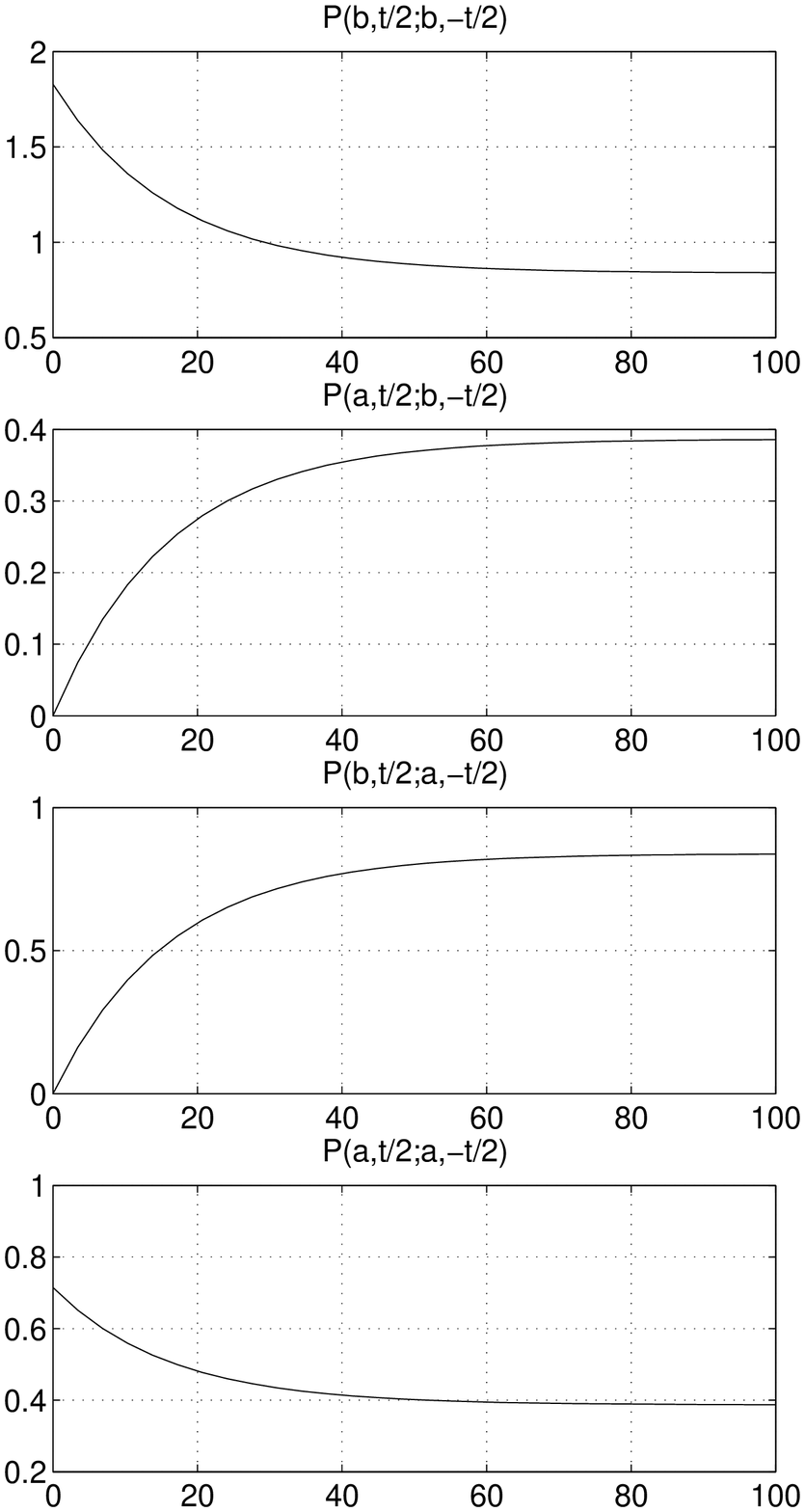}
   \end{minipage}
\par
\bigskip \centering
\begin{minipage}[b]{0.40\textwidth}
    \centering
    \includegraphics[height=5.cm]{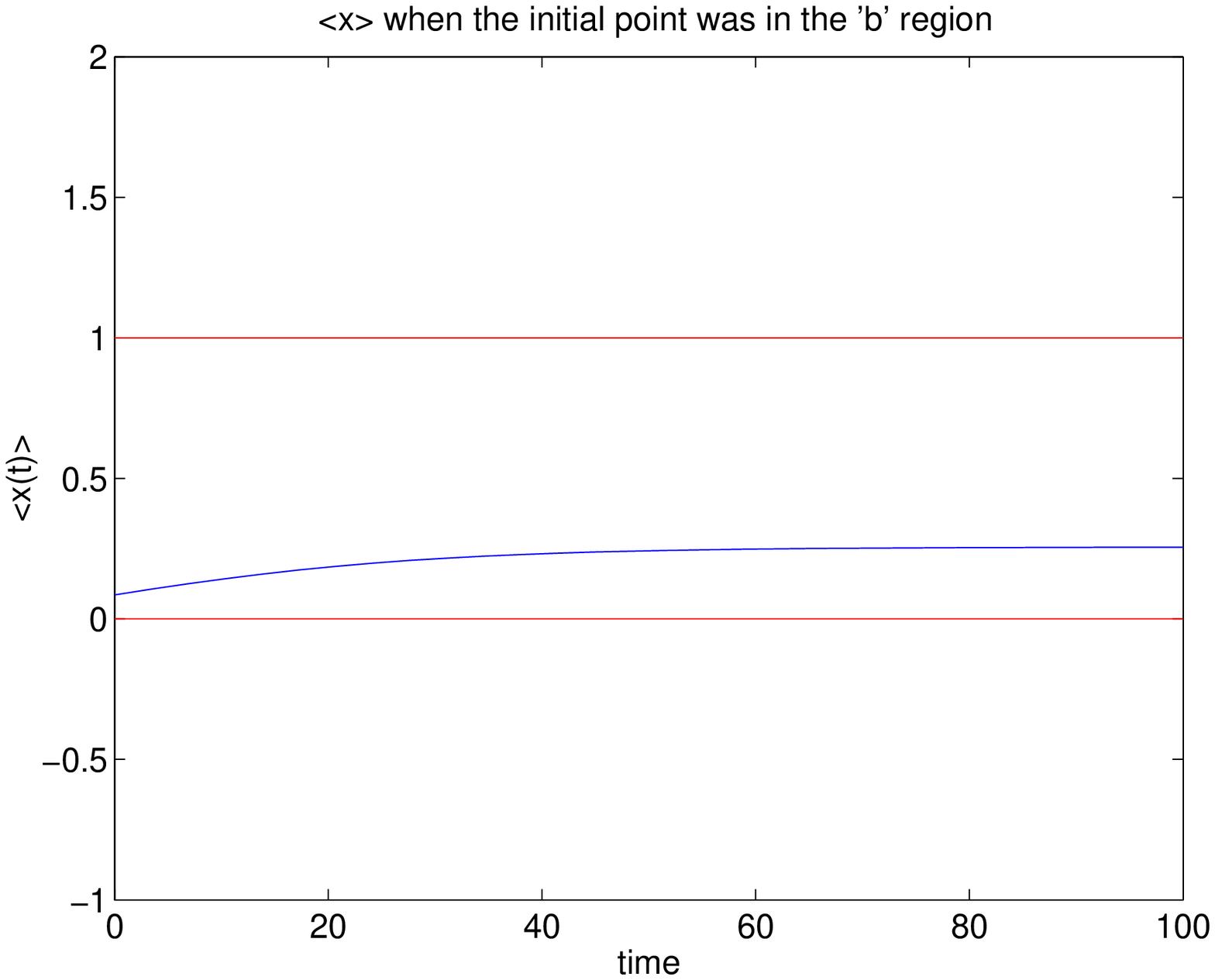}
   \end{minipage} \hspace{0.1\textwidth} 
\begin{minipage}[b]{0.40\textwidth}
    \centering
    \includegraphics[height=5.cm]{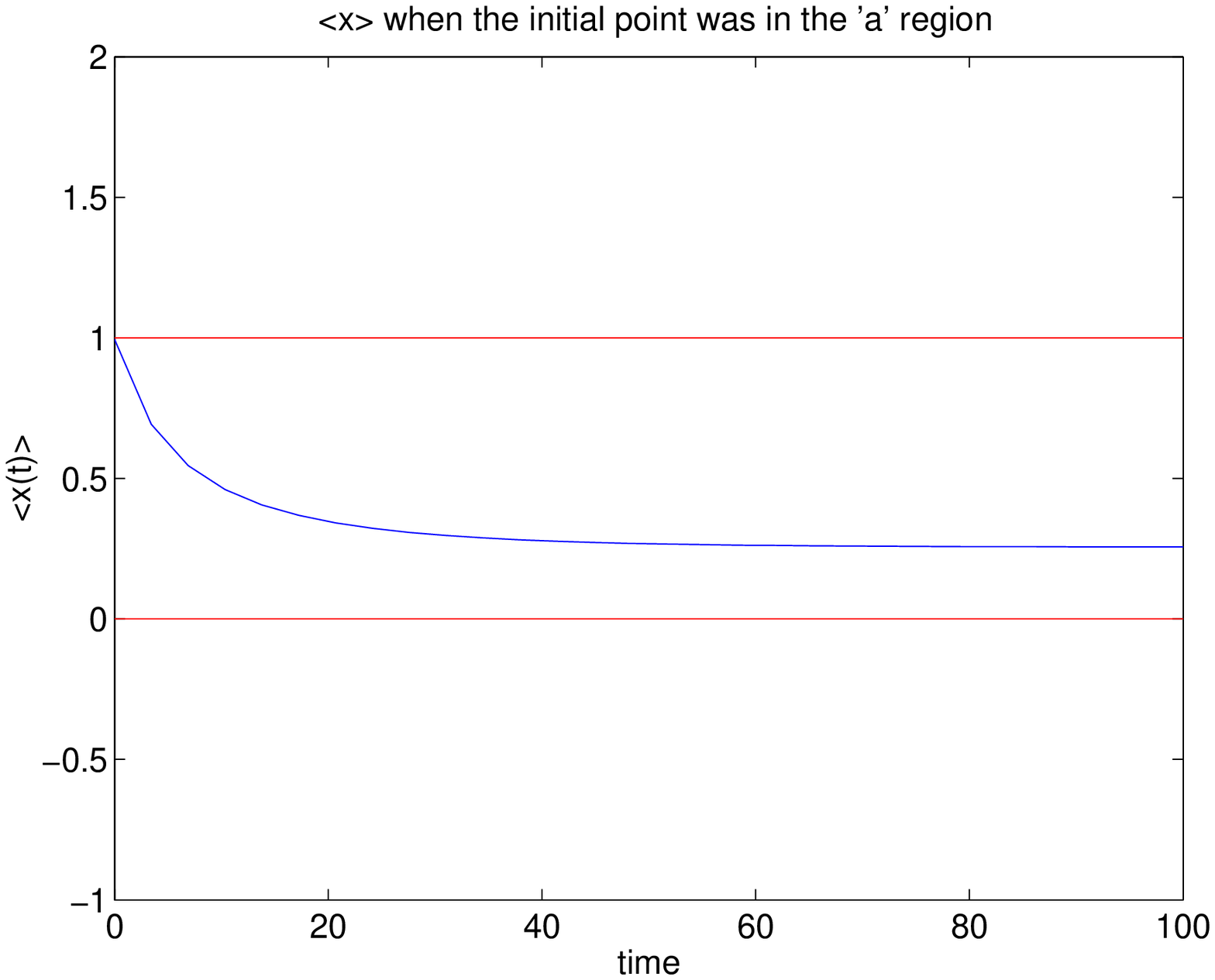}
   \end{minipage}
\end{figure}

\begin{figure}[htb]
\centering
\includegraphics[width=11cm]{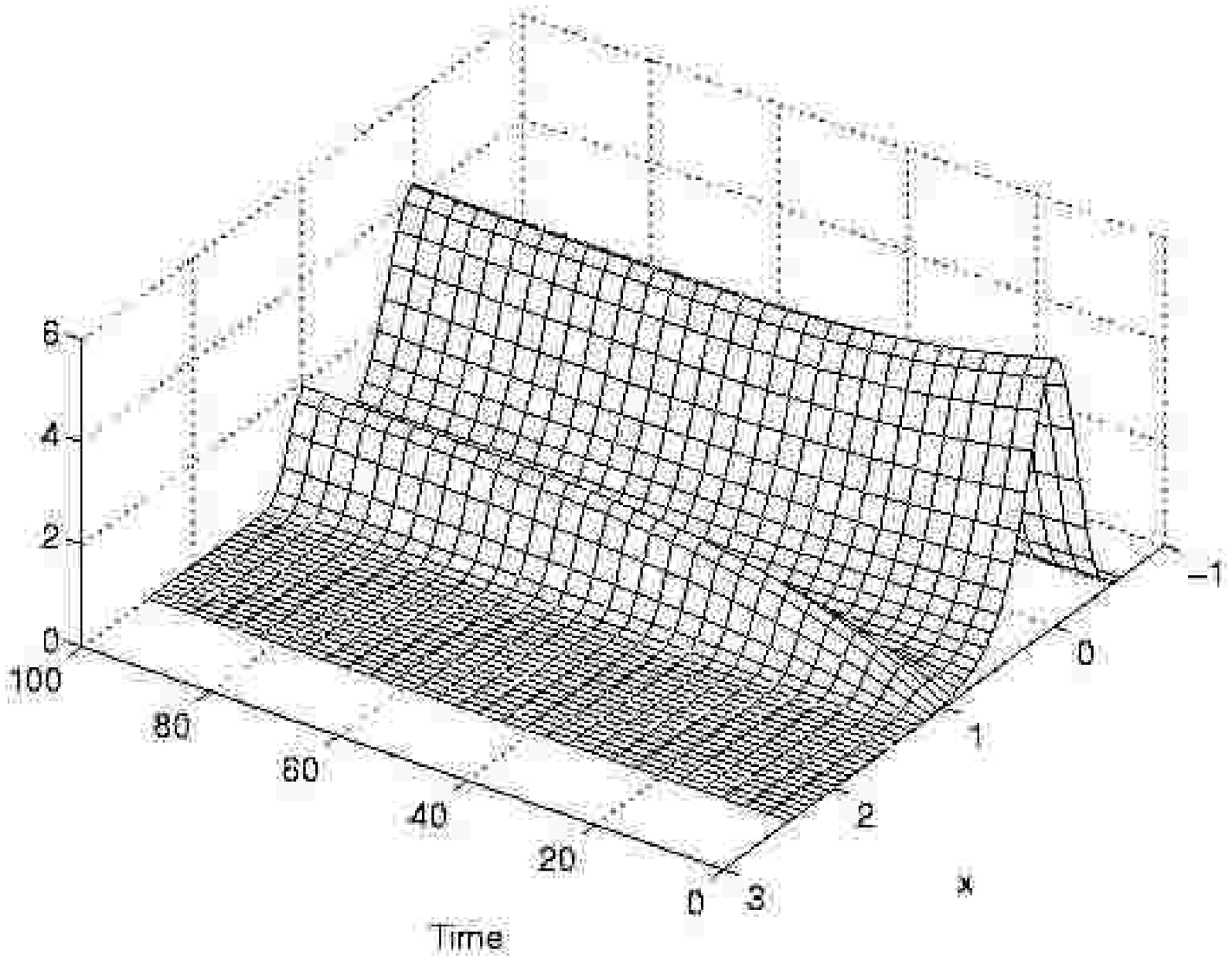}
\end{figure}
\begin{figure}[htbp]
\centering
\includegraphics[width=11cm]{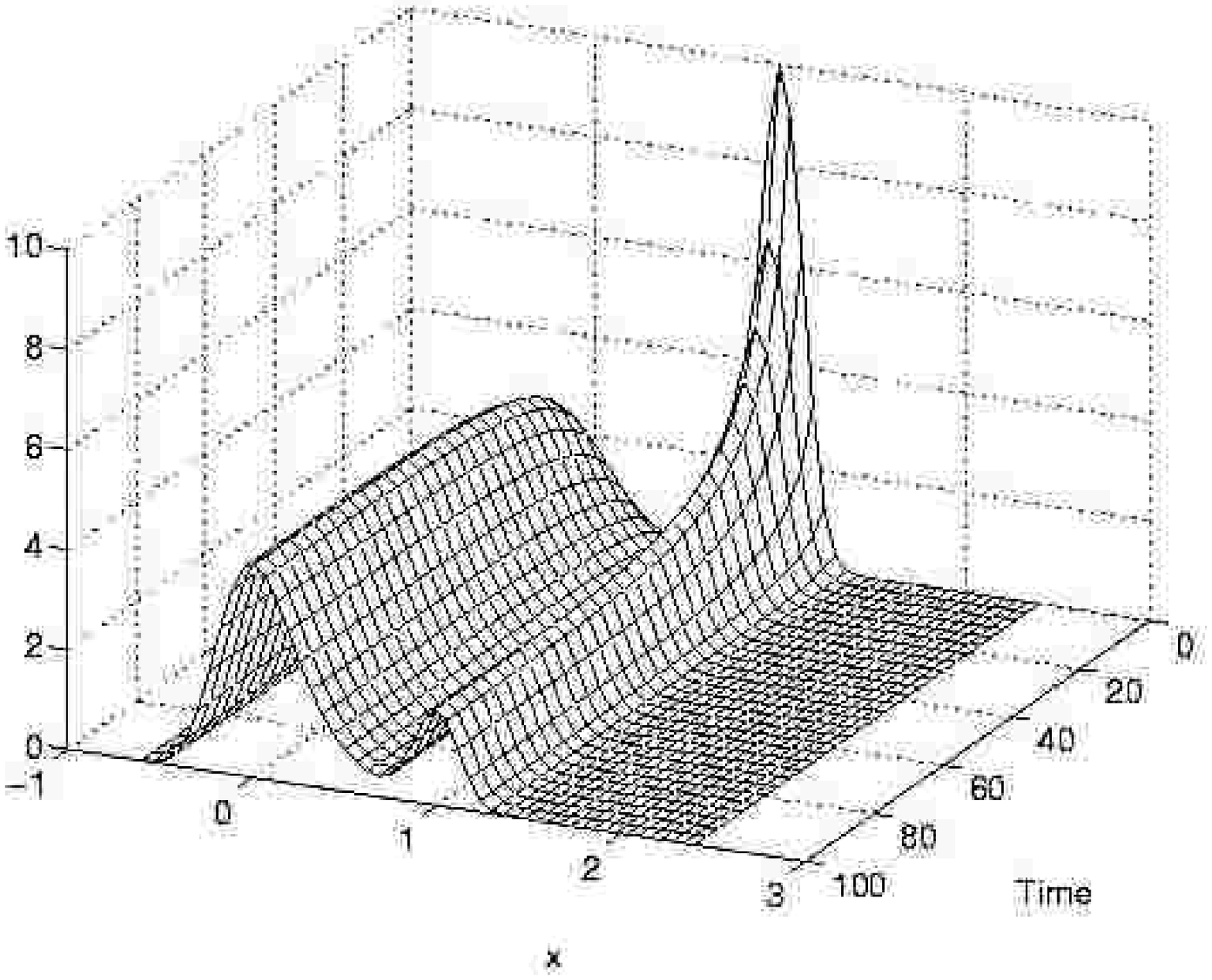}
\caption{Parameters: $a = 0.75$ and $D = 0.05$}
\label{Figa0.75D0.05_6}
\end{figure}

\clearpage

\begin{figure}[htb]
\centering
\begin{minipage}[b]{0.40\textwidth}
    \centering
    \includegraphics[height=11cm]{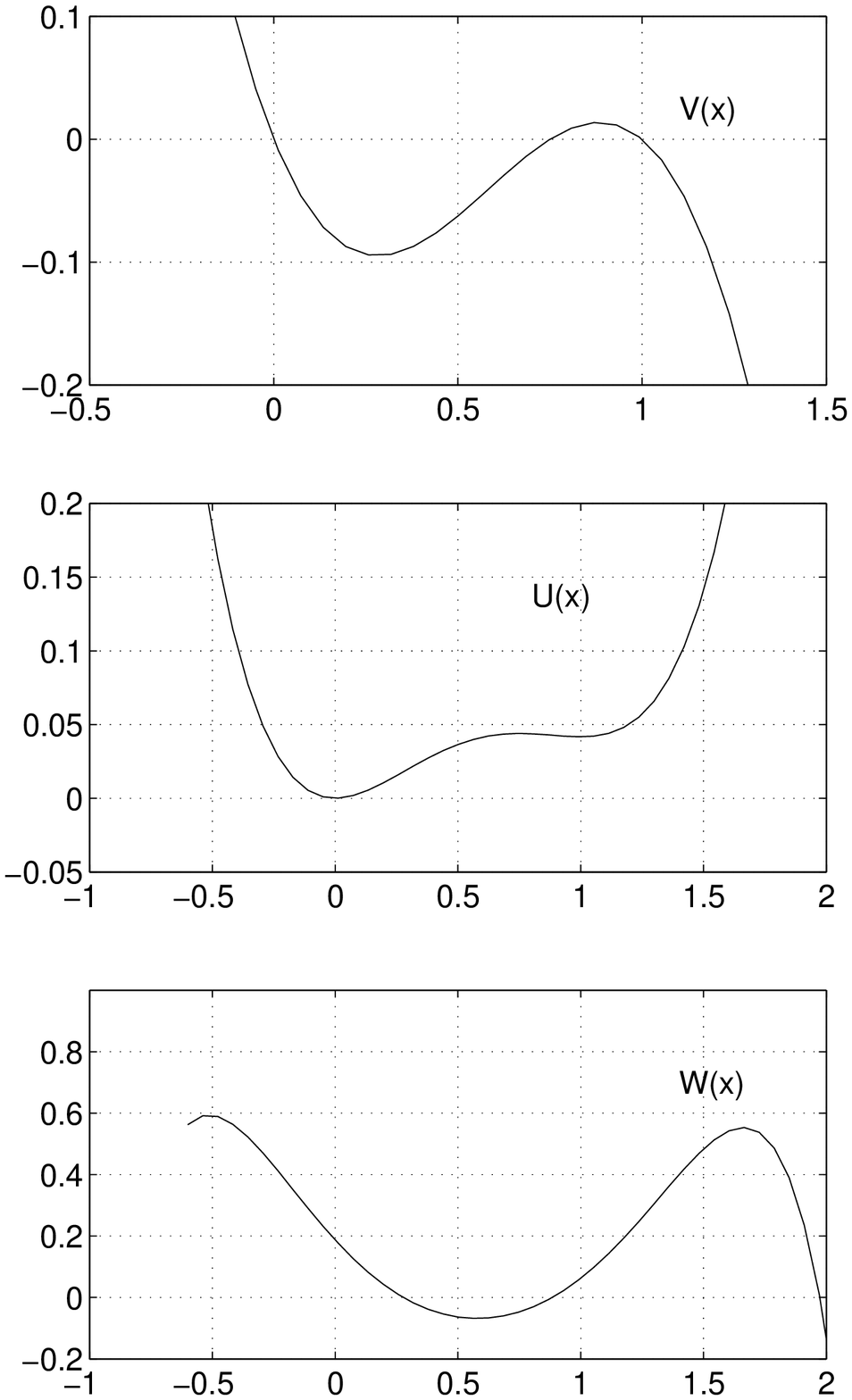}
   \end{minipage} \hspace{0.1\textwidth} 
\begin{minipage}[b]{0.40\textwidth}
    \centering
    \includegraphics[height=11.5cm]{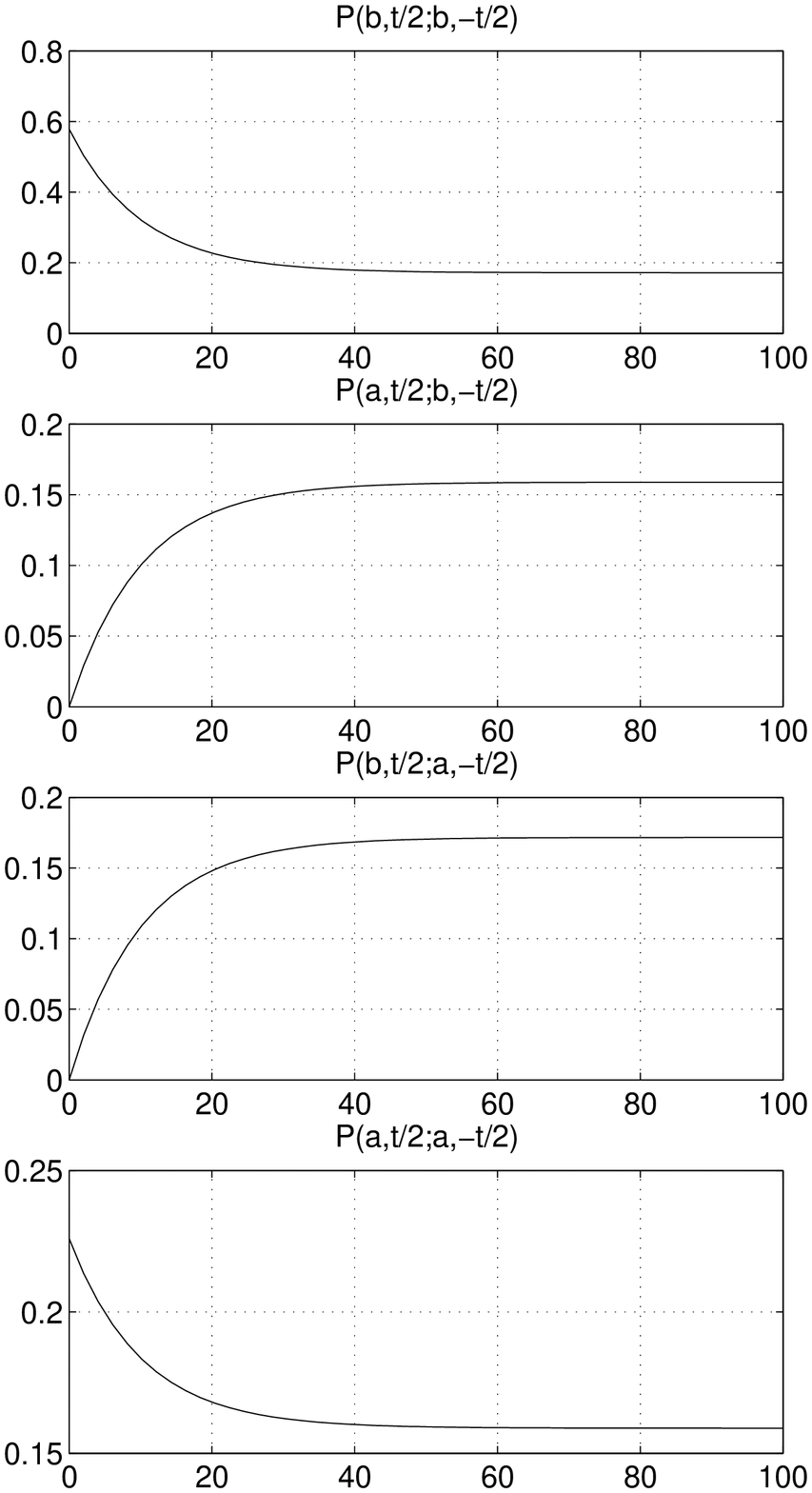}
   \end{minipage}
\par
\bigskip \centering
\begin{minipage}[b]{0.40\textwidth}
    \centering
    \includegraphics[height=5.cm]{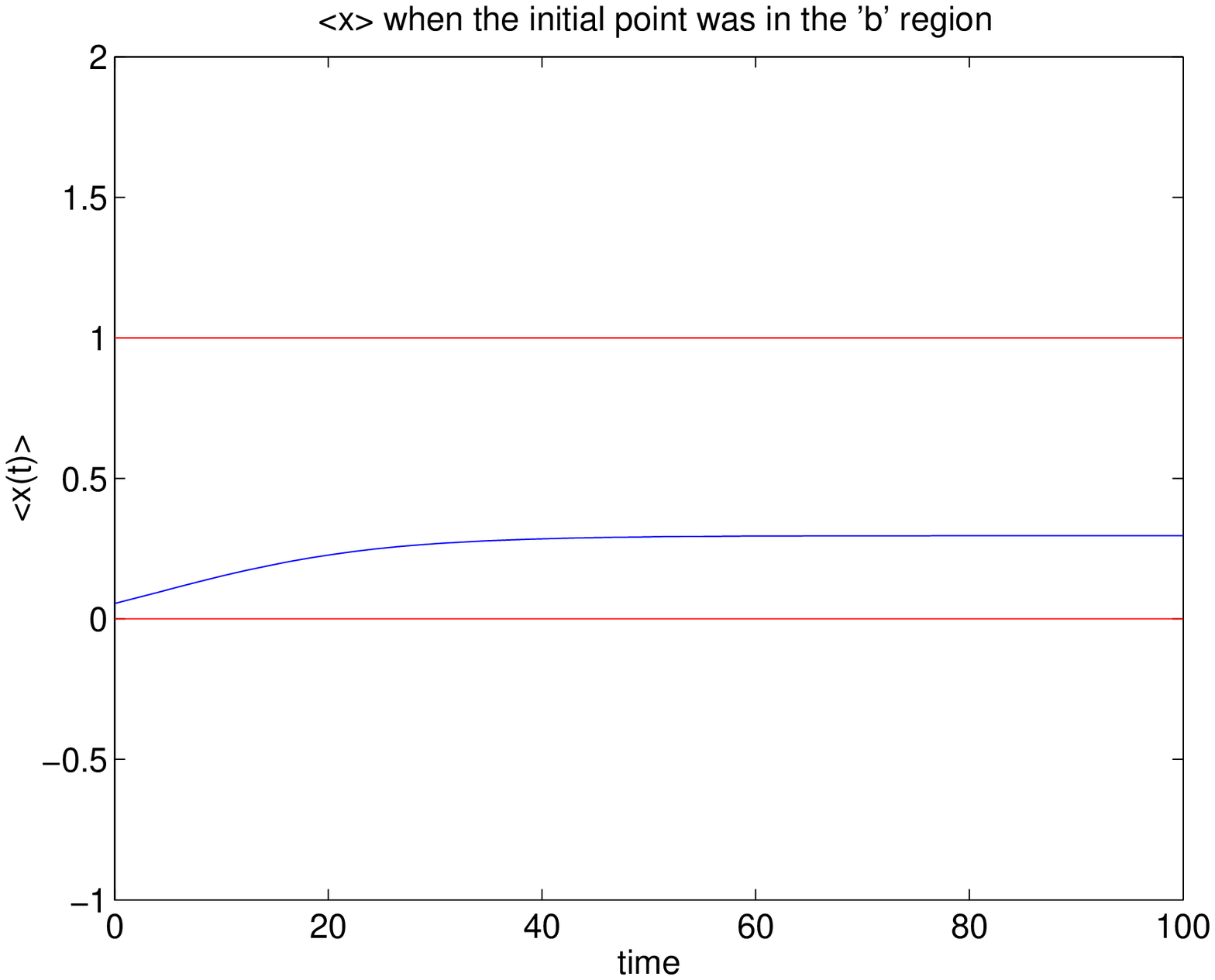}
   \end{minipage} \hspace{0.1\textwidth} 
\begin{minipage}[b]{0.40\textwidth}
    \centering
    \includegraphics[height=5.cm]{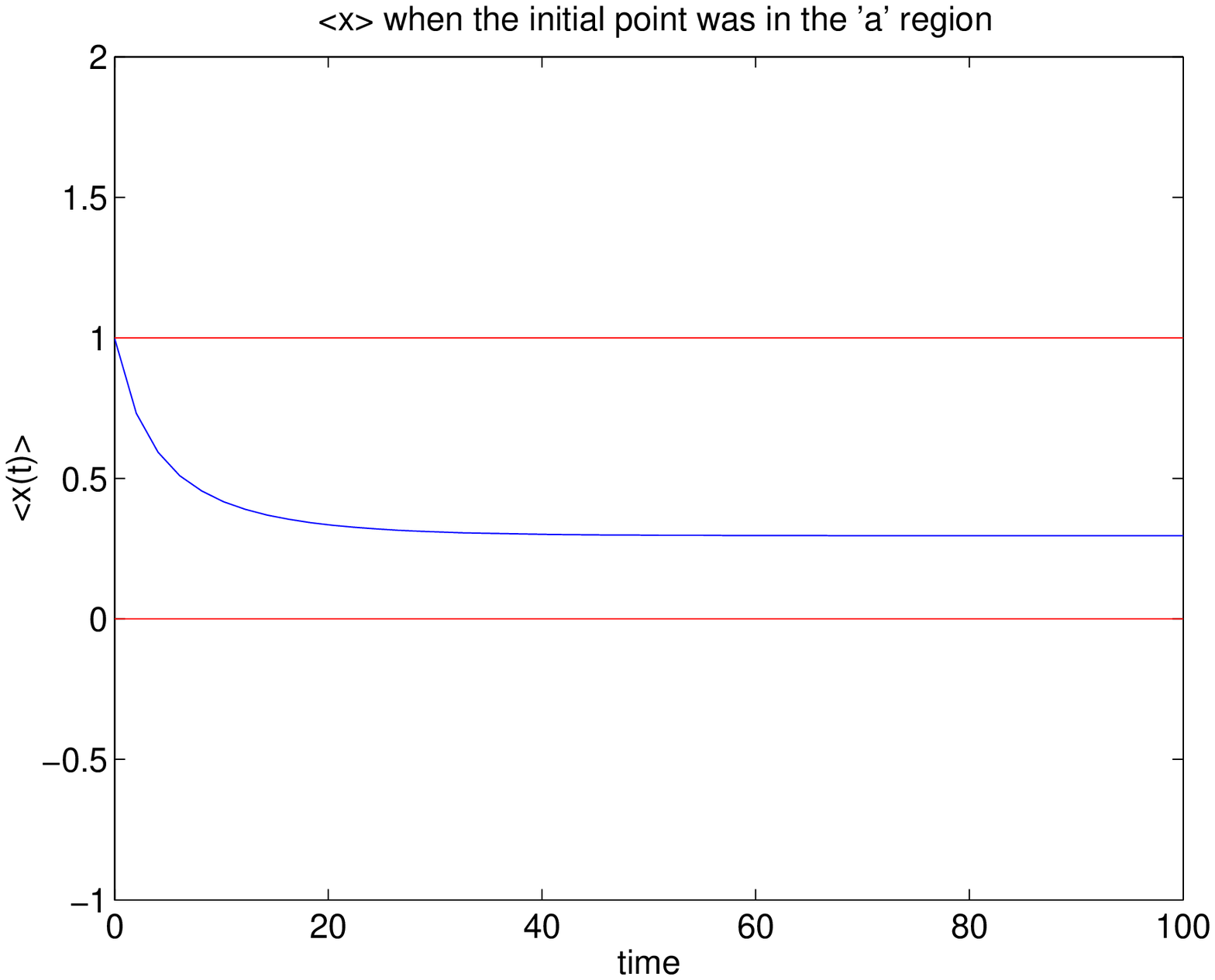}
   \end{minipage}
\end{figure}

\begin{figure}[htb]
\centering
\includegraphics[width=11cm]{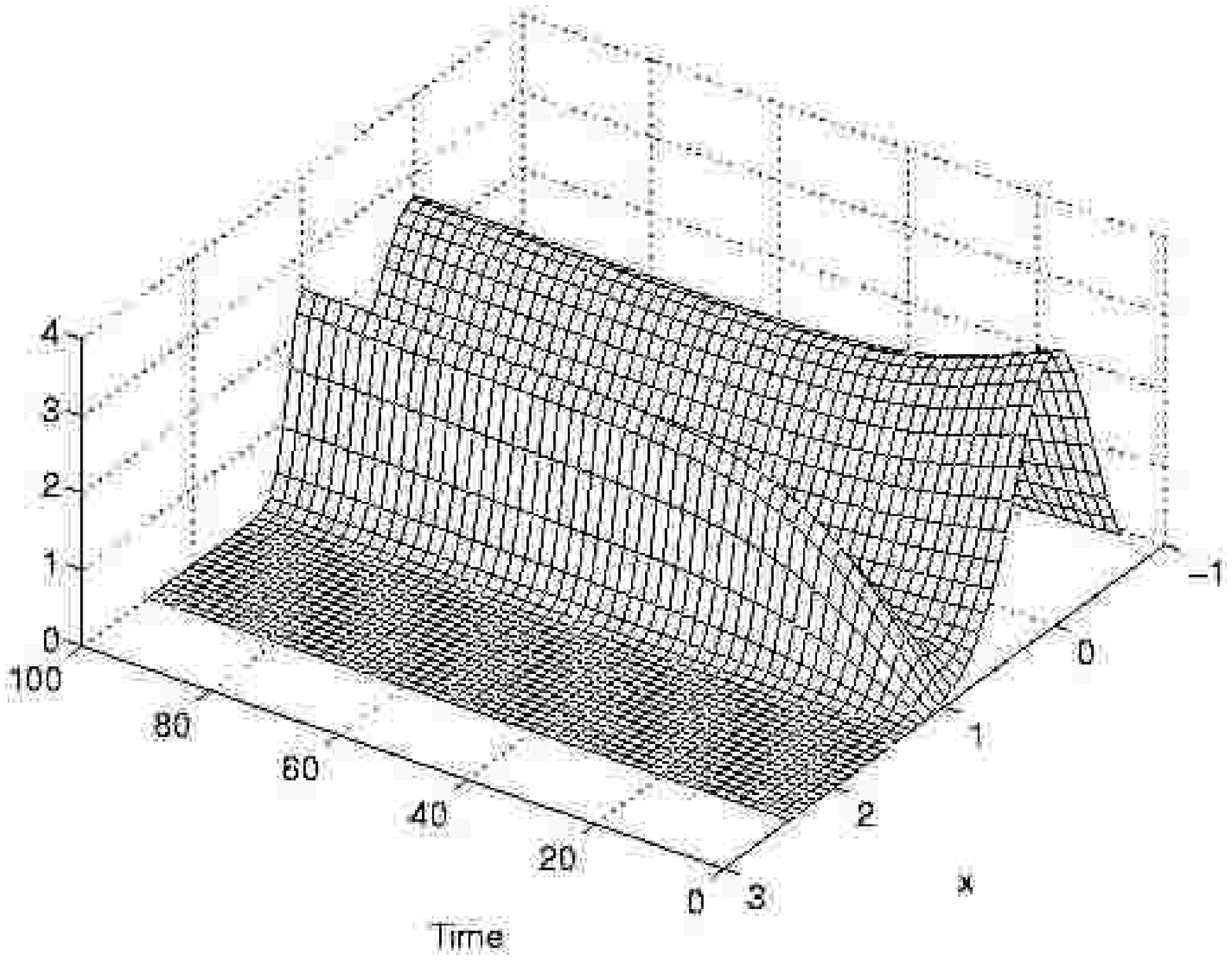}
\end{figure}
\begin{figure}[htbp]
\centering
\includegraphics[width=11cm]{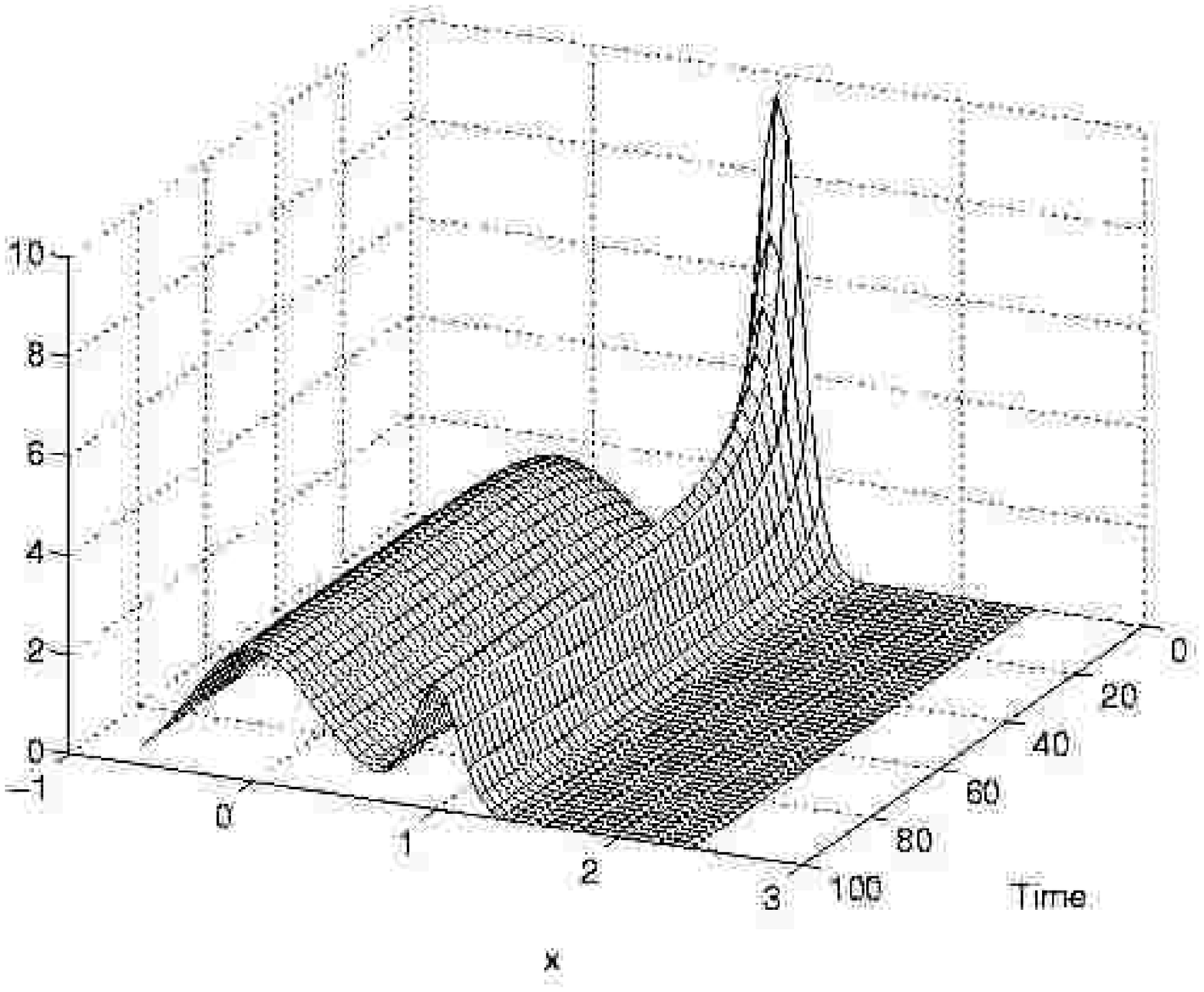}
\caption{Parameters: $a = 0.75$ and $D = 0.5$}
\label{Fig13}
\end{figure}

\clearpage

\begin{figure}[htb]
\centering
\begin{minipage}[b]{0.40\textwidth}
    \centering
    \includegraphics[height=11cm]{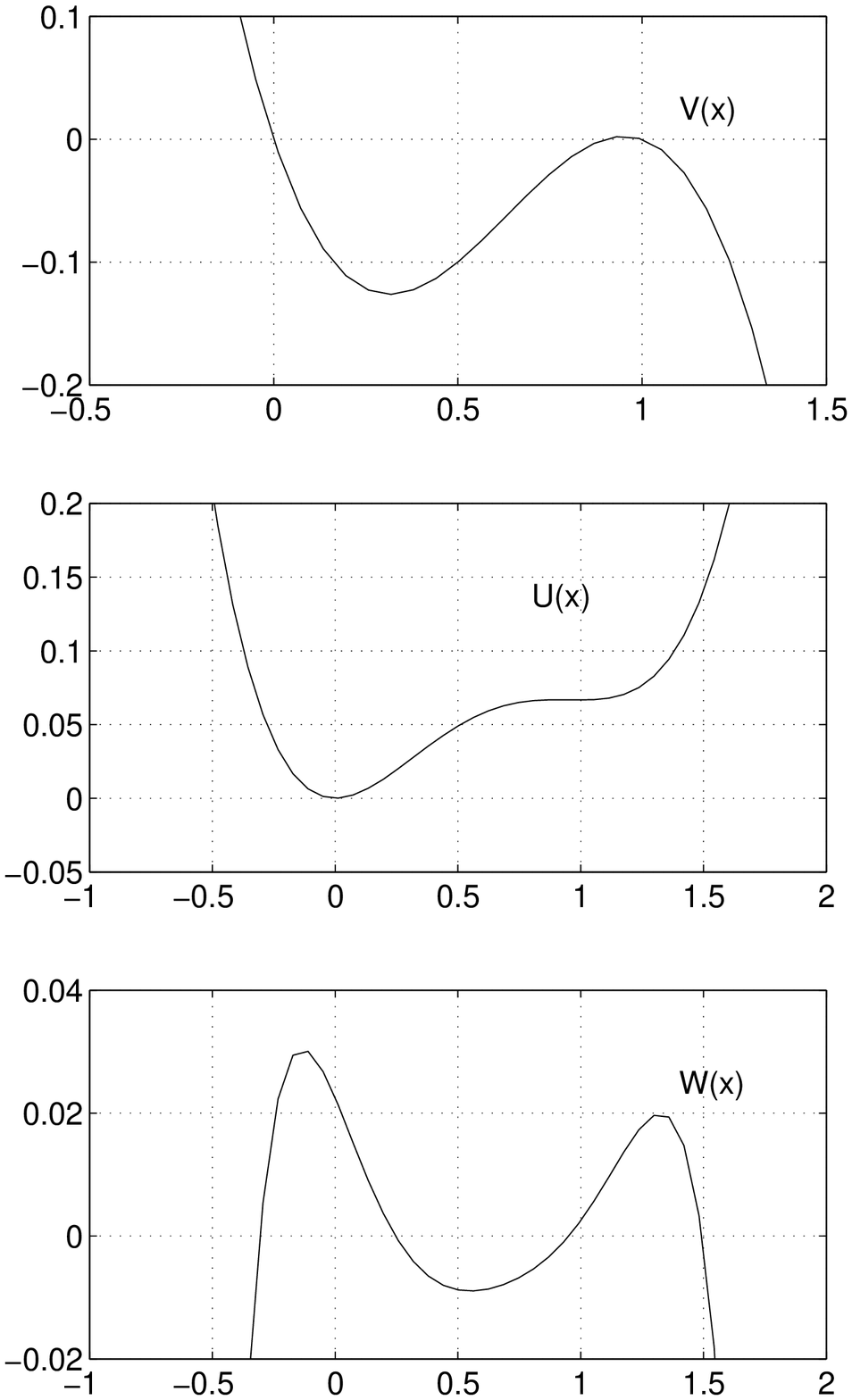}
   \end{minipage} \hspace{0.1\textwidth} 
\begin{minipage}[b]{0.40\textwidth}
    \centering
    \includegraphics[height=11.5cm]{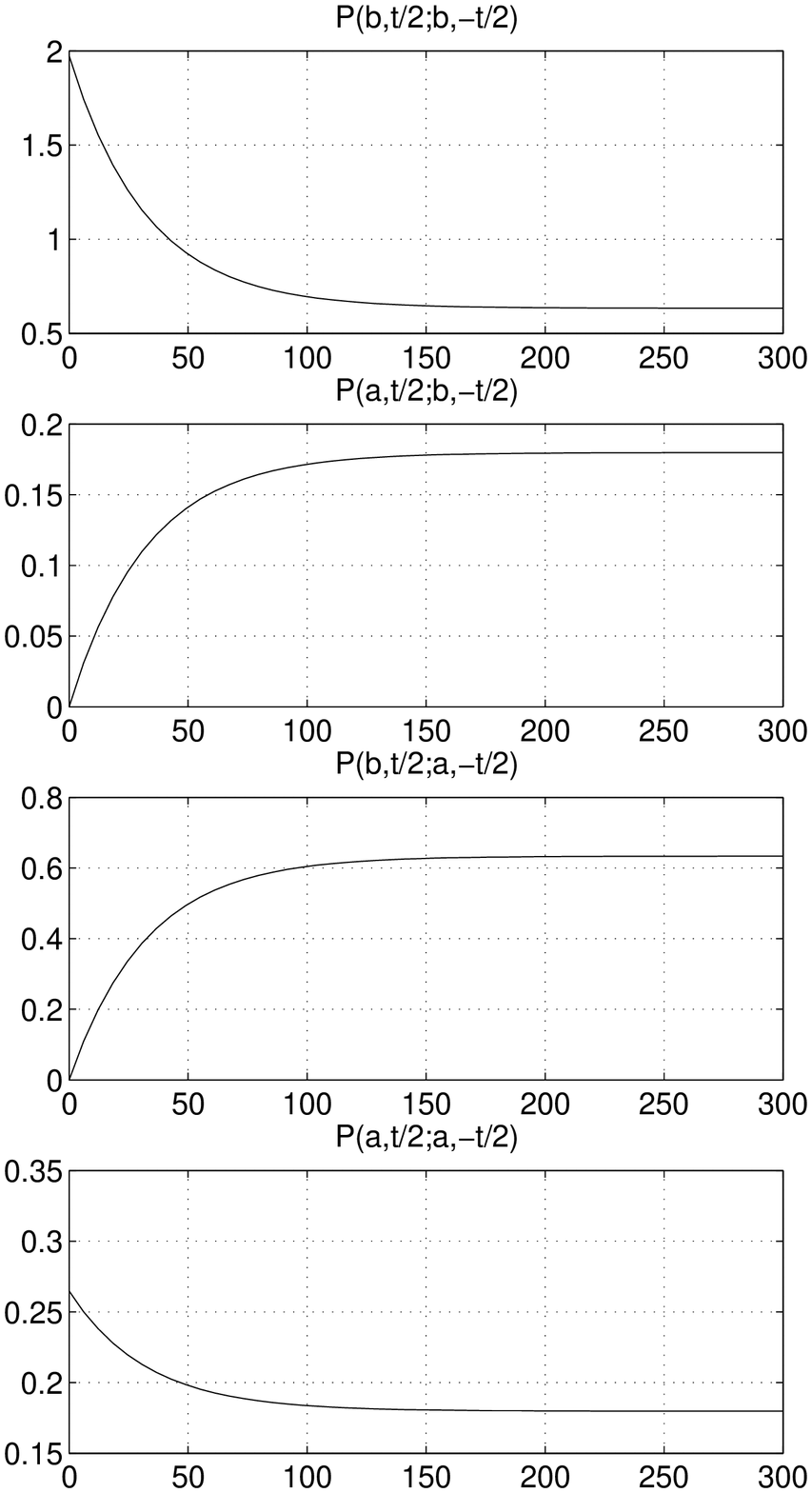}
   \end{minipage}
\par
\bigskip \centering
\begin{minipage}[b]{0.40\textwidth}
    \centering
    \includegraphics[height=5.cm]{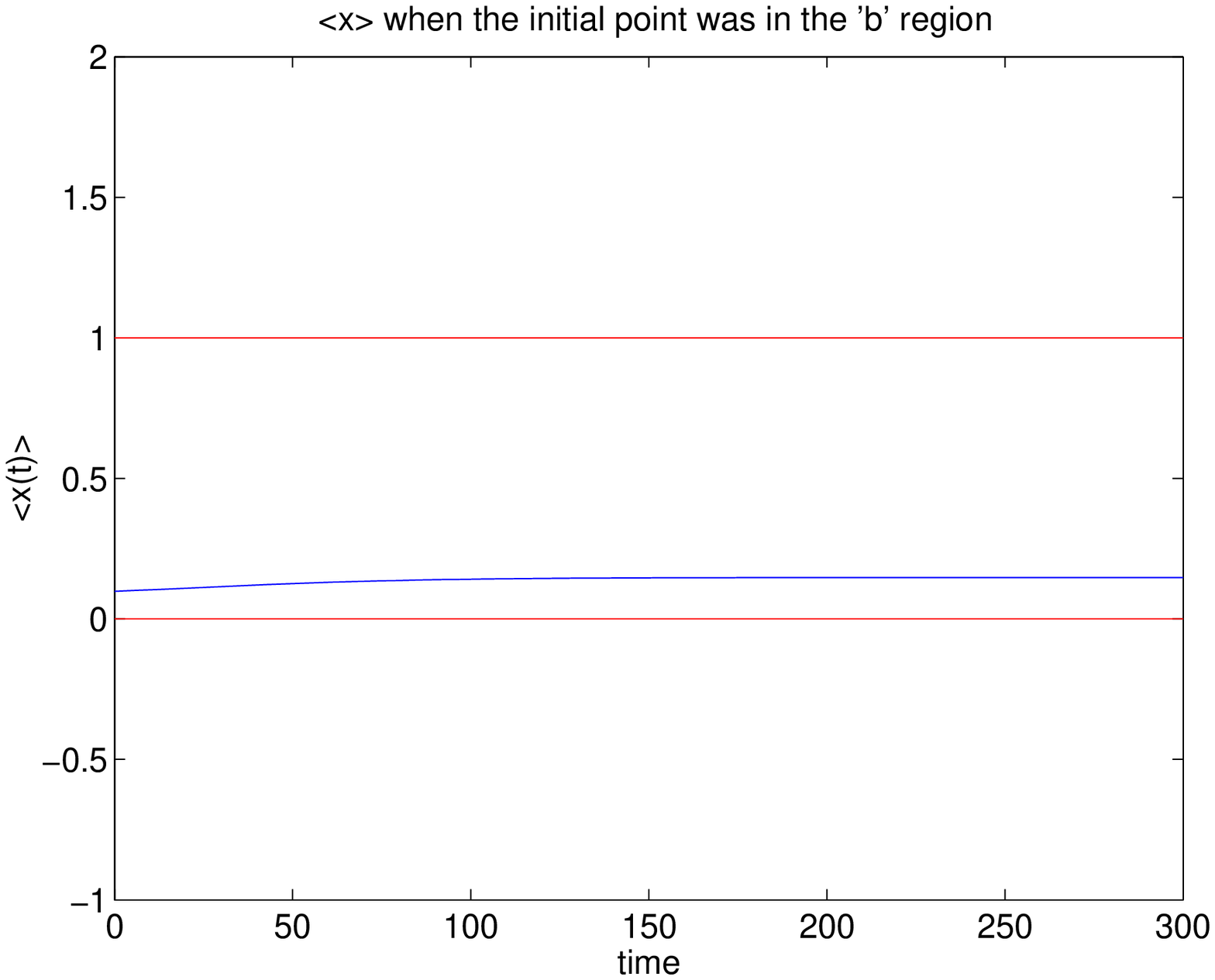}
   \end{minipage} \hspace{0.1\textwidth} 
\begin{minipage}[b]{0.40\textwidth}
    \centering
    \includegraphics[height=5.cm]{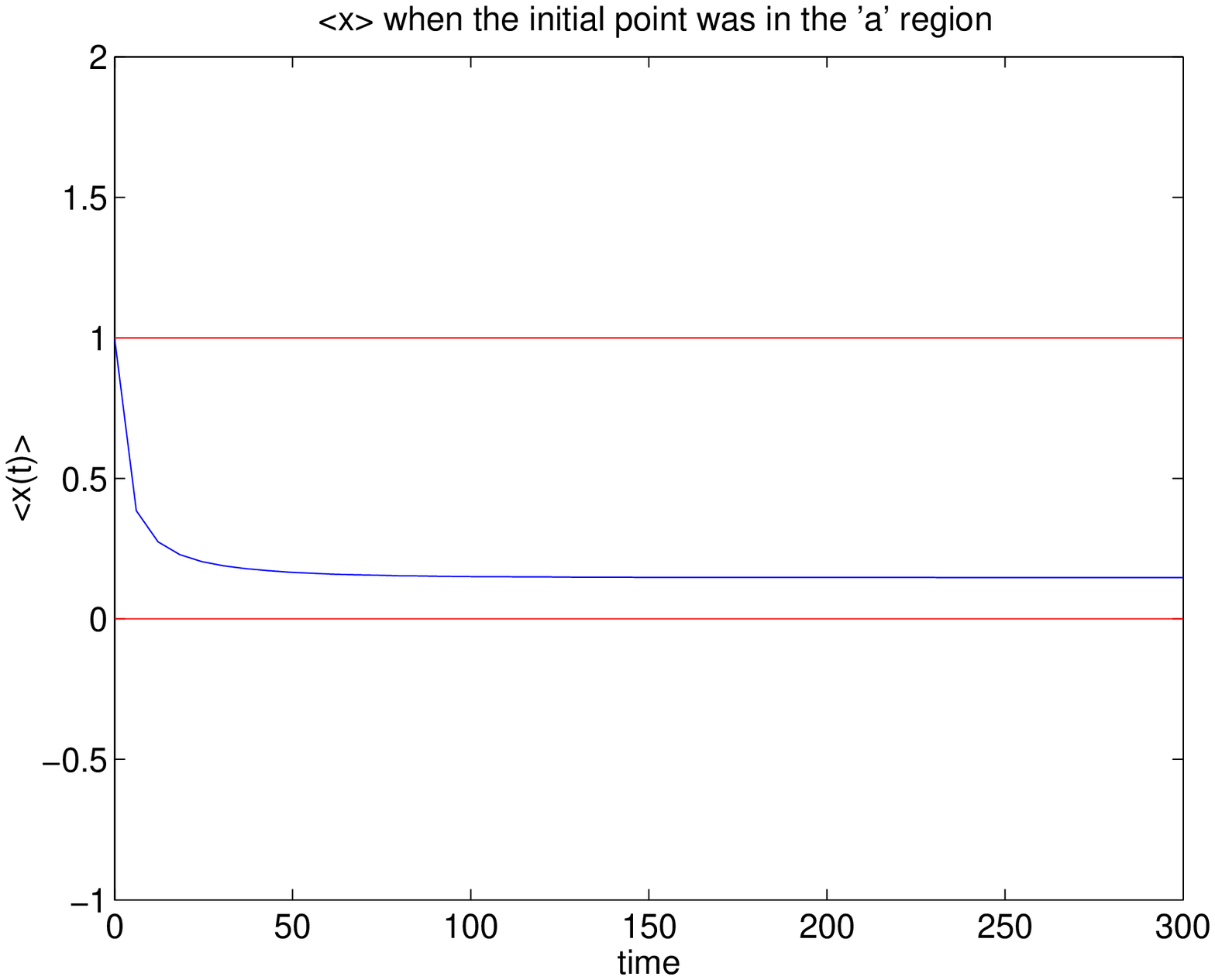}
   \end{minipage}
\end{figure}

\begin{figure}[htb]
\centering
\includegraphics[width=11cm]{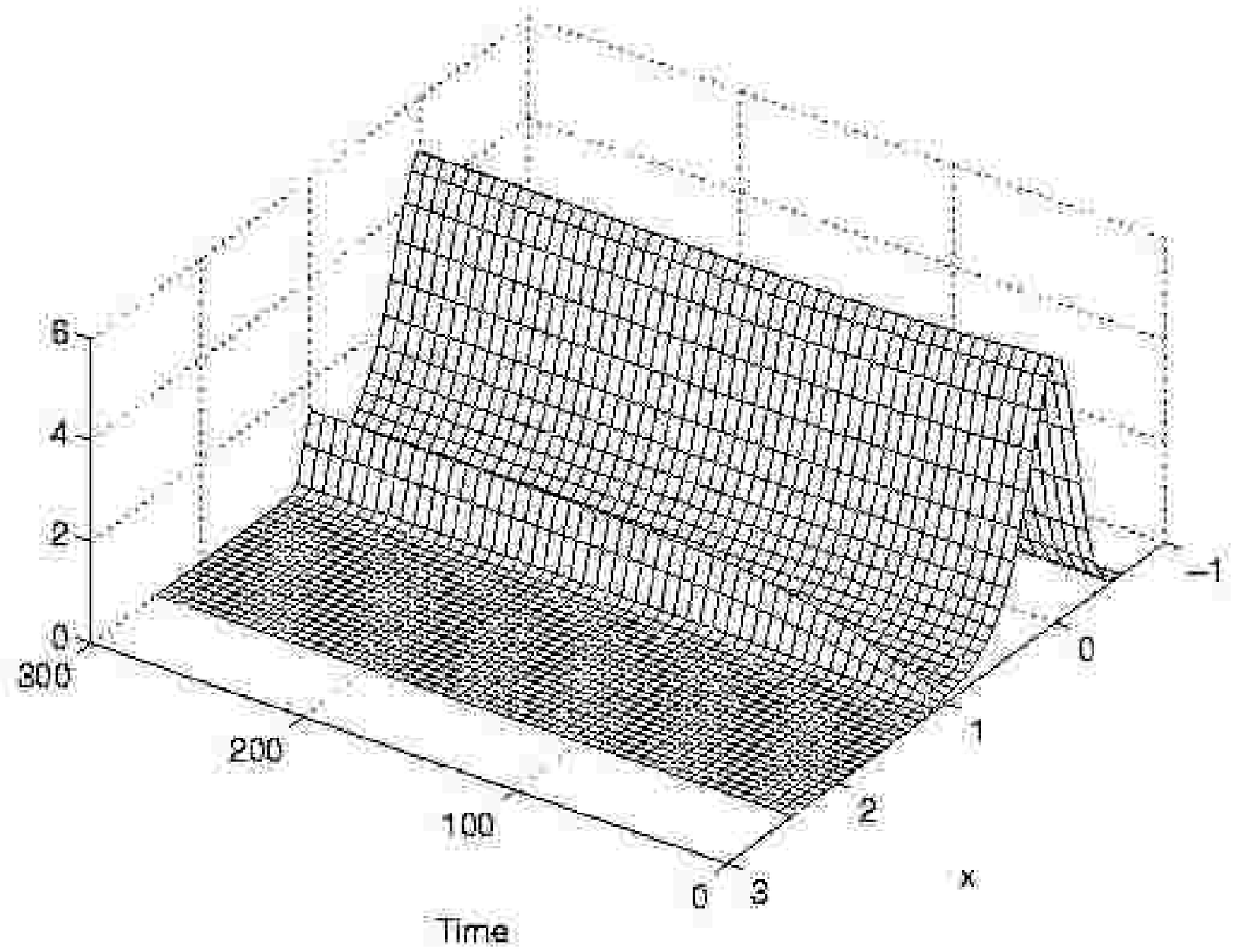}
\end{figure}
\begin{figure}[htbp]
\centering
\includegraphics[width=11cm]{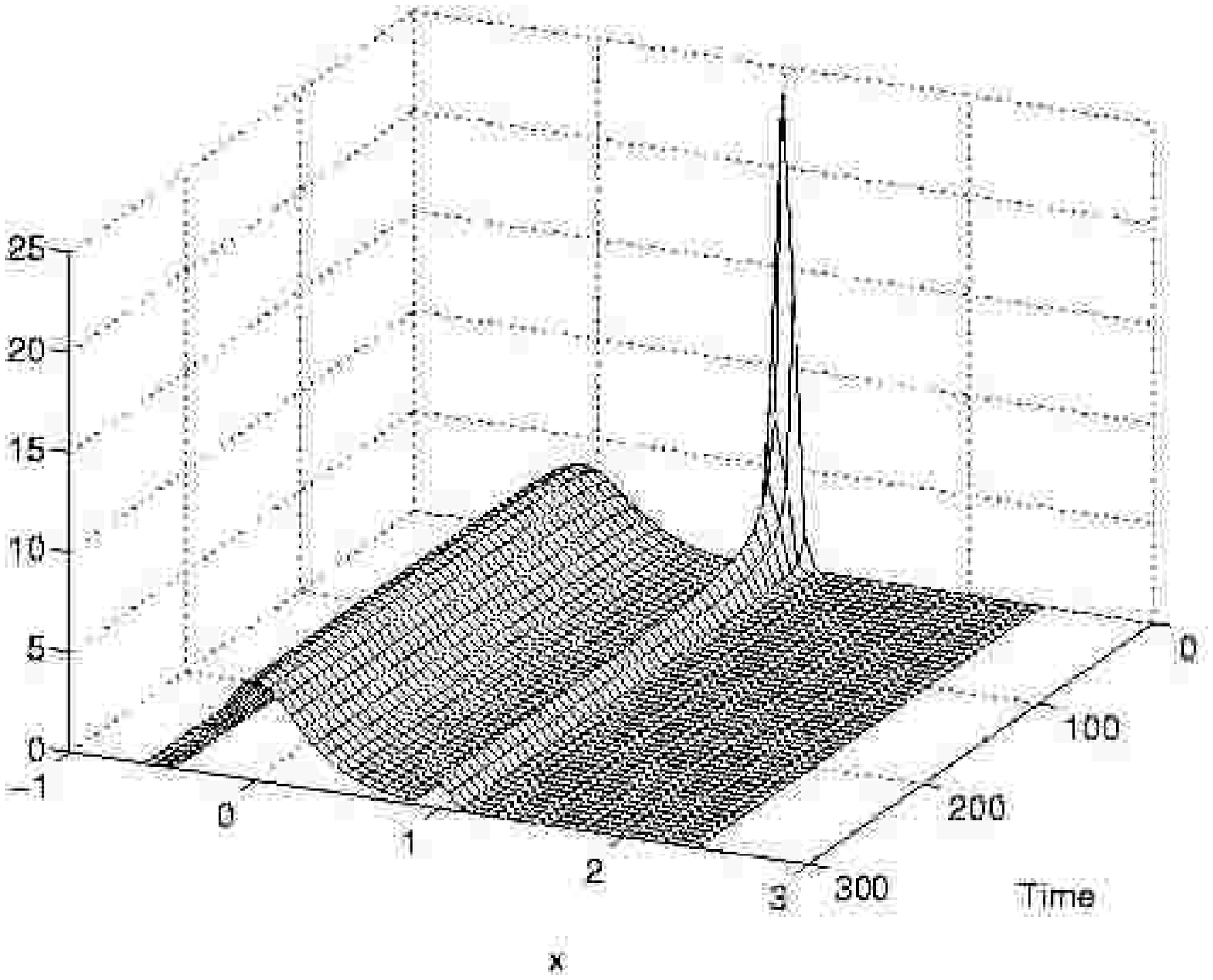}
\caption{Parameters: $a = 0.9$ and $D = 0.05$}
\label{Figa0.9D0.05_5}
\end{figure}
\clearpage

\section*{Appendix A : Time variation for general form of the potential}

\appendix
\renewcommand{\theequation}{A.\arabic{equation}} \setcounter{equation}{0}

The stochastic motion is described in terms of the \emph{conditional
probability} $P\left( x,t;x_{i},0\right) $ that the particle initially $%
\left( t=0\right) $ at $x_{i}$ to be at the point $x$ at time $t$. We will
use the notation that suppress the $0$ as the initial time, $P\left(
x,t;x_{i}\right) $. The conditional probability obeys the following
Fokker-Planck equation 
\begin{equation*}
\frac{\partial P}{\partial t}=-\frac{\partial }{\partial x}\left[ U^{\prime
}\left( x\right) P\right] +D\frac{\partial ^{2}P}{\partial x^{2}}
\end{equation*}
where the velocity function is here derived from the potential $U$%
\begin{equation*}
F\left[ x\left( t\right) ,t\right] \equiv -U^{\prime }\left( x\right)
\end{equation*}
since there is no drive. The initial condition for the probability function
is 
\begin{equation*}
P\left( x,0;x_{i}\right) =\delta \left( x-x_{i}\right)
\end{equation*}

The simple form of the potential $U\left( x\right) $ allows the introduction
of two constants 
\begin{eqnarray*}
\omega _{0} &\equiv &\left| U^{\prime \prime }\left( x=0\right) \right| \\
\omega _{1} &\equiv &U^{\prime \prime }\left( x=1\right)
\end{eqnarray*}
The following orderings are assumed 
\begin{eqnarray*}
D &\ll &\omega _{0} \\
D &\ll &\omega _{1}
\end{eqnarray*}

The solution of the Fokker-Planck equation is given in terms of the
following \emph{path-integral} 
\begin{equation*}
P\left( x,t;x_{i}\right) =\frac{\exp \left[ -U\left( x\right) /2D\right] }{%
\exp \left[ -U\left( x_{i}\right) /2D\right] }K\left( x,t;x_{i}\right)
\end{equation*}
where 
\begin{equation}
K\left( x,t;x_{i}\right) =\int \emph{D}\left[ x\left( \tau \right) \right]
\exp \left[ -\frac{S}{D}\right]  \label{kfunc}
\end{equation}
where the functional integration is done over all trajectories that start at 
$\left( x_{i},0\right) $ and end at $\left( x,t\right) $. The \emph{action
functional} is given by 
\begin{equation*}
S\left( x,t;x_{i}\right) =\int_{0}^{t}d\tau \left[ \frac{1}{4}\overset{\cdot 
}{x}^{2}+W\left( x\right) \right]
\end{equation*}
with the notation 
\begin{equation*}
W\left( x\right) \equiv \frac{1}{4}U^{\prime 2}\left( x\right) -\frac{D}{2}%
U^{\prime \prime }\left( x\right)
\end{equation*}

In order to calculate explicitely the functional integral we look first for
the paths $x\left( t\right) $ that extremize the action. They are provided
by the Euler-Lagrange equations, which reads 
\begin{equation*}
\frac{\delta S}{\delta x_{c}}=-\frac{1}{2}\overset{\cdot \cdot }{x}%
_{c}+W^{\prime }\left( x_{c}\right) =0
\end{equation*}
with the boundary condition 
\begin{eqnarray*}
x_{c}\left( 0\right) &=&x_{i} \\
x_{c}\left( t\right) &=&x
\end{eqnarray*}

The first thing to do after finding the extremizing paths is to calculate
the \emph{action} functional along them, $S_{c}$. After obtaining these
extremum paths we have to consider the contribution to the functional
integral of the paths situated in the neighborhood and this is done by
expanding the action to second order. The argument of the expansion is the
difference between a path from this neigborhood and the extremum path. The
functional integration over these differences can be done since it is
Gaussian and the result is expressed in terms of the \emph{determinant} of
the operator resulting from the second order expansion of $S$. Then the
expression (\ref{kfunc}) becomes 
\begin{equation}
K=N\frac{1}{\left| \det \widehat{L}\left( x_{c}\right) \right| ^{1/2}}\exp
\left( -\frac{S_{c}}{D}\right)  \label{kexp}
\end{equation}
where 
\begin{equation}
\widehat{L}\left( x_{c}\right) \equiv \left. \frac{\delta ^{2}S}{\delta
x\left( \tau \right) ^{2}}\right| _{x\left( \tau \right) =x_{c}\left( \tau
\right) }=-\frac{1}{2}\frac{d^{2}}{d\tau ^{2}}+W^{\prime \prime }\left(
x_{c}\right)  \label{d2s}
\end{equation}
The constant $N$ is for normalization.

In order to find the determinant, one has to solve the eigenvalue problem 
\begin{equation}
\left[ \widehat{L}\left( x_{c}\right) -\lambda _{n}\right] \phi _{n}\left(
\tau \right) =0  \label{eig}
\end{equation}
with boundary conditions for the \emph{differences} between the trajectories
in the neighborhood and the extremum trajectory 
\begin{equation*}
\phi _{n}\left( 0\right) =\phi _{n}\left( t\right) =0
\end{equation*}

It can be shown that 
\begin{equation*}
\frac{N^{2}}{\left| \det \widehat{L}\left( x_{c}\right) \right| }=\frac{1}{%
2\pi D\overset{\cdot }{x}_{c}\left( 0\right) }\frac{\partial E}{\partial x}
\end{equation*}
where $E$ is the energy of the path $x_{c}\left( \tau \right) $. (\textbf{%
see Dashen and Patrascioiu}).

There is a fundamental problem concerning the use of the equation (\ref{kexp}%
). It depends on the possibility that all the eigenvalues from the Eq.(\ref
{eig}) are determined. However, a path connecting a point in the neigborhood
of the maximum of $U\left( x\right) $, 
\begin{equation*}
x_{i}\sim 0+O\left( \sqrt{D}\right)
\end{equation*}
with a point in the neighborhood of one of the minima 
\begin{equation*}
x\sim \pm 1+O\left( \sqrt{D}\right)
\end{equation*}
is a kinklike solution. This solution (which will be replaced in the
expression of the operator in Eq.(\ref{eig}) contains a parameter, the
``center'' of the kink, $t_{0}$. This is an arbitrary parameter since the
moment of traversation is arbitrary. The equation for the eigenvalues has
therefore a symmetry at time translations of $t_{0}$ and has as a
consequence the appearence of a \emph{zero} eigenvalue. Then the expression
of the determinant would be infinite and the rate of transfer would vanish.
Actually, the time translation invariance is treated by considering the
arbitrary moment $t_{0}$ as a new variable and performing a change of
variable, from the set of functions $\phi _{n}$ to the set $\left(
t_{0},\phi _{n,n\neq 0}\right) $. \ The trajectory that extremizes the
action is a kinklike solution $\overline{x}\left( \tau ,t_{0}\right) $ (is
not exactly a kink since the shape of the potential is not that which
produces the $\tanh $ solution) connecting the points : 
\begin{eqnarray*}
\overline{x}\left( 0,t_{0}\right) &=&x_{i} \\
\overline{x}\left( t,t_{0}\right) &=&x
\end{eqnarray*}
This treatment then consists of considering $t_{0}$ as a collective
coordinate (\textbf{see Rajaraman and Coleman and Gervais \& Sakita}). The
new form of Eq.(\ref{kexp}) is 
\begin{equation}
K=N\int_{0}^{t}dt_{0}J\frac{\exp \left( -\overline{S}/D\right) }{\left|
\det^{\prime }\widehat{L}\left( \overline{x}\right) \right| ^{1/2}}
\label{kexpbar}
\end{equation}
where $\overline{S}$ is the action of the path $\overline{x}\left( \tau
,t_{0}\right) $. The fact that the zero eigenvalue has been eliminated is
indicated by the $^{\prime }$.

The factor in the integral is approximately 
\begin{equation*}
J^{2}=\frac{1}{2\pi D}\left[ U\left( x_{i}\right) -U\left( x\right) \right]
\end{equation*}

Two limits of time are important. The first is 
\begin{equation*}
t_{s}=\frac{1}{2\omega _{0}}\ln \left( \frac{\omega _{0}}{D}\right)
\end{equation*}
For the times 
\begin{equation*}
0\leq t\lesssim t_{s}
\end{equation*}
for a particle initially in the region of the \emph{unstable} state, $%
x_{i}\sim O\left( \sqrt{D}\right) $, the region around the \emph{stable}
state, $\pm 1$, or: $x\sim \pm 1+O\left( \sqrt{D}\right) $ is insignificant
for the calculation of the probability $P\left( x,t;x_{i}\right) $. Then the
expression (\ref{kexp}) can be used.

Much more important is the subsequent time regime 
\begin{equation*}
t>t_{s}
\end{equation*}
where the particles leave the \emph{unstable} point and the equilibrium
state with density around the \emph{stable} positions $\pm 1$ is approached.
Then the general formula (\ref{kexpbar}) should be used.

\bigskip

The expression of the potential in general does not allow an explicit
determination of the extremizing trajectories. Approximations are necessary.

The energy $E$ of the path $x_{c}\left( \tau \right) $ connecting the points 
$\left( x_{i}=0,0\right) $ and $\left( x,t\right) $ can be approximated from
the expression (\textbf{see Caroli}) 
\begin{equation*}
t=\int_{0}^{x}dx^{\prime }\frac{1}{\left[ 2\left( W_{0}\left( x^{\prime
}\right) +E\right) \right] ^{1/2}}-\Theta _{0}\left( x,0\right)
\end{equation*}
where 
\begin{equation*}
\Theta _{0}\left( z,y\right) =\int_{y}^{z}dx^{\prime }\left( \frac{1}{\left|
U_{0}^{\prime }\left( x^{\prime }\right) \right| }-\frac{1}{\left| U^{\prime
}\left( x^{\prime }\right) \right| }\right)
\end{equation*}
The notations are 
\begin{equation*}
W_{0}=\frac{1}{4}U_{0}^{\prime 2}-\frac{D}{2}U_{0}^{\prime \prime }
\end{equation*}
\begin{equation*}
U_{0}^{\prime }\left( x\right) =-\omega _{0}x
\end{equation*}
The last expressions are the harmonic approximation of $V_{0}$ and the
potential around the \emph{unstable} point $x=0$.

The additional term $\Theta \left( x,0\right) $ is the amount of time
accounting for the deviation of $V$ from the parabolic profile, the
anharmonic deviation $V-V_{0}$.

The result for this simpler case is 
\begin{equation*}
P\left( x,t;0,0\right) =\frac{\omega _{0}}{\sqrt{\pi }}\frac{1}{\left|
U^{\prime }\left( x\right) \right| }\sqrt{G}\exp \left( -G\right)
\end{equation*}
where 
\begin{equation*}
G=\frac{\omega _{0}}{D}\frac{x^{2}}{\exp \left[ 2\omega _{0}T_{0}\left(
x\right) \right] -1}
\end{equation*}
\begin{equation*}
T_{0}\left( x\right) \equiv t+\Theta _{0}\left( x,0\right)
\end{equation*}

Two time regimes are intersting. First, the very short time, less than the
period of the particle in the potential, 
\begin{equation*}
0\leq t\lesssim \frac{1}{2\omega _{0}}
\end{equation*}
the result reduces to the harmonic problem around $x=0$.

For longer times, 
\begin{equation*}
\frac{1}{2\omega _{0}}\ll t\lesssim t_{s}
\end{equation*}
the result is 
\begin{equation*}
P\left( x,t;0,0\right) =\frac{1}{\sqrt{\pi }}F^{\prime }\exp \left(
-F^{2}\right)
\end{equation*}
where 
\begin{equation*}
F=\left( \frac{\omega _{0}}{D}\right) ^{1/2}x\,\exp \left[ -\omega
_{0}T_{0}\left( x\right) \right]
\end{equation*}
This is the scaling solution of Suzuki.

\bigskip

The general expression of the function $K$ in the case where we include the
time regimes beyond the limits given above. It has the form of a convolution 
\begin{equation*}
K=\int_{0}^{t}dt_{0}\overset{\cdot }{\overline{x}}_{0}\left( t_{0}\right)
K\left( x,t_{1};x_{m}\right) K\left( x_{m},t_{0};x_{i}\right) 
\end{equation*}
Here the approximations for the ratio $N/\left[ \det \widehat{L}\left(
x_{c}\right) \right] ^{1/2}$ are obtained by the same method as before. The
relation between the energy and the time is used for the two main regions:
around the \emph{stable} and the \emph{unstable} (initial) positions 
\begin{equation*}
t_{0}=\int_{x_{i}}^{x_{m}}dx^{\prime }\frac{1}{\left[ 2\left(
W_{0}+E_{0}\right) \right] ^{1/2}}-\Theta _{0}\left( x_{m},x_{i}\right) 
\end{equation*}
\begin{equation*}
t_{1}=\int_{xm}^{x}dx^{\prime }\frac{1}{\left[ 2\left( W_{1}+E_{1}\right) %
\right] ^{1/2}}-\Theta _{1}\left( x,x_{m}\right) 
\end{equation*}

\end{document}